\documentclass[a4paper, 11pt]{article}
\usepackage[T1]{fontenc}
\usepackage[utf8]{inputenc}
\usepackage[english]{babel}
\usepackage{amsmath}
\usepackage{amssymb}
\usepackage{braket}
\usepackage{mathtools}
\usepackage{dsfont}
\usepackage{subcaption}
\usepackage{array}
\usepackage{booktabs}
\usepackage{yfonts}
\usepackage{setspace}
\usepackage{verbatim}

\usepackage[usenames,dvipsnames]{color}

\numberwithin{equation}{section}

\def\be{\begin{equation}}
\def\ee{\end{equation}}
\def\rme{{\rm e}}
\newcommand{\nn}{\nonumber}
\newcommand{\diff}{\mathrm{d}}

\def\aa{\mathtt{a}}
\def\cc{\mathtt{c}}
\newcommand{\ext}[1]{\mathsf{#1}}
 
\usepackage{jheppub}                   

\makeatletter
\gdef\@fpheader{\ }                    
\makeatother

\title{Higher-derivative corrections to flavoured BPS black hole thermodynamics and holography}

\author[a]
{Davide Cassani,}
\author[c]
{Alejandro Ruip\'erez,}
\author[a,b]
{Enrico Turetta,}
\emailAdd{davide.cassani@pd.infn.it, alejandro.ruiperez@roma2.infn.it, enrico.turetta@phd.unipd.it}

\affiliation[a]{INFN, Sezione di Padova, Via Marzolo 8, 35131 Padova, Italy}
\affiliation[b]{Dipartimento di Fisica e Astronomia ``Galileo Galilei'', Universit\`a di Padova,\\Via Marzolo 8, 35131 Padova, Italy}
\affiliation[c]{Dipartimento di Fisica, Universit\`a di Roma ``Tor Vergata'' \& Sezione INFN Roma 2,\\ Via della Ricerca Scientifica 1, 00133, Roma, Italy}

\abstract{A Cardy-like regime of the four-dimensional superconformal index has been shown to be governed by 't Hooft anomalies and to  single out a large-$N$ saddle carrying the Bekenstein-Hawking entropy of dual supersymmetric black holes in AdS$_5$. For the universal index where no flavour fugacities are turned on, this correspondence has been improved by matching the first subleading corrections to the saddle-point action with the four-derivative corrections to the black hole action in minimal gauged supergravity, as well as the respective corrected entropies. Here, we extend this match by including flavour symmetries. We consider five-dimensional gauged supergravity with vector multiplet and four-derivative couplings, and provide an effective theory reproducing the 't Hooft anomalies of the R- and flavour symmetries of generic holographic superconformal field theories at next-to-leading order in the large-$N$ expansion. Then we focus on a specific model dual to $\mathbb{C}^3/\mathbb{Z}_\nu$ quiver gauge theories, where the 't Hooft anomaly coefficients receive simple but sufficiently generic corrections. In this model, we evaluate the four-derivative corrections to the on-shell action of the supersymmetric multi-charge black hole, showing agreement with the flavoured Cardy-like formula from the index. We give a prediction for the corrected entropy of the supersymmetric black hole and discuss the general validity of our results. Taking the limit of infinite AdS$_5$ radius, we also obtain four-derivative corrections to the action and entropy of supersymmetric asymptotically flat black holes.
 }

\begin{document}

\maketitle

\section{Introduction}

Holography allows us to understand quantum gravity in asymptotically Anti de Sitter (AdS) space via a dual conformal field theory. Recently, starting with \cite{Cabo-Bizet:2018ehj,Choi:2018hmj,Benini:2018ywd}, new evidence has been provided that the supersymmetric quantum gravity partition function with suitable  AdS$_5$ boundary conditions is computed by the superconformal index \cite{Kinney:2005ej,Romelsberger:2005eg} of a dual four-dimensional superconformal field theory (SCFT). In particular, a family of large-$N$ saddles of the index has been shown to correspond to supersymmetric black holes, with the Bekenstein-Hawking entropy being encoded in the saddle-point action.
It is expected that studying the full series of corrections to the large-$N$ saddle should in principle allow to determine the exact quantum black hole entropy.   On the gravity side, these corrections should correspond  to higher-derivative and quantum corrections from supergravity and stringy modes. Determining how precisely these arise and what is their contribution to the complete answer provides an intriguing opportunity to advance our understanding of quantum gravity in the controlled setup of supersymmetric holography.

In the present paper, we make some steps in the direction outlined above: we study the first subleading corrections to black hole thermodynamics  in five-dimensional gauged supergravity including vector multiplet and higher-derivative couplings, and match them holographically. This extends the previous work~\cite{Cassani:2022lrk,Bobev:2022bjm} on minimal gauged supergravity to the case where multiple electric charges are turned on in the solution. While black hole solutions to minimal gauged supergravity are universal (in the sense that they can be embedded in any compactification admitting a supersymmetric AdS$_5$ vacuum, and any holographic SCFT$_4$ with a weakly-coupled gravity dual has a corresponding large-$N$ saddle), solutions carrying multiple electric charges rely on the existence of flavour symmetries in the dual $\mathcal{N}=1$ SCFT and are therefore sensitive to the details of the holographic pair considered.
 
We start our analysis in field theory. We consider a four-dimensional $\mathcal{N}=1$ SCFT on the spatial manifold $S^3$.  We choose a supercharge $\mathcal{Q}$ satisfying the commutation relations 
\be\label{eq:comm_charges}
[J_{1},\mathcal{Q}]  = [J_{2},\mathcal{Q}] =  \frac12  \mathcal{Q}\,,\qquad [Q_I,\mathcal{Q}] = -r_I \mathcal{Q}\,,
\ee
where $J_1$, $J_2$ are the angular momenta generating the Cartan subalgebra of the ${\rm SO}(4)$ isometry of $S^3$, while $Q_I$, $I=1,\ldots,n+1$, are conserved charges made of linear combinations of the superconformal R-charge and the $n$ Abelian flavour charges that the theory may admit. The eigenvalues $r_I$ vanish if $Q_I$ is a flavour symmetry and  take the value $r_I = 1$ if it is a canonically-normalized R-symmetry; they may also be assigned different non-vanishing values. 
The refined superconformal index can be defined as 
\be\label{our_flavored_index}
\mathcal{I} \,=\,   {\rm Tr}\,  \rme^{\pi i (1+n_0) F} \, \rme^{-\beta \{\mathcal{Q},\overline{\mathcal{Q}}\}+
\omega_1  J_1 + \omega_2  J_2 + \varphi^{I} Q_{I} } \,,
\ee
with the constraint
\be\label{eq:linearconstraint_gen} 
\omega_1+ \omega_2  - 2 r_I \varphi^I \,=\, 2\pi i  n_0\,,
\ee 
where the trace is taken over the Hilbert space of the theory on $S^3$, $F$ is the fermion number, $n_0$ is an integer and the complex variables $\beta$, $\omega_1$, $\omega_2$, $\varphi^I$ are chemical potentials for the respective charges. The index $\mathcal{I}$ does not depend on $\beta$ and is a holomorphic function of $\omega_1,\omega_2,\varphi^I$, subject to the constraint \eqref{eq:linearconstraint_gen}. The latter is a supersymmetry condition, ensuring that the combination $\rme^{\pi i n_0 F} \, \rme^{\omega_1  J_1 + \omega_2  J_2 + \varphi^{I} Q_{I} }$ entering in \eqref{our_flavored_index} commutes with the supercharge. The integer $n_0$ was introduced in~\cite{Cabo-Bizet:2018ehj,Cabo-Bizet:2019osg}: although it can be removed by shifting e.g.\ $\omega_1\to\omega_1+2\pi i n_0$ so as to reach the standard definition of the index with a $(-1)^F$ insertion as originally formulated in~\cite{Romelsberger:2005eg,Kinney:2005ej},  we find it convenient to keep it as, when set to $n_0=\pm 1$, it makes it manifest that the index can be obtained as a continuous limit of a non-supersymmetric, thermal partition function lacking the $(-1)^F$ insertion. This will be useful when comparing with the gravitational partition function.


At  large $N$, the superconformal index displays an intricate structure of complex saddles, including one that carries the entropy of the dual supersymmetric black hole in AdS$_5$, which may therefore be called the black hole saddle. This has been verified using different methods (including the Bethe ansatz method where the dominant configurations do not immediately arise as a saddle), see e.g.\ \cite{Benini:2018ywd,GonzalezLezcano:2019nca,Lanir:2019abx,Cabo-Bizet:2019eaf,Cabo-Bizet:2020nkr,Benini:2020gjh,Copetti:2020dil,Cabo-Bizet:2020ewf,Choi:2021rxi,Aharony:2021zkr,Choi:2023tiq}.
A convenient way to  isolate the contribution of the black hole saddle is to take a Cardy-like limit of small chemical potentials  $\omega_1,\omega_2 \to 0$ after setting $n_0=\pm 1$ and before taking the large-$N$ limit.
It has been shown in a number of papers \cite{Choi:2018hmj,Honda:2019cio,ArabiArdehali:2019tdm,Kim:2019yrz,Amariti:2019mgp,Cabo-Bizet:2019osg,Gadde:2020bov,GonzalezLezcano:2020yeb,Goldstein:2020yvj,Amariti:2020jyx,Amariti:2021ubd,Cassani:2021fyv,ArabiArdehali:2021nsx} with progressively increasing accuracy, that the index in this regime is controlled by the cubic and linear 't Hooft anomalies of the SCFT conserved global currents. In this paper we extend the universal Cardy-like formula given in~\cite{Cassani:2021fyv} to the case where flavour chemical potentials are turned on. 
We show that the asymptotic formula for the index in this regime reads
\be\label{eq:logI}
\log \mathcal{I} \,=\, -I + \ldots\,,
\ee
\be\label{eq:index_asympt}
I \, =\, \frac{k_{IJK}\, \varphi^I\varphi^J\varphi^K}{6\,\omega_1\omega_2} - k_I \varphi^I \,\frac{\omega_1^2+\omega_2^2-4\pi^2}{24\,\omega_1\omega_2} \,,
\ee
where the choice $n_0=\pm 1$ has been made in the constraint \eqref{eq:linearconstraint_gen},
\be\label{eq:linearconstraint} 
\omega_1+ \omega_2  - 2 r_I \varphi^I \,=\, \pm 2\pi i \,,
\ee  
and where
\begin{equation}\label{def_anomaly_coeff}
k_{IJK}= {\rm{Tr}}\, (Q_I  Q_J Q_K)\,, \hspace{1cm} k_I={\rm Tr}\, Q_I\, 
\end{equation}
are the cubic and linear 't Hooft anomaly coefficients for the charges $Q_I$. Eq. \eqref{eq:index_asympt} should provide the exact action of the saddle of interest.
The dots in \eqref{eq:logI} indicate that we are omitting terms  that are exponentially suppressed in the limit $\omega_1,\omega_2 \to 0$, as well as a logarithmic term $\log |\mathcal{G}|$, where $|\mathcal{G}|$ is the order of a discrete one-form symmetry $\mathcal{G}$ that the theory may have, which is related to the degeneracy of saddles.
The sign choice in \eqref{eq:linearconstraint} corresponds to two equivalent saddles.

We will give a derivation of \eqref{eq:index_asympt} via equivariant integration of the anomaly polynomial, following the approach of~\cite{Ohmori:2021dzb}. 
The expression is valid at finite $N$ and should apply both to Lagrangian and non-Lagrangian theories, not necessarily holographic.
For a class of holographic theories whose 't Hooft anomaly coefficients satisfy certain requirements that we specify, we are able to evaluate the Legendre transform of~\eqref{eq:index_asympt} at first subleading order in  the large-$N$ limit, obtaining in this way a prediction for the corrected entropy of the dual supersymmetric black hole as a function of the electric charges and angular momenta. 

Among the field  theories falling in the class we consider, there are $\mathcal{N}=4$ SYM with gauge group ${\rm SU}(N)$ and the $\mathcal{N}=1$ $\mathbb{C}^3/\Gamma$ quiver gauge theories, namely the theories describing the low-energy dynamics of D3-branes probing a $\mathbb{C}^3/\Gamma$ singularity, where $\Gamma$ is a discrete subgroup of ${\rm SU}(3)$. For these theories the  corrections to the large-$N$ results are suppressed by a factor of $1/N^2$.  For $\mathcal{N}=4$ SYM the cancellations due to maximal supersymmetry set $k_I=0$ and restrict the correction of $k_{IJK}$  to an $N^2 \to N^2-1$  renormalization of the overall factor (the exact value in our basis for the charges 
 being $k_{IJK}= \frac{N^2-1}{2}|\epsilon_{IJK}|$, $I=1,2,3$), which makes it straightforward to obtain the corrected entropy from \eqref{eq:index_asympt}. However, in the case of $\mathbb{C}^3/\Gamma$ theories both 't Hooft anomaly coefficients $k_{IJK}$ and $k_I$ receive simple but sufficiently generic corrections to the leading-order result, that make these theories an interesting class to study. Focussing on the case $\Gamma=\mathbb{Z}_\nu$, we give our prediction for the corrected entropy of a dual supersymmetric black hole in section~\ref{subsec:entropy_orbifolds}. 

A further motivation for considering the orbifold theories has to do with the gravity side of the correspondence. There, we face the issue that, perhaps surprisingly, very few asymptotically AdS$_5$ black hole solutions carrying multiple electric charges and admitting an uplift to string theory or M-theory are explicitly known, though more are expected to exist. In fact, the only  examples are solutions carrying at most three independent electric charges, uplifting to type IIB supergravity on $S^5$ or its quotients $S^5/\Gamma$,
which are dual to $\mathcal{N}=4$ SYM or the $\mathbb{C}^3/\Gamma$ theories we consider~\cite{Gutowski:2004yv,Cvetic:2004ny,Chong:2005da,Kunduri:2006ek,Mei:2007bn,Wu:2011gq}. 
 
We then turn to supergravity  and aim at obtaining a precise holographic match of the SCFT results described above. In order to determine the suitable supergravity theory in five dimensions, we first fix the Chern-Simons terms that capture the dual SCFT 't Hooft anomalies playing a role in the formula~\eqref{eq:index_asympt}. It is well-known that field theory anomalies arise as boundary terms by varying suitable Chern-Simons terms, and that in holography this is precisely the mechanism by which gravity matches the dual field theory global anomalies~\cite{Witten:1998qj}. 
The Chern-Simons terms that reproduce the cubic and linear 't Hooft anomalies are two-derivative and four-derivative terms, and involve as many vector fields as there are conserved charges. They are proportional to 
\be\label{CSterms_intro}
\ \frac{1}{24\pi^2}\left( k_{IJK} A^I \wedge \diff A ^J \wedge \diff A^K - \frac{1}{8}  k_I A^I \wedge \mathcal{R}_{ab} \wedge \mathcal{R}^{ab} \right)\,,
\ee
where $A^I$ are Abelian gauge fields, $\mathcal{R}_{ab}$ is the Riemann curvature two-form and the dictionary with the supergravity couplings will be specified in section~\ref{sec:holographicdictionary}.
We then look for  the supersymmetrization of~\eqref{CSterms_intro}, that is a four-derivative supergravity coupled to vector multiplets, and implement a gauging of the R-symmetry so that the theory admits an AdS vacuum.
In general, this supergravity theory should be understood neither as a consistent truncation (at least not with the usual definition of consistent truncations as properties of the classical equations of motion, since the higher-derivative terms may be generated quantum mechanically in the compactification), nor as a low-energy effective action (since we are not including all massless modes), it rather is a supersymmetric effective action for the SCFT 't Hooft anomalies.

Following the approach we already adopted in the minimal case~\cite{Cassani:2022lrk}, we start from off-shell supergravity including four-derivative invariants and work at linear order in the couplings governing the latter. This makes it possible to easily eliminate the auxiliary fields by solving algebraic equations of motion (which become dynamical, and thus much harder to solve, in the non-linear theory), and to perform field redefinitions that simplify the resulting Lagrangian. Clearly, in the present case the computations are more complicated than in~\cite{Cassani:2022lrk} due to the many terms involving the scalar fields belonging to the vector multiplets.

Besides these technical complications, we encounter a more fundamental issue: we find that the four-derivative off-shell invariants available in the literature~\cite{Hanaki:2006pj,Ozkan:2013nwa,Ozkan:2016csy} are in general not sufficient to achieve a perfect match with the corrections to the dual 't Hooft anomaly coefficients.
For instance, the issue arises when considering the five-dimensional gravity dual of $\mathcal{N}=4$ SYM, by which here we mean a five-dimensional supergravity theory that at the two-derivative level uplifts to type IIB supergravity on $S^5$ and that reproduces the full  ${\rm U}(1)^3$ 't Hooft anomalies of ${\rm SU}(N)$ $\mathcal{N}=4$ SYM. As already noted, in our basis the corrections   just shift the overall factor in the cubic coefficient $k_{IJK}$ as $N^2\to N^2-1$, however we have found no way to reproduce this starting from the known off-shell invariants.
We overcome this difficulty by proposing a simple modification of the two-derivative theory that does the job (and also reproduces the corrections obtained starting from the known invariants and going on-shell at linear order). 
The supergravity model capturing the anomalies of the orbifold theories, on the other hand, requires two different sets of corrections, which involve genuinely four-derivative terms.

Our results for the bosonic sector of the four-derivative corrections to gauged supergravity coupled to an arbitrary number of vector multiplets is given in  section~\ref{sec:final_Lagr}. Of course, it contains much more than Chern-Simons terms and, as anticipated, it is considerably more involved than the minimal gauged supergravity case we studied previously. We show that it is sufficient for reproducing any 't Hooft anomaly coefficient of a dual SCFT at next-to-leading order in the large-$N$ expansion.  This also allows us to complete the discussion of  \cite{Tachikawa:2005tq,Hanaki:2006pj} for the gravity dual of $a$-maximization by including general next-to-leading order corrections.

Next we specialize to the supergravity model reproducing the\ 't Hooft anomalies of the $\mathbb{C}^3/\Gamma$ quiver theories. 
We compute the corrected supersymmetric black hole on-shell action within this model using the method of~\cite{Cabo-Bizet:2018ehj}, and match it  with the SCFT formula~\eqref{eq:index_asympt}. Due to the intrinsic complication of the calculation, in order to check this equivalence we partially resort to numerics and assume equality of the two a priori independent angular velocities, $\omega_1=\omega_2$. (Since we are already committed to a specific saddle, here we do not need to take a limit of small angular velocities). The result justifies why the entropy obtained in section~\ref{subsec:entropy_orbifolds} from the SCFT formula is a prediction for the corrected black hole entropy.
 Our result generalizes the match obtained at the two-derivative level in \cite{Cassani:2019mms}, as well as the universal four-derivative result of~\cite{Cassani:2022lrk,Bobev:2022bjm}.

We also discuss the ungauged supergravity limit of the corrected black hole on-shell action. In this limit, the two-derivative solution is the asymptotically flat black hole of \cite{Cvetic:1996xz} with three electric charges. We consider its supersymmetric non-extremal version, which admits a regular Euclidean section, and give the expression for its corrected on-shell action  in terms of the supersymmetric chemical potentials. We also take the Legendre transform and obtain the corrected entropy. We briefly comment on its interpretation as the saddle of a gravitational index.

The rest of the paper is organized as follows. In section~\ref{sec:field_th_section} we derive the Cardy-like SCFT formula~\eqref{eq:index_asympt}, while in section~\ref{sec:Legendre_transf_gen} we evaluate its Legendre transform at first subleading order in the large-$N$ expansion, obtaining our prediction for the corrected black hole entropy with flavour charges turned on. In section~\ref{sec:orbifold_section} we apply our results to $\mathcal{N}=1$ orbifold theories. In section~\ref{sec:fourder_action} we present our (bosonic)  Lagrangian for supergravity including vector multiplet and four-derivative couplings. In section~\ref{sec:onshellaction} we specify the supergravity model dual to the orbifold SCFT's and evaluate the corrected  on-shell action of supersymmetric black hole solutions,  matching \eqref{eq:index_asympt}. In section~\ref{sec:ungaugedlimit} we discuss the ungauged limit of the black hole action. We conclude in section~\ref{sec:Conclusions}. In Appendix~\ref{sec:amaxim_section} we discuss $a$-maximization and  its gravity dual, including the corrections and completing the analysis of \cite{Tachikawa:2005tq,Hanaki:2006pj} at the four-derivative level. In Appendices~\ref{app:offshell_invariants} and~\ref{app:field_redefinitions} we give the off-shell supersymmetry invariants and some field redefinitions needed in section~\ref{sec:fourder_action}.

\section{The multi-charge Cardy-like formula from equivariant integration}\label{sec:field_th_section}

In this section we provide a derivation of the flavoured Cardy-like formula \eqref{eq:index_asympt}.
This formula generalizes the expression given in~\cite{Cassani:2021fyv} 
 for the universal case where no flavour chemical potential is turned on.\footnote{It also extends the flavoured formula given in~\cite{Kim:2019yrz} by including the non-divergent, polynomial terms in the regime of small $\omega_1,\omega_2$.
}
  Indeed, \eqref{eq:index_asympt}  maps into the formula given there if one makes the replacement $\varphi^I Q_I \to \varphi Q_R $, where $Q_R$ denotes the R-charge and $\varphi = r_I \varphi^I$ is the R-symmetry chemical potential.\footnote{This replacement is derived by first going to a basis that isolates the R-symmetry from the flavour symmetries. Given any R-charge $Q_R = s^I Q_I$, with $r_Is^I=1$, we can apply the projectors \eqref{projectors_Rsymm} and  obtain the decomposition $Q_I = r_I Q_R + \widetilde{Q}_I$, where the $\widetilde{Q}_I = (\delta_I{}^J - r_I s^J)Q_J$ are flavour charges. Analogously, we can decompose the chemical potentials as $\varphi^I = s^I \varphi + \widetilde\varphi^I$, with  $\varphi = r_I \varphi^I$ and $\widetilde{\varphi}^I = (\delta^I{}_J - s^I r_J)\varphi^J$. It follows that $\varphi^I Q_I = \varphi Q_R + \widetilde \varphi^I \widetilde{Q}_I$. By turning off the flavour potentials, $\widetilde{\varphi}^I= 0$, we obtain the replacement above.\label{foot:change_basis}}
One way to prove~\eqref{eq:index_asympt} is therefore to extend the three-dimensional effective field theory approach of~\cite{Cassani:2021fyv,ArabiArdehali:2021nsx} to the flavoured case: the twisted supersymmetric reduction on a small Euclidean time circle discussed there can also be performed in the presence of background vector multiplets coupling to flavour currents; this leads to additional supersymmetric Chern-Simons contact terms in three dimensions~\cite{Ardehali:2021irq,reftoPieter}.
Here we choose a different route and present a quicker, though more formal, way to reach the same result, which extends the equivariant integration of the anomaly polynomial presented in \cite{Ohmori:2021dzb} (see also \cite{Nahmgoong:2019hko}) to the flavoured case. 
Equivariant integration of anomaly polynomials is a technique that has already proven effective for different scopes, such as  obtaining the anomaly polynomial of lower-dimensional theories~\cite{Benini:2009mz,Alday:2009qq}, or reproducing the supersymmetric Casimir energy~\cite{Bobev:2015kza}. 
Since the extension we present is straightforward, we will focus on the essential steps of the procedure and make a few comments on its rationale, while referring to the above papers for details.\footnote{Recently, starting with~\cite{BenettiGenolini:2023kxp,Martelli:2023oqk}, equivariant integration has  been recognized to also play an important role on the gravity side of supersymmetric holography. We expect it should be possible to derive the formula \eqref{eq:index_asympt} -- which in the present paper will be matched with a supersymmetric black hole on-shell action -- using the techniques introduced there.
}

We place our SCFT in a Euclidean background $\mathcal{M}_4$ comprising Abelian gauge fields $\hat A^I$, $I=1,\ldots,n+1$, coupling to the ${\rm U}(1)^{n+1}$ global currents. The anomaly polynomial $\ext{P}$ is a six-form defined on an extension $\mathcal{Y}_6$ of $\mathcal{M}_4$, that reads: 
\be\label{anomaly_poly}
\ext{P} \,=\, \frac{1}{6}\,{\rm Tr}\,\bigg( \frac{\ext{F}}{2\pi}\bigg)^3 - \frac{1}{24}\,  {\rm Tr}\,\bigg( \frac{\ext{F}}{2\pi}\bigg) \, \ext{p}_1(T\mathcal{Y}_6)\,,
\ee 
where $\ext{F} = \diff \ext{A}$ and $\ext{A} = \ext{A}^I Q_I$ is the extension to $\mathcal{Y}_6$ of the  ${\rm U}(1)^{n+1}$   connection $\hat A = \hat A^IQ_I$ on $\mathcal{M}_4$,
 while
 \be
   \ext{p}_1 = -\frac{1}{8\pi^2} \ext{R}_{ab}\wedge \ext{R}^{ba}
 \ee 
 is the first Pontryagin class defined out of the Riemann curvature two-form $\ext{R}_{ab}$ of $\mathcal{Y}_6$.

We next specify the essential features of the background of interest to us, namely its topology and the group action. We take a space that topologically is $\mathcal{M}_4= S^3\times S^1$. 
 Then we choose a smooth six-dimensional extension with topology $\mathcal{Y}_6 = \mathcal{N}_4 \times D_2$, such that 
   $\partial \mathcal{N}_4 = S^3$ and $\partial D_2 = S^1$. $D_2$ is given the shape of a cigar.
  The gauge fields $\ext{A}^I$ also need to be regular on $\mathcal{Y}_6$. In addition to the ${\rm U}(1)^{n+1}$ symmetry bundle, we consider the ${\rm U}(1)^2$ isometries corresponding to rotation in $S^3$ and the ${\rm U}(1)_{\mathsf T}$ isometry along $S^1$ (where ``${\mathsf T}$'' stands for ``thermal''); the circles defined by the orbits of the respective Killing vectors are non-trivially fibred, and the connection of the fibration contributes to the curvature two-form $\ext{R}_{ab}$. Overall, we thus have a ${\rm U}(1)^2\times {\rm U}(1)_{\mathsf T}\times {\rm U}(1)^{n+1}$ group action on $\mathcal{Y}_6$. We assume that $\mathcal{Y}_6$ can be chosen so that the only fixed point of this action is at the origin of the six-dimensional space; this is possible because $\mathcal{Y}_6$ has {\it two} dimensions more than $\mathcal{M}_4$, so that both $S^3$ and $S^1$ can be made cobordant to the empty set.

The final step is to implement equivariant integration of the anomaly polynomial on $\mathcal{Y}_6$. In order to do so, we assume we can promote the characteristic classes appearing in~\eqref{anomaly_poly} to equivariant classes with respect to the group action. Then we associate complex equivariant parameters $\omega_1,\omega_2$ to the ${\rm U}(1)^2$ rotations, $\mu_{\mathsf T}$ to the ${\rm U}(1)_{\mathsf T}$ shifts and $\varphi^I$ to the ${\rm U}(1)^{n+1}$ action. 
It follows that the Killing vector $K_{\rm eq}$ appearing in the equivariant differential $\diff+ 2\pi \, \iota_{K_{\rm eq}}$ reads
$K_{\rm eq} = \mu_{\mathsf T} \partial_\tau+\omega_1\partial_{\varphi_1} + \omega_2\partial_{\varphi_2}$, where $\partial_{\varphi_1}$, $\partial_{\varphi_2}$ generate the ${\rm U}(1)^2$ rotations in $S^3$ while $\partial_\tau$ advances the coordinate $\tau$ parameterizing $S^1$; here all angular coordinates are taken $2\pi$-periodic. The idea, that it would be nice to establish more rigorously, is that the existence of the equivariant action is a consequence of supersymmetry. This implies that we should identify the vector $K_{\rm eq}$ specifying the equivariant action with the Killing vector $K$ obtained by taking suitable bilinears of the Killing spinor ensuring supersymmetry of the background. Up to an irrelevant proportionality constant, the supersymmetric Killing vector in the background of interest reads $K = - 2\pi i\, \partial_\tau+\omega_1\partial_{\varphi_1} + \omega_2\partial_{\varphi_2}$, where $\omega_1,\omega_2$ are precisely the chemical potentials appearing in the definition of the superconformal index~\cite{Cabo-Bizet:2018ehj,Cassani:2021fyv}. Therefore we see that we should fix $\mu_{\mathsf T}=- 2\pi i$, while the remaining equivariant parameters are identified with the chemical potentials appearing in the superconformal index.

We then apply the Atiyah-Bott-Berline-Vergne fixed point theorem, stating that the integral of an equivariantly closed form only receives contributions from the fixed points of the group action, to evaluate the integral
\be\label{integration_anomaly_p}
I_{\rm eq}\equiv -2\pi i \int_{\mathcal{Y}_6}\ext{P} \,=\, -2\pi i\, \frac{\ext{P}|_0}{\ext{e}(T\mathcal{Y}_6)|_0}\,,
\ee
where $\ext{e}(T\mathcal{Y}_6)$ is the equivariant Euler class of $\mathcal{Y}_6$ and $|_0$ denotes the zero-form contribution of the equivariant class at the fixed point.
Close to the fixed point, $\mathcal{Y}_6$ can be modelled as $\mathbb{R}^6$, with the ${\rm U}(1)^2\times {\rm U}(1)_{\mathsf T}$ action rotating the three orthogonal planes. We can then evaluate the equivariant classes in~\eqref{anomaly_poly} using the standard moment map and symplectic form on $\mathbb{R}^6$ (see e.g.~\cite[App.\:A]{Bobev:2015kza}).
This boils down to replacing the Chern roots of the characteristic classes with the equivariant parameters, i.e.\ implementing the rules
\be
{\rm Tr}\,\bigg( \frac{\ext{F}}{2\pi}\bigg)^3\,\bigg|_0\,=\,  k_{IJK}\,\varphi^I\varphi^J\varphi^K\,,  \qquad\quad {\rm Tr}\,\bigg( \frac{\ext{F}}{2\pi}\bigg)\bigg|_0\,=\,  k_I\varphi^I \,,
\ee
\be
\ext{p}_1(T\mathcal{Y}_6)|_0 \,=\, \omega_1^2+\omega_2^2+\mu_{\mathsf T}^2\,,\qquad\quad  \ext{e}(T\mathcal{Y}_6)|_0 \,=\,  \omega_1\omega_2\mu_{\mathsf T}\,,
\ee
where we used the definitions~\eqref{def_anomaly_coeff} of the 't Hooft anomaly coefficients.
Plugging this in \eqref{integration_anomaly_p} and recalling that we are setting $\mu_{\mathsf T} = -2\pi i$, we find that $I_{\rm eq}$ precisely reproduces the expression for $I$ in \eqref{eq:index_asympt}. 
The requirement that the Killing spinor on $\mathcal{M}_4$ extends to a well-defined spinor on $\mathcal{Y}_6$  (that in particular is anti-periodic at the tip of the cigar $D_2$) leads to the constraint~\eqref{eq:linearconstraint}, the argument being analogous to the one given in~\cite{Cabo-Bizet:2018ehj} for the five-dimensional supergravity bulk filling of $\mathcal{M}_4$. This concludes the derivation.

\section{Corrected entropy via Legendre transform}\label{sec:Legendre_transf_gen}

In this section, we obtain a prediction for the corrected entropy of supersymmetric multi-charge AdS${}_5$ black holes by taking the Legendre transform of the formula for $I$ given in \eqref{eq:index_asympt}. Doing this in full generality is a difficult task, hence we make some convenient assumptions on the 't Hooft anomaly coefficients that we specify next.

\subsection{Assumptions}\label{sec:assumptions}

The formula \eqref{eq:index_asympt}  holds at finite $N$ and independently of whether the SCFT is holographic. However, for a holographic theory we can study it in the large-$N$ expansion. In this paper we will focus on (the leading and) the next-to-leading terms in the large-$N$ expansion, assuming the theory has a weakly-coupled holographic dual. Then we can write 
\be\label{eq:expansion_k's}
k_{IJK} = k^{(0)}_{IJK}+k^{(1)}_{IJK} + \ldots\,, \qquad\ k_I = 0 + k^{(1)}_I +\ldots\,, 
\ee
where the dots denote possible higher-order terms.
 At leading-order, eq.~\eqref{eq:index_asympt} reduces to
\be\label{eq:index_asympt_leading}
I^{(0)} \, =\, \frac{k^{(0)}_{IJK}\, \varphi^I\varphi^J\varphi^K}{6\,\omega_1\omega_2} \,,
\ee
where $k^{(0)}_{IJK}$ denotes the leading-order cubic 't Hooft anomaly. This agrees with a number of existing leading-order results, starting from the conjectured formula in Appendix~A of~\cite{Hosseini:2018dob}.

Our first assumption, that has already been used in~\cite{Cassani:2019mms}, regards the leading-order cubic anomaly $k^{(0)}_{IJK}$. We assume the existence of a constant fully-symmetric tensor $k^{(0)}{}^{IJK}$ such that
\begin{equation}\label{eq:cubic_relation}
k^{(0)}{}^{IJK}k^{(0)}_{J(LM}k^{(0)}_{NP)K}=\gamma \, \delta^{I}_{(L}k^{(0)}_{MNP)}\,,
\end{equation}
where $\gamma$ is some coefficient. Without loss of generality, we can fix the convenient normalization
\be\label{eq:normalization_k}
8k^{(0)}{}^{IJK}r_Ir_J r_K =1\,.
\ee
Then one can prove that\footnote{The proof goes as follows. Contracting~\eqref{eq:cubic_relation} with $r_I\bar s^L\bar s^M\bar s^N\bar s^P $ and decomposing the indices along the R-symmetry and the flavour symmetries by means of the projections~\eqref{projectors_Rsymm}, one obtains
$$
\gamma \,=\, \frac{1}{8}\, k^{(0)}_{RRR} + 2\, r_I r_J\, k^{(0)IJ L} \,\widetilde k^{(0)}_{L RR}  +  \frac{r_I \,k^{(0)I  J L} \,\widetilde k^{(0)}_{ J RR}\, \widetilde k^{(0)}_{ L RR}}{k^{(0)}_{RRR}}\,,
$$
where the index $R$ denotes the projection along the R-symmetry while $\widetilde k_{IRR}$  has the  $I$ index projected on the flavour directions, see Appendix~\ref{amax_at_largeN} for details. Around eq.~\eqref{eq:extremization_a_leading_order} we also show that $ \widetilde k^{(0)}_{I RR}  = 0$ for holographic theories with a weakly-coupled gravity dual, which leads us to~\eqref{expr_gamma}.} 
\be\label{expr_gamma}
\gamma\,=\, \frac{1}{8}k^{(0)}_{RRR} \,=\, \frac{4}{9}\, \aa^{(0)}\,,
\ee
where $\aa^{(0)}$ is the leading-order term of the Weyl anomaly coefficient $\aa$.
 We should note that property \eqref{eq:cubic_relation} is non-generic, for instance the cubic 't Hooft anomalies for the four global symmetries of the conifold theory do not satisfy it~\cite{Amariti:2019mgp}. For the five-dimensional supergravity which matches these global anomalies holographically, \eqref{eq:cubic_relation} holds when the scalar manifold is a symmetric space~\cite{Gunaydin:1983bi}.

Our second assumption regards the corrections: we assume the following relation between the cubic and linear coefficients,
\be\label{relation_k3_k1}
k_{IJK} \,=\,   k^{(0)}_{IJK}  + k_{(I} r_J r_{K)}  \,.
\ee
This condition implies a relation between the first-order corrections to the cubic and linear 't Hooft anomaly coefficients for the superconformal R-symmetry, $k^{(1)}_{RRR}=k^{(1)}_R$ (in order to see this one has to use $a$-maximization, see Appendix~\ref{amax_at_largeN}).
 The condition is satisfied quite generally by the four-dimensional $\mathcal{N}=1$ quiver gauge theories which describe D3-branes probing the tip of a Calabi-Yau conical singularity, whose gravity dual is given by type IIB string theory on the Sasaki-Einstein base of the Calabi-Yau cone.  
 These are $\mathcal{N}=1$ quiver gauge theories made of $\nu$ SU$(N)$ nodes connected by chiral superfields, and \eqref{relation_k3_k1} holds as long as there are bifundamental chiral fields but {\it no} adjoint ones.  
 In order to see this, we note that for these theories the 't Hooft anomaly coefficients \eqref{def_anomaly_coeff} are made of a $\mathcal{O}(N^2)$ term and a $\mathcal{O}(1)$ term (that is, a term independent of $N$). Given the dimensions of the respective representations,  bifundamental fermions contribute with $N^2$ to the anomaly while adjoint fermions contribute with $N^2-1$. If no matter fields transform in the adjoint, then only the gaugini contribute to the $\mathcal{O}(1)$ correction. Note that the gaugino has charge $r_I$ under $Q_I$, that is the same charge as the supersymmetry parameter.
Then the form of the 't Hooft anomaly coefficients for these theories is:
\be\label{eq:exp_anomaly_coeff}
k_{IJK}  \,=\,  k^{(0)}_{IJK}  - \nu \, r_I r_J r_K \,,\qquad
k_I  \,=\,  - \nu\, r_I\,,
\ee
where the explicit expression of the leading $\mathcal{O}(N^2)$ cubic coefficient $k^{(0)}_{IJK}$ depends on the details of the quiver.
 On the other hand, it is a general fact that for quiver gauge theories of the type considered, $k_I$ contains no $\mathcal{O}(N^2)$ term provided the symmetry is non-anomalous, namely the $Q_I$-gauge-gauge anomaly vanishes as we assume here.\footnote{The argument is reviewed e.g.\ in Appendix B of~\cite{Cabo-Bizet:2020nkr}; there it is given for an R-symmetry but extension to flavour symmetries is straightforward. When the quivers contains adjoint chiral superfields, 
the corrections to the anomaly coefficients are less universal as they depend on the charges of the adjoint fields: one has
$$
\begin{aligned}
k_I \,&=\, -\nu r_I - \sum_{\alpha\in{\rm adjoints}} (q_{I,\alpha}-r_I)\,,\\
k_{IJK} \,&=\, k_{IJK}^{(0)} - \nu\, r_Ir_Jr_K - \sum_{\alpha\in{\rm adjoints}} (q_{I,\alpha}-r_I)(q_{J,\alpha}-r_J)(q_{K,\alpha}-r_K)\,,
\end{aligned}
$$ 
where $q_{I,\alpha}$ is the charge of the adjoint chiral superfield $\alpha$ under $Q_I$.
In this case we do not have a general relation between $k_I$ and the $\mathcal{O}(1)$ term in $k_{IJK}$.
So we would have to resort to a case by case analysis. In this paper will just discuss the case of $\mathcal{N}=4$ SYM, which is very simple and does not require a general formulation. Other examples of theories with adjoint matter fields and a known gravity dual are the $\mathbb{C}^3/\mathbb{Z}_2$ $\mathcal{N}=2$ orbifold theory, the Suspended Pinch Point (SPP) and the $L^{a,b,a}$ class of quivers.
}
Clearly, \eqref{eq:exp_anomaly_coeff} implies \eqref{relation_k3_k1}.
Note that these corrections can be understood as the consequence of decoupling a ${\rm U}(1)$ vector multiplet at each node while passing from the $({\rm U}(N))^{\nu}$ gauge theory (which would give no corrections since both the adjoint and bifundamental representations have dimension $N^2$) to the $({\rm SU}(N))^{\nu}$ theory in the infrared.

Projecting onto the R-symmetry as discussed in Appendix~\ref{amax_at_largeN}, we obtain the corrections to the R-symmetry anomaly coefficients, $k_{RRR} = k_{RRR}^{(0)}-\nu$, $k_R = -\nu$. Then, recalling the relation between the R-symmetry anomaly coefficients and the Weyl anomaly coefficients $\aa$, $\cc$ given in~\eqref{relacTrR}, we find the corrections to the latter:
\be\label{Weyl_coeffs_gen_quivers}
{\mathtt a}\,=\,\aa^{(0)} -\frac{3\nu}{16}\,,  \qquad  {\mathtt c}\,=\,\aa^{(0)} -\frac{\nu}{8}\,,
\ee
where by $\aa^{(0)} = \cc^{(0)}$ we denote the leading-order term in the large-$N$ expansion.

In the following we keep calling $\nu$ the parameter controlling the corrections. For  the class of quivers specified above, $\nu$ denotes the number of gauge groups. More generally, we impose \eqref{relation_k3_k1} and denote $\nu=-k^{(1)}_{R}=-k^{(1)}_{RRR}\,$.

\subsection{Legendre transform}

The Legendre transform consists of the extremization principle
\begin{equation}\label{eq:Legendretransform}
\mathcal{S}= {\rm{ext}}_{\left\{\varphi^I,\, \omega_{1}, \,\omega_{2},\, \Lambda\right\}}\left[-I-\omega_1 J_1-\omega_2 J_2-\varphi^I Q_I -\Lambda \left(\omega_1+\omega_2-2r_I \varphi^I- 2\pi i\right)\right]\, ,
\end{equation}
which gives the microcanonical form of the entropy (namely, the entropy as a function of the charges and angular momenta). This will be evaluated by extending to the present higher-derivative case the method of  \cite{Hosseini:2017mds,Cabo-Bizet:2018ehj,Cassani:2019mms}. 
The extremization equations are
\begin{equation}\label{eq:extremization_eqs}
-\frac{\partial I}{\partial \varphi^I}=Q_I-2 r_I \Lambda\,, \hspace{1cm} -\frac{\partial I}{\partial \omega_{1}}=J_{1}+\Lambda\,, \hspace{1cm} -\frac{\partial I}{\partial \omega_{2}}=J_{2}+\Lambda\,, 
\end{equation}
together with the linear constraint~\eqref{eq:linearconstraint} which follows from the variation with respect to the Lagrange multiplier $\Lambda$. For definiteness we have made the upper sign choice in \eqref{eq:linearconstraint}; making the other choice leads essentially to the same computations, in particular it gives the same reality condition for the entropy and the same final expression for it.
It is convenient for our purposes to rewrite the expression \eqref{eq:index_asympt} for $I$ using the linear constraint \eqref{eq:linearconstraint} in such a way that it reads:
\begin{equation}\label{eq:rewriting_index}
I= \frac{\left(k_{IJK}-k_Ir_Jr_K\right)\varphi^I \varphi^J \varphi^K} {6\, \omega_1 \omega_2}-\frac{k_I\varphi^I}{12}\left(\frac{\omega_1}{\omega_2}+\frac{\omega_2}{\omega_1}+1\right) \,+\frac{k_Ir_J\varphi^I\varphi^J}{6}\left(\frac{1}{\omega_1}+\frac{1}{\omega_2}\right) ,
\end{equation}
since now it is manifestly a homogeneous function of degree one with respect to $\varphi^I$, $\omega_1$, $\omega_2$. Euler's theorem then implies that the entropy is simply given by the extremum value of the Lagrange multiplier,
\begin{equation}\label{eq:Legendre}
\mathcal{S}=2\pi i \Lambda|_{\rm{ext}}\, .
\end{equation}

Next we use our assumption \eqref{relation_k3_k1}.  
 Remarkably, this implies that the ${\cal O}(1)$ corrections in the first term of \eqref{eq:rewriting_index} cancel.
Then the effective action becomes:
\begin{equation}\label{eq:index_quiver}
I= \frac{k^{(0)}_{IJK}\varphi^I \varphi^J \varphi^K} {6 \,\omega_1 \omega_2}-\frac{k_I\varphi^I}{12}\left(\frac{\omega_1}{\omega_2}+\frac{\omega_2}{\omega_1}+1\right) \,+\frac{k_Ir_J\varphi^I\varphi^J}{6}\left(\frac{1}{\omega_1}+\frac{1}{\omega_2}\right) \,.
\end{equation}

\paragraph{Leading contribution to the entropy.}
As a useful warm-up, we start by recalling how the Legendre transform  is implemented at leading-order in the large-$N$ expansion~\cite{Cassani:2019mms}. We then consider  \eqref{eq:index_asympt_leading}.
  Using our assumption \eqref{eq:cubic_relation} on the leading-order coefficients, it is not hard to see that it satisfies
\begin{equation}\label{eq:I}
k^{(0)}{}^{IJK}\frac{\partial I^{(0)} }{\partial \varphi^I}\frac{\partial I^{(0)} }{\partial \varphi^J}\frac{\partial I^{(0)} }{\partial \varphi^K}- 2\aa^{(0)}  \frac{\partial I^{(0)} }{\partial \omega_1}\frac{\partial I^{(0)} }{\partial \omega_2}=0\, .
\end{equation}
After using the extremization equations \eqref{eq:extremization_eqs}, this becomes a polynomial equation for $\Lambda$:
\begin{equation}\label{eq:Lambda}
\Lambda^3+p_2 \Lambda^2+p_1 \Lambda+p_0=0\, ,
\end{equation}
where 
\begin{equation}\label{eq:p_coeffs}
\begin{aligned}
p_2\,&=\, -12 k^{(0)}{}^{IJK}r_Ir_J Q_K- 2\aa^{(0)} \, ,\\[1mm]
p_1\,&=\, 6 k^{(0)}{}^{IJK}r_IQ_J Q_K-    2\aa^{(0)}  \left(J_1+J_2\right)\, ,\\[1mm]
p_0\,&=\, -k^{(0)}{}^{IJK}Q_IQ_J Q_K- 2\aa^{(0)} J_1J_2\, .
\end{aligned}
\end{equation}
From \eqref{eq:Legendre} we see that for the entropy to be real, $\Lambda$ must be a purely imaginary number. This implies a condition on the charges and angular momenta. In terms of the coefficients of the polynomial equation, such condition reads
\begin{equation}
 p_0 = p_1 p_2\,.
\end{equation}
Then \eqref{eq:Lambda} factorizes as
\begin{equation}
 \left(\Lambda^2+p_1\right)\left(\Lambda+ p_2\right)=0\,.
\end{equation}
Taking the purely imaginary root $\Lambda=-i \sqrt{p_1}$ (assuming $p_1>0$), we find that the supersymmetric extremal entropy is given by
\begin{equation}\label{eq:BPS_entropy_0thorder}
\mathcal{S}^{(0)}\,=\, 2\pi \sqrt{6 k^{(0)}{}^{IJK}r_IQ_J Q_K- 2\aa^{(0)}\left(J_1+J_2\right)}\, .
\end{equation}

\paragraph{First-order corrections.}
Now we perform the Legendre transform keeping the corrections in \eqref{eq:index_quiver}. We work at linear order in $k_I$. In this approximation, it is possible to check that the corrections to \eqref{eq:I} are the following,
\begin{equation}\label{eq:I_corrected}
\begin{aligned}
&\left(k^{(0)}{}^{IJK}\frac{\partial I}{\partial \varphi^I}\frac{\partial I}{\partial \varphi^J}\frac{\partial I}{\partial \varphi^K}- 2 \aa^{(0)} \frac{\partial I}{\partial \omega_1}\frac{\partial I}{\partial \omega_2}\right)\frac{\partial I}{\partial \omega_1}\frac{\partial I}{\partial \omega_2} + \frac{1}{4}k^{(0)}{}^{IJK}\frac{\partial I}{\partial \varphi^I}\frac{\partial I}{\partial \varphi^J}k_K\left[3\frac{\partial I}{\partial \omega_1}\frac{\partial I}{\partial \omega_2}+\right.\\[1mm]
&+\,\left(\frac{\partial I}{\partial \omega_1}-\frac{\partial I}{\partial \omega_2}\right)^2+\frac{3}{\aa^{(0)}   }k^{(0)}{}^{LMN}\frac{\partial I}{\partial \varphi^L}\frac{\partial I}{\partial \varphi^M}r_N \left(\frac{\partial I}{\partial \omega_1}+\frac{\partial I}{\partial \omega_2}\right)\bigg]=0\, .
\end{aligned}
\end{equation}
As before it boils down to a polynomial equation for $\Lambda$, now of order five:
\begin{equation}\label{eq:Lambda2}
\begin{aligned}
{\cal P}_{5}\left(\Lambda\right)\equiv&\left(\Lambda+J_1\right)\left(\Lambda+J_2\right)\left(\Lambda^3+p_2 \Lambda^2+p_1 \Lambda+p_0\right)+  \frac{1}{4} {\cal P}_2\left(\Lambda; k_I\right)\Big[3 \left(\Lambda+J_1\right)\left(\Lambda+J_2\right)\\[1mm]
& + \left(J_1-J_2\right)^2-\frac{3}{\aa^{(0)}}{\cal P}_2\left(\Lambda; r_I\right)\left(2\Lambda+J_1+J_2\right)\Big]=0\,,
\end{aligned}
\end{equation}
where $p_0, p_1, p_2$ are still given by \eqref{eq:p_coeffs} and we have introduced
\begin{equation}
{\cal P}_2\left(\Lambda; v_I\right)=k^{(0)}{}^{IJK}  v_I\left(2r_J\Lambda -Q_J\right)\left(2r_K\Lambda -Q_K\right)\,.
\end{equation}
From now on we specify the $k_I$ as in the second of \eqref{eq:exp_anomaly_coeff}. Though this is not really necessary in order to work out the Legendre transform, it makes the final expressions slightly simpler.  Then ${\cal P}_5\left(\Lambda\right)$ in \eqref{eq:Lambda2} becomes
\begin{equation}
\begin{aligned}
{\cal P}_{5}\left(\Lambda\right)\equiv&\left(\Lambda+J_1\right)\left(\Lambda+J_2\right)\left(\Lambda^3+p_2 \Lambda^2+p_1 \Lambda+p_0\right)- \frac{\nu}{4}\,{\cal P}_2\left(\Lambda; r_I\right)\left[3 \left(\Lambda+J_1\right)\left(\Lambda+J_2\right)\right.\\[1mm]
&\left.+ \left(J_1-J_2\right)^2-\frac{3}{\aa^{(0)}}{\cal P}_2\left(\Lambda; r_I\right)\left(2\Lambda+J_1+J_2\right)\right]=0\,.
\end{aligned}
\end{equation}
We also note that
\begin{equation}
{\cal P}_2\left(\Lambda; r_I\right)=\frac{1}{2}\Lambda^2+ \frac{1}{3}\left(p_2 + 2\aa^{(0)}\right)\Lambda+\frac{p_1}{6}+\frac{1}{3}\aa^{(0)} \left(J_1+J_2\right)\, .
\end{equation}
As before, in order to obtain a real entropy, we must impose the factorization of the polynomial, which in this case takes the form
\begin{equation}\label{eq:factorization}
{\cal P}_{5}\left(\Lambda\right)=\left(\Lambda^2+X\right)\left(Y_0+Y_1 \Lambda +Y_2 \Lambda^2+Y_3\Lambda^3\right)\, ,
\end{equation}
where $X, Y_{0}, Y_1, Y_2$ and $Y_3$ are just coefficients. This factorization translates into a condition on the coefficients of ${\cal P}_{5}\left(\Lambda\right)$, and eventually on the charges $Q_I$ and angular momenta $J_1$,$J_2$.

\paragraph{Solution in the $J_1=J_2$ case.} Let us illustrate in detail the case where there is only one independent angular momentum, $J_1=J_2\equiv J$, as this will be the case for which we will actually calculate the on-shell action on the gravity side. We note that in this case the polynomial \eqref{eq:Lambda} factorizes as ${\cal P}_5(\Lambda)=\left(\Lambda+J\right){\cal P}_{4}(\Lambda)$, where
\begin{equation}
{\cal P}_4\left(\Lambda\right)=\left(\Lambda+J\right)\left(\Lambda^3+p_2 \Lambda^2+p_1 \Lambda+p_0\right)-\nu\, {\cal P}_2\left(\Lambda; r_I\right)\left[\frac{3}{4}\left(\Lambda+J\right)-\frac{3}{2\aa^{(0)}}{\cal P}_2\left(\Lambda; r_J\right)\right]\,.
\end{equation}
So the factorization condition can be written as
\begin{equation}
{\cal P}_{4}\left(\Lambda\right)=\left(\Lambda^2+X\right)\left(Y_0+Y_1 \Lambda +Y_2 \Lambda^2\right)\, .
\end{equation}
Comparing the last two expressions and working perturbatively, one finds the solution 
\be
\begin{aligned}
Y_2\,&=\, 1+\frac{3\nu }{8 \aa^{(0)}}\,,\\[1mm]
Y_1\, &=\, p_2+ J  +\frac{\nu}{2} \left(\frac{5}{4}+\frac{p_2}{\aa^{(0)}}\right)\,,\\[1mm]
Y_0 \, &=\, J p_2+   \frac{\nu}{8}   \left[ 5 J +\frac{p_1^2+4 J^2 p_2^2+J p_1\left(8p_2-3J  \right)}{3 \aa^{(0)} \left(p_1+J^2  \right)}\right]\,,\\[1mm]
\label{eq:X_J1=J2}
X\,&=\, p_1 + \frac{\nu}{6} \left[\aa^{(0)} + \frac{5}{2} p_2 -\frac{ p_1\left(p_1 -p_2^2+2 J p_2 \right)}{ \aa^{(0)}   \left(p_1+J^2  \right)}\right]\,,
\end{aligned}
\ee
and the factorization condition leading to the non-linear constraint among the charges reads
\begin{equation}\label{eq:nonlinearconstraint_J1=J2}
p_0 -p_1p_2\,=\,\frac{\nu}{6} \left[\frac{5}{2}\left(p_1 +p_2^2\right)+ \aa^{(0)}\left(p_2-J  \right)+\frac{ p_1\left(p_2-J  \right)\left(p_1 + p_2^2  \right)}{\aa^{(0)}  \left(p_1+J^2  \right)}\right]\,.
\end{equation}
From the expression for $X$ we immediately obtain the entropy:
\be\label{eq:entropy_equalJ}
\mathcal{S}\,=\, 2\pi \sqrt{p_1 + \frac{\nu}{12} \left[2 \aa^{(0)}+ 5 p_2 -\frac{2 p_1\left(p_1 -p_2^2+2 J p_2 \right)}{\aa^{(0)}   \left(p_1+J^2  \right)}\right]}\,.
\ee
We emphasize that this expression can only be trusted at linear order in the correction, even if we have not explicitly linearized the square root (the reason for not doing so being that the way it is derived suggests that the form $\mathcal{S} = 2\pi i\Lambda = 2\pi \sqrt{X}$ of the entropy may hold beyond linear order).

\paragraph{General solution $J_1\neq J_2$.} In the case of two unequal angular momenta, the expression for the entropy receives additional corrections and reads
\begin{equation}\label{eq:entropy_J1neqJ2}
\begin{aligned}
\mathcal{S}&=\,2\pi\sqrt{p_1}\left\{1+\frac{\nu}{24}\left[\frac{\left(p_{1}+J_{+}^2\right)\left(\aa^{(0)} \left(p_{1}+J_{+}^2\right)\left(2\aa^{(0)}+5 p_2\right)-2p_1\left(p_1-p_2^2+2 J_+ p_2\right)\right)}{{\aa^{(0)}}p_1\left[p_1+\left(J_++J_-\right)^2\right]\left[p_1+\left(J_+-J_-\right)^2\right]}\right.\right.\\[1mm]
&\left.\left.+\,\frac{J_-^2\left[\left(p_2+2\aa^{(0)}\right)\left(2p_1\left(p_2+\aa^{(0)}+2J_+\right)+\aa^{(0)} J_-^2\right)-2\aa^{(0)} J_+^2\left(3p_2+ 2\aa^{(0)}\right)-2p_1^2\right]}{{\aa^{(0)}}p_1\left[p_1+\left(J_++J_-\right)^2\right]\left[p_1+\left(J_+-J_-\right)^2\right]}\right]\right\},
\end{aligned}
\end{equation}
where $J_{\pm}=\frac{J_1\pm J_2}{2}$. In turn, the non-linear constraint reads
\begin{equation}\label{eq:nonlinearconstraint_J1neqJ2}
\begin{aligned}
p_0-p_1p_2=\,&\frac{\nu}{6}\left\{\aa^{(0)}\left(p_2-J_+\right)+\frac{\left(p_1+p_2^2\right)\left(p_1+J_+^2\right)\left[5 \aa^{(0)}\left(p_1+J_+^2\right)+2p_1\left(p_2-J_+\right)\right]}{2\, \aa^{(0)} \left[p_1+\left(J_++J_-\right)^2\right]\left[p_1+\left(J_+-J_-\right)^2\right]}\right.\\[1mm]
&\quad\left.+\,\frac{\left(p_1+p_2^2\right)J_-^2\left[ \aa^{(0)} \left(6p_1-6J_+^2+J_-^2\right)+2p_1(p_2+J_+)\right]}{2\, \aa^{(0)} \left[p_1+\left(J_++J_-\right)^2\right]\left[p_1+\left(J_+-J_-\right)^2\right]}\right\}\,.
\end{aligned}
\end{equation}

\paragraph{Recovering the universal case.} 

As a sanity check, we verify that the results above are in agreement with those obtained in \cite{Cassani:2022lrk,Bobev:2022bjm} for the universal case where all flavour charges are turned off and one is left only with the R-charge $Q_R$ and  the angular momenta $J_1,J_2$. 
This case is reached by setting $Q_I = r_I Q_R$ for all $I$'s, see footnote~\ref{foot:change_basis}.
Then, recalling \eqref{eq:normalization_k},
the coefficients \eqref{eq:p_coeffs} reduce to
\begin{equation}\label{eq:coeffs_minimalcase}
\begin{aligned}
p_2\,&=\, -\frac{3}{2}Q_R-   2\aa^{(0)} \,, \\[1mm]
p_1\,&=\, \frac{3}{4}Q_R^2 - 2\aa^{(0)} \left(J_1+J_2\right)\,, \\[1mm]
p_0\,&=\, -\frac{1}{8}Q_R^3-  2\aa^{(0)} J_1 J_2\,. \\[1mm]
\end{aligned}
\end{equation}
Plugging these into \eqref{eq:entropy_J1neqJ2} and using the non-linear constraint \eqref{eq:nonlinearconstraint_J1neqJ2}, one can verify that the expression for the entropy reduces to
\begin{equation}
\begin{aligned}
\mathcal{S}\,
&=\, \pi \sqrt{3Q_R^2-8\aa \left(J_1+J_2\right)-16\aa^{(0)} \left(\aa-\cc\right)\frac{\left(J_1-J_2\right)^2}{Q_R^2-2\aa^{(0)}\left(J_1+J_2\right)}}\,,
\end{aligned}
\end{equation}
where we have used \eqref{Weyl_coeffs_gen_quivers} to introduce the corrected Weyl anomaly coefficients.
This agrees with our previous results \cite{Cassani:2022lrk}. The non-linear constraint \eqref{eq:nonlinearconstraint_J1=J2} also reduces correctly to the one given there.

\section{Application to $\mathcal{N}=1$ orbifold theories}\label{sec:orbifold_section}

Given the general field theory results discussed in the previous section, we would now like to write down explicit expressions in some concrete examples, to be then studied on the gravity side. 
A limitation with matching these results with a dual gravitational computation is represented by the restricted number of known asymptotically AdS$_5$ multi-charge black hole solutions uplifting to ten- or eleven-dimensional supergravity. Indeed, besides the universal case that involves just the R-charge and applies to any compactification admitting a supersymmetric AdS$_5\times M$ solution -- whose corrections have already been discussed in~\cite{Bobev:2022bjm,Cassani:2022lrk} -- the only known such solution has three independent electric charges and uplifts to type IIB supergravity on $S^5$ or the $S^5/\Gamma$ orbifold, dual to ${\rm SU}(N)$ $\mathcal{N}=4$ SYM or the $\mathbb{C}^3/\Gamma$ orbifold theories. In the case of $\mathcal{N}=4$ SYM, the Cardy asymptotics of the index involve extra simplifications due to the underlying maximal supersymmetry,  specifically one has the exact expressions $k_{IJK} =\frac{N^2-1}{2}|\epsilon_{IJK}|$ and $k_I=0$. The orbifold theories, instead, have a more interesting set of corrections, which as such offer a 
more ``realistic'' view on the corrections of generic $\mathcal{N}=1$ SCFT's. This provides our main motivation for considering such theories here. We start by briefly recalling the features that will be relevant for us and obtain the anomaly coefficients.

\subsection{Anomaly coefficients}

The structure of the orbifold theories describing the low-energy limit of a stack of D3-branes probing a $\mathbb{C}^3/\Gamma$ singularity has been discussed long ago~\cite{Douglas:1996sw,Douglas:1997de,Kachru:1998ys,Lawrence:1998ja}.\footnote{The superconformal index of orbifold theories has been studied in~\cite{Nakayama:2005mf,Arai:2019wgv}.}
 In order to preserve $\mathcal{N}=1$ supersymmetry, we require  the finite group $\Gamma$ to be a subgroup of $ {\rm SU}(3)\subset {\rm SU}(4)$, with ${\rm SU}(4)\simeq{\rm SO}(6)$ corresponding to rotations in the $\mathbb{R}^6\simeq\mathbb{C}^3$ space transverse to the branes.
 
\begin{figure}
	\centering
	\includegraphics[width=0.7\textwidth]{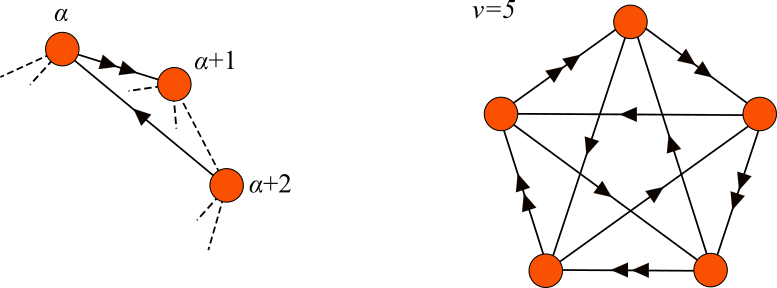}
	\caption{\it The quivers describing the low-energy limit of D3-branes probing a $\mathbb{C}^3/\mathbb{Z}_\nu$ singularity.  Left: a generic ${\rm SU}(N)$ node $\alpha$ is connected to the node $(\alpha+1)$ by a doublet of outgoing arrows  
	and to the node $(\alpha+2)$ by an incoming arrow 
	 (an arrow pointing from node $\alpha$ to node $\beta$ denotes a chiral superfield in the $({\bf N}, {\bf \bar N})$ bifundamental representation of ${\rm SU}(N)_{\alpha}\times {\rm SU}(N)_{\beta}$). Right: the case $\nu = 5$.}
	\label{fig:quivers}
\end{figure}

  We will focus on the $\Gamma=\mathbb{Z}_\nu$ orbifolds whose action on the $(z_1,z_2,z_3)$ coordinates of $\mathbb{C}^3$ is generated by the element 
  \be
 \Theta ={\rm diag} (\rme^{\frac{2\pi i}{\nu}},\, \rme^{\frac{2\pi i}{\nu}},\,\rme^{-\frac{4\pi i}{\nu}})\,.
  \ee 
 We take $\nu\geq 3$ so that the quotient preserves exactly $\mathcal{N}=1$ supersymmetry. For $\nu=2$ one has $\Gamma \subset {\rm SU}(2)$, so the quotient preserves $\mathcal{N}=2$ supersymmetry and, when described in $\mathcal{N}=1$ language, involves chiral superfields in the adjoint representation (from the decomposition of the $\mathcal{N}=2$ vector multiplets into $\mathcal{N}=1$  multiplets), a case falling out of our assumptions in section~\ref{sec:assumptions} (though it would not be hard to study it separately).\footnote{We could also consider other $\mathbb{C}^3/\Gamma$ orbifolds with $\Gamma\subset{\rm SU}(3)$, including the more general Abelian case $\Gamma=\mathbb{Z}_{\nu_1}\times \mathbb{Z}_{\nu_2}$  
  as well as non-Abelian cases, see e.g.~\cite{Hanany:1998sd}. 
  We expect that the study of these more complicated examples does not involve qualitatively new features.
}
Note that $\Gamma$ commutes with the ${\rm SU}(2)$ acting on $(z_1,z_2)$, so a ${\rm U}(1)$ global symmetry enhances to ${\rm SU}(2)$.
The resulting theories 
 are quiver gauge theories of the type discussed in the previous section, namely they contain $\nu$ ${\rm SU}(N)$ nodes, connected by bifundamental chiral superfields. Each node $\alpha$ is connected to the node $(\alpha+1)$ by a doublet of chiral fields transforming in the $({\bf N}, {\bf \bar N})$ bifundamental representation of ${\rm SU}(N)_\alpha\times {\rm SU}(N)_{\alpha+1}$ and to the node $(\alpha+2)$ by a chiral field transforming in the  $({\bf \bar N}, {\bf N})$ representation of ${\rm SU}(N)_{\alpha}\times {\rm SU}(N)_{\alpha+2}$. In figure~\ref{fig:quivers} we show the generic structure at a node (to be repeated for all nodes) and the quiver for $\Gamma = \mathbb{Z}_5$ as an example.

For odd $\nu$ the orbifold action only has the origin of $\mathbb{C}^3$ as its fixed point, hence the base space $S^5/\mathbb{Z}_\nu$ is smooth and the low-energy spectrum of type IIB string theory on this space is simply given by the orbifold projection of the supergravity modes on $S^5$. On the other hand, for even $\nu$ there is a $\mathbb{Z}_2\subset\mathbb{Z}_\nu$ subgroup generated by the element $\Theta^{\nu/2}$ which leaves the complex line in $\mathbb{C}^3$ parameterized by $z_3$ invariant. This translates in an invariant circle in $S^5$, implying that the resulting orbifold space $S^5/\mathbb{Z}_\nu$ is singular and leads to a light twisted sector for the string modes localized at the invariant circle. 

The Abelian global symmetries are the R-symmetry and two ${\rm U}(1)$ flavour symmetries, whose generators span the Cartan subalgebra of ${\rm SO}(6)$. When $\nu$ is even there is also a non-anomalous ${\rm U}(1)$ baryonic symmetry;  on the gravity side this  acts in the twisted sector and is not visible at the level of type IIB supergravity. Because of this we will switch off the baryonic charge for now and come back to it at the end.

We choose a basis where the global charges $Q_I$ are all R-charges with $r_I=\frac{1}{2}$, $I=1,2,3$  (since this is the basis that is naturally obtained when reducing type IIB supergravity on $S^5$). It follows that the fermion $\psi$ in a chiral multiplet with charge $q_I$ has charge $q_I - \frac{1}{2}$, while the gaugino has charge $+\frac{1}{2}$ under all $Q_I$'s. The charge assignement for the fermions in the theory are given in table~\ref{table:charges_fermions}.
\begin{table}
\centering
\begin{tabular}{|c|c|c|c|c|c|}
 \hline
 Field & multiplicity & $Q_1$ & $Q_2$ & $Q_3$ \\
 \hline
$\phantom{\Big[} \!\!\!\psi_1$  & $ \nu N^2$  & $ \frac{1}{2}$ &  $-\frac{1}{2}$  &  $-\frac{1}{2}$    \\[2mm]
$\psi_2$ & $ \nu N^2$  &  $-\frac{1}{2}$  &  $\frac{1}{2}$  &  $-\frac{1}{2}$   \\[2mm]   
$\psi_3$   &  $\nu N^2$  & $-\frac{1}{2}$  &  $-\frac{1}{2}$  &  $\frac{1}{2}$   \\[2mm]  
gaugini   &  $\nu (N^2-1)$ &  $\frac{1}{2}$  & $\frac{1}{2}$  & $\frac{1}{2}$   \\[1mm]  
 \hline
\end{tabular}
\caption{Multiplicities and charge assignments for the fermion fields in the $\mathbb{C}^3/\mathbb{Z}_\nu$ quiver theories. $\psi_1,\psi_2,\psi_3$ are the fermion fields belonging to the $\nu$ triplets of bifundamental chiral multiplets.}
\label{table:charges_fermions}
\end{table}
 
Evaluating the 't Hooft anomaly coefficients using their definition \eqref{def_anomaly_coeff} one finds:
\be
\begin{aligned}\label{tHooft_coeff_orbifolds}
k_{IJK}   \, &=\, k^{(0)}_{IJK } + k^{(1)}_{IJK}\,,\quad\ \text{with}\quad\ 
k^{(0)}_{IJK }  \, =\, \frac{\nu N^2}{2} |\epsilon_{IJK}|\,,\quad\ k^{(1)}_{IJK} = -\frac{\nu}{8}\,,\\[1mm]
k_I  \,&=\, - \frac{\nu}{2}\,.
\end{aligned}
\ee
 These satisfy relation \eqref{relation_k3_k1} since $r_I=\frac{1}{2}$ for all charges.
The R-symmetry is given by the exact relation
\be
\mathcal{R}\,=\, \frac{2}{3}\left(Q_1+Q_2+Q_3\right)\,.
\ee
It follows from~\eqref{relacTrR} that the (exact) 
 Weyl anomaly coefficients read
\begin{equation}\label{a_c_orbifolds}
{\mathtt a}\,=\,\frac{\nu N^2}{4}-\frac{3\nu}{16}\,,\qquad\quad {\mathtt c}\,=\,\frac{\nu N^2}{4}-\frac{\nu}{8}\, .
\end{equation}
The value of $\cc-\aa= \frac{\nu}{16}$ has been matched with a supergravity (string theory) computation  in~\cite{ArabiArdehali:2013jiu}.

\subsection{Corrected entropy}\label{subsec:entropy_orbifolds}

 For the  tensor $k^{(0)}{}^{IJK}$ we take
\begin{equation}\label{eq:specifyingkIJK}
 k^{(0)}{}^{IJK}=\frac{1}{6}|\epsilon^{IJK}|\,,
\end{equation}
where $\epsilon_{IJK}$ is the Levi-Civita symbol and $\epsilon^{IJK}=\delta^{II'}\delta^{JJ'}\delta^{KK'}\epsilon_{I'J'K'}$. Note that $ k^{(0)}{}^{IJK}$ satisfies the normalization condition~\eqref{eq:normalization_k} as well as the cubic relation \eqref{eq:cubic_relation}, with $\gamma = \frac{4}{9}\,{\mathtt a}^{(0)} = \frac{\nu N^2}{9}$.
 Since all assumptions are satisfied, we can use the general results of section~\ref{sec:Legendre_transf_gen}.
The coefficients \eqref{eq:p_coeffs} read
\be
\begin{aligned}
p_2 &= -(Q_1+Q_2+Q_3)-2 \aa^{(0)} \,,\\[1mm]
p_1 &= Q_1Q_2+Q_2Q_3+Q_3Q_1 - 2 \aa^{(0)}(J_1+J_2) \,,\\[1mm]
p_0 &= -Q_1Q_2Q_3 -2 \aa^{(0)}J_1J_2\,.
\end{aligned}
\ee
Then from \eqref{eq:BPS_entropy_0thorder} one finds that the expression for the leading-order entropy reads:
\begin{equation}\label{2der_S_explicit_model}
\mathcal{S}^{(0)}\,=\,  2\pi \sqrt{Q_1 Q_2+Q_2 Q_3 +Q_1 Q_3-  2 \, {\mathtt a}^{(0)}\left(J_1+J_2\right)}\, .
\end{equation}
Apart from the multiplicative factor of $\nu$ hidden in the anomaly coefficient ${\mathtt a}^{(0)}$, this expression is the same as the one that is obtained for ${\rm SU}(N)$ $\mathcal{N}=4$ SYM. 
However, when we include the corrections things become more interesting: while for $\mathcal{N}=4$ SYM the replacement $\aa^{(0)}= \frac{N^2}{4}\to \aa = \frac{N^2-1}{4}$ in \eqref{2der_S_explicit_model} accounts for all $1/N^2$ corrections to the black hole entropy (in the Cardy limit),
 for the orbifold theories we obtain a more complicated expression. We provide the explicit form of the entropy for the slightly simpler case of $J_1=J_2\equiv J$:
\begin{equation}\label{eq:entropy_orbifolds_eqJ}
\mathcal{S}= 2\pi \sqrt{Q_1 Q_2+Q_2 Q_3 +Q_1 Q_3- 4 \, {\mathtt a}J+\frac{2({\mathtt c}-{\mathtt a})}{3{\mathtt a}}\frac{  \mathcal{U}(1,2,3) + \mathcal{U}(2,3,1)+ \mathcal{U}(3,1,2)}{Q_1 Q_2+Q_2 Q_3 +Q_1 Q_3-4 \, {\mathtt a}J+J^2}}\,,
\end{equation}
where
\be
\mathcal{U}(1,2,3) \,=\, \left[ Q_1 Q_2 -J(Q_3+2\aa) \right](Q_1-Q_2)^2 \,,
\ee
and $\aa$, $\cc$ have been given in eq.~\eqref{a_c_orbifolds}. It is understood that the result is only valid at first order in the $1/N^2$ corrections. 
 The constraint \eqref{eq:nonlinearconstraint_J1=J2} can be written as
\be
\begin{aligned}
&\left[Q_1+Q_2+Q_3 +2(2\aa-\cc)\right] (Q_1Q_2+Q_2Q_3+Q_3Q_1-4\cc J) - Q_1Q_2Q_3 - 2(3\cc-2\aa)J^2\\[1mm]
&+ \frac{2(\cc-\aa)}{3\aa}\,\frac{ \mathcal{T}(1,2,3) + \mathcal{T}(2,3,1)  + \mathcal{T}(3,1,2) }{{Q_1 Q_2+Q_2 Q_3 +Q_1 Q_3-4 {\mathtt a}J+ J^2}}  \,=\,0\,,
\end{aligned}
\ee
with
\be\label{eq:Tfunction_eqJ}
\mathcal{T}(1,2,3)  \,=\, \left[(3 Q_1 + 3 Q_2 - 2 Q_3 - 2 J) Q_3 - 6 \aa J \right] (Q_1+Q_2+2\aa) (Q_1-Q_2)^2.
\ee

\subsection{Including the baryonic symmetry} \label{baryonic_sec}
We now include the non-anomalous baryonic charge that is admitted by the $\mathbb{C}^3/\mathbb{Z}_\nu$ theories when $\nu$ is even.\footnote{The $\mathbb{C}^3/\mathbb{Z}_{\nu =2p}$ theories are also the $Y^{p,p}$ members of the $Y^{p,q}$ family \cite{Benvenuti:2004dy}.}
The 't Hooft anomaly coefficients involving the baryonic charge can be computed by recalling that
 half of the bifundamental fermions $\psi_1$ carries baryonic charge $\frac{\nu}{2}$ while the other half carries baryonic charge $-\frac{\nu}{2}$, and the same holds for $\psi_2$, while the $\psi_3$ fermions are neutral. 
Using this information one can see that all anomaly coefficients involving the baryonic symmetry vanish, except for
\be\label{eq:tHooftAnomaliesBaryonic}
 k_{BBI} \,=\, -\frac{\nu^3N^2}{4} \,\delta_I^3\,,
\ee
where $B$ is the baryonic index while  $I=1,2,3$ labels the other symmetries as above. Recalling that $r_B=0$ since the baryonic charge preserves the supercharge, we see that relation \eqref{relation_k3_k1} continues to be satisfied even after including the baryonic direction. We also checked that the property \eqref{eq:cubic_relation} continues to hold after including \eqref{eq:tHooftAnomaliesBaryonic}, with $k^{(0)\,IJK} =\frac{1}{6}|\epsilon^{IJK}|$, $k^{(0)\,3BB} =-\frac{1}{3\nu^2} $ and all other components vanishing. 
 Since all requirements are satisfied, we can implement the Legendre transform of the action \eqref{eq:index_asympt} as illustrated above and conclude that the corrected entropy takes the form~\eqref{eq:entropy_J1neqJ2}, where the $p$-coefficients now read:
\be
\begin{aligned}
p_2 \,&=\, -(Q_1+Q_2+Q_3)-2 \aa^{(0)} \,,\\[1mm]
p_1 \,&=\, Q_1Q_2+Q_2Q_3+Q_3Q_1 -\tilde{Q}_B^2 - 2 \aa^{(0)}(J_1+J_2) \,,\\[1mm]
p_0\, &=\, -Q_1Q_2Q_3 + Q_3 \,\tilde Q_B^2 -2 \aa^{(0)}J_1J_2\,,
\end{aligned}
\ee 
where we defined $\tilde Q_B = \frac{Q_B}{\nu}$, with $Q_B$ the baryonic charge.
Let us provide the explicit expressions for the entropy and the non-linear constraint in the simpler case where all flavour charges are switched off, that is $Q_1=Q_2=Q_3\equiv \frac{1}{2}Q_R$, and where  $J_1=J_2\equiv J$. 
Using that $\aa$ and  $\cc$ are still given by~\eqref{a_c_orbifolds},
we find that \eqref{eq:entropy_equalJ} reduces to 
\begin{equation}
\mathcal{S}\,=\,\pi \sqrt{3Q_R^2- 4\tilde Q_B^2-16 {\mathtt a}J+\frac{16(\cc-\aa)}{\aa}\tilde Q_B^2\frac{Q_R^2-\frac{8}{3}\tilde Q_B^2-2Q_RJ-8{\mathtt a}J}{3Q_R^2- 4 \tilde Q_B^2  -16 {\mathtt a}J+4J^2}}\, .
\end{equation}
This provides the corrections to the leading-order results for the entropy given in~\cite{Kim:2019yrz,Amariti:2019mgp}.
The non-linear constraint between the charges in this case reads
\begin{equation}
\begin{aligned}
&Q_R^3+16\left(3{\mathtt c}-2{\mathtt a}\right)J^2- 4 Q_R \tilde Q_B^2 -\left[3Q_R+4\left(2{\mathtt a}-{\mathtt c}\right)\right]\big(3Q_R^2- 4 \tilde Q_B^2 -16 {\mathtt c} J\big)=\\[1mm]
&=\,\frac{64 (\aa-\cc) }{3\aa}\tilde{Q}_B^2  \frac{6\left(Q_R+2{\mathtt a}\right)\left(-Q_R^2+Q_R J+6{\aa} J\right)+ \tilde Q_B^2(20\aa-4J+9Q_R)}{3Q_R^2- 4 \tilde Q_B^2   -16 {\mathtt a}J+4J^2}\,.
\end{aligned}
\end{equation}

\section{Four-derivative ${\cal N}=2$ U(1)$_{R}$-gauged supergravity in five dimensions}\label{sec:fourder_action}

The details of the field theory enter in the formula \eqref{eq:index_asympt} only through the anomaly coefficients. This suggests that in order to reproduce such formula via a holographic computation it is enough to consider a matter-coupled five-dimensional supergravity which reproduces the anomalies. The goal of this section is to construct such theory.

The main ingredients of the gravitational theory will be the Chern-Simons terms, as these are the terms which holographically match the global anomalies of the dual field theories. The standard two-derivative Chern-Simons term of five-dimensional supergravity $\epsilon^{\mu\nu\rho\sigma\lambda}C_{IJK}F^{I}_{\mu\nu} F^J_{\rho\sigma} A^{K}_{\lambda}$ matches the cubic `t Hooft anomaly controlled by $\text{Tr}\,(Q_I Q_J Q_K)$. In turn, the mixed anomaly controlled by $\text{Tr}\, Q_I$ (which is subleading in the large-$N$ limit) is matched by the four-derivative Chern-Simons term $\epsilon^{\mu\nu\rho\sigma\lambda}R_{\mu\nu\alpha\beta}R_{\rho\sigma}{}^{\alpha\beta} A^I_{\lambda}$. Thus, our goal here is to write down a suitable four-derivative effective action containing the supersymmetrizations of the aforementioned Chern-Simons terms. More specifically, given the applications that we have in mind, we shall consider a four-derivative extension of ${\cal N}=2$ five-dimensional gauged supergravity coupled to an arbitrary number $n$ of abelian vector multiplets, with gauge group consisting of a U(1)$_{R}$ subgroup of the SU(2) R-symmetry group and $n$ additional U(1) isometries of the scalar manifold. We shall not consider the coupling to hyper- or tensor multiplets. 

Our starting point in this section will be an off-shell formulation of ${\cal N}=2$ supergravity that includes the relevant supersymmetric four-derivative invariants. While strictly speaking we are not obliged to pass through the off-shell formulation, it turns out to be highly convenient for practical purposes, as dealing with supersymmetric higher-derivative invariants is much easier if supersymmetry is realized off-shell \cite{Hanaki:2006pj, Bergshoeff:2011xn, Ozkan:2013uk, Ozkan:2013nwa, Gold:2023ymc,Gold:2023ykx}. Eventually we will integrate out the auxiliary fields (working at linear order in the corrections) to obtain a four-derivative effective action for the propagating degrees of freedom, further exploiting the possibility of performing perturbative field redefinitions to reduce the number of independent terms in the action. The procedure  mimics the one we followed in \cite{Cassani:2022lrk} in the case of minimal supergravity, see also \cite{Hanaki:2006pj, Baggio:2014hua,Bobev:2021qxx,Liu:2022sew}.

The plan of this section is the following. We start in section~\ref{sec:2der_action} introducing the basics of off-shell five-dimensional ${\cal N}=2$ supergravity; in particular reviewing how the on-shell theory is recovered once the auxiliary fields have been integrated out. Then in section~\ref{sec:4der_action} we repeat the same process including the relevant four-derivative off-shell invariants, treating them as a perturbation. We conclude in section~\ref{sec:final_Lagr} summarizing the final form of the Lagrangian. The reader can safely skip the first two parts if not interested in the derivation of the results.

\subsection{Two-derivative ${\cal N}=2$ gauged supergravity in five dimensions}\label{sec:2der_action}

${\cal N}=2$, $D=5$ off-shell Poincar\'e supergravity can be obtained from superconformal methods \cite{Bergshoeff:2001hc, Fujita:2001kv, Kugo:2002js, Bergshoeff:2002qk, Bergshoeff:2004kh} after fixing the redundant gauge symmetries. The procedure is however not unique. To begin with, one can make use of two inequivalent Weyl multiplets: the so-called standard and dilaton Weyl multiplets \cite{Bergshoeff:2001hc}. Even when one of these has been chosen, one still has the freedom to choose the multiplets which will act as compensators. Here we follow the construction in \cite{Ozkan:2013nwa} based on the standard Weyl multiplet and using as compensators one vector multiplet and one linear multiplet (instead of one hyper-multiplet, as in \cite{Bergshoeff:2004kh}). After fixing the gauge redundancies one gets an off-shell supergravity theory whose bosonic field content is the following:
\begin{itemize}
\item the vielbein $e^a{}_\mu$, a scalar $D$, an antisymmetric tensor $T_{ab}=-T_{ba}$ and a triplet of SU(2) vector fields $V_{\mu}{}^{ij}$ ($i, j=1, 2$). All these fields originally belonged to the standard Weyl multiplet of the superconformal theory.
\item $n+1$ vector fields $A^{I}_{\mu}$, scalars $X^I$ and SU(2) triplets $Y^{I}{}^{ij}$, all belonging originally to the vector multiplets (one of which acts as a compensating multiplet).
\item a scalar $N$ and a vector $P_{\mu}$, which originally belonged to the compensating linear multiplet.
\end{itemize}
Here we follow the conventions of \cite{Bergshoeff:2004kh} for the SU(2) indices. Any SU(2) triplet $A_{i}{}^{j}$ can be expanded in terms of the Pauli matrices ${\vec \sigma}_{i}{}^{j}$ as follows
\begin{equation}
{\mathnormal A}_{i}{}^{j} \,=\, i\,{\vec A}\cdot {\vec \sigma}_{i}{}^{j}\, ,
\end{equation}
and the indices $i, j$ are raised (lowered) with $\varepsilon^{ij}$ ($\varepsilon_{ij}$), following the NW-SE convention:
\begin{equation}
A^{ij}=\varepsilon^{ik}A_{k}{}^{j}\, , \hspace{1cm} A_{ij}=A_{i}{}^{k}\varepsilon_{kj}\, .
\end{equation}
Since $A^{[ij]}=0$, we can always split $A^{ij}$ into its traceless ${A'}^{ij}$ and trace $A$ contributions, in a way such that
\begin{equation}
A^{ij}={A'}^{ij}+\frac{1}{2} \,\delta^{ij} A\, .
\end{equation}

The two-derivative off-shell supergravity Lagrangian ${\cal L}_{2\partial}^{\rm{off-shell}}$ is given by \cite{Ozkan:2013nwa}\footnote{In this section we set $16\pi G=1$.}
\begin{equation}\label{eq:offshellLagrangian2d}
\begin{aligned}
{\cal L}_{2\partial}^{\rm{off-shell}}=\,&\frac{1}{4}\left({\cal C}+3\right)R+\frac{2}{3}\left(104 {\cal C}-8\right)T_{\mu\nu}T^{\mu\nu}+8\left({\cal C}-1\right)D-2\,N^2-2\,P_{\mu}P^{\mu}+2\,{V'}_{\mu}^{ij}{V'}_{ij}^{\mu}\\
&-2\sqrt{2}\,P_{\mu}V^{\mu}+\frac{1}{4}\,{\cal C}_{IJ}\,F^{I}_{\mu\nu}F^{J}{}^{\mu\nu}+\frac{1}{2}\,{\cal C}_{IJ}\,\partial_{\mu} X^I \partial^{\mu} X^J-{\cal  C}_{IJ}{Y^I}_{ij} {Y^J}^{ij}\\
&-24{X}_I F^I_{\mu\nu} T^{\mu\nu}+\frac{1}{4}\,\epsilon^{\mu\nu\rho\sigma\lambda}C_{IJK}F^{I}_{\mu\nu}F^{J}_{\rho\sigma}A^{K}_{\lambda} -3\sqrt{2}\,g_{I}Y^{I}_{ij}\delta^{ij}-6\,g_I \,A_{\mu}^{I}P^{\mu}\\
&-6\,g_IX^I N\,,
\end{aligned}
\end{equation}
where ${\cal C}$, ${\cal C}_{IJ}$ and $X_I$ are defined as
\begin{equation}
{\cal C}=C_{IJK}\,X^I X^J X^K\,, \hspace{1cm} {\cal C}_{IJ}=6\,C_{IJK}\,X^K\,, \hspace{1cm}X_{I}=C_{IJK}\,X^J X^K\,,
\end{equation}
being $C_{IJK}$ a totally symmetric constant tensor which will specify the \emph{very special geometry} of the scalar manifold. The gauging parameters $g_{I}$ select the linear combination of the vector fields $A^I_{\mu}$ that gauges the U(1) R-symmetry. 

Let us now integrate out the auxiliary fields in order to obtain a Lagrangian for the propagating degrees of freedom. This amounts to solving their equations of motion and plugging the solution back into \eqref{eq:offshellLagrangian2d}. The solution to the equations of motion of the auxiliary fields $P_{\mu}, V^{ij}_{\mu}, T_{\mu\nu}, N$ and $Y^I_{ij}$ is
\begin{equation}\label{eq:solauxfields}
\begin{aligned}
P_{\mu}=\,&0\,, \hspace{5mm} {V'}_{\mu}^{ij}=\,0\,,  \hspace{5mm} V_{\mu}=\,-\frac{3}{\sqrt{2}}g_{I}A^{I}_{\mu}\, , \hspace{5mm} T_{\mu\nu}=\,\frac{3}{16}\,{X}_I F^{I}_{\mu\nu}\,,\\[1mm]
N=\,& -\frac{3}{2}\,g_{I}X^I\,, \hspace{5mm} Y^{I}_{ij}=\, -\frac{3}{\sqrt{2}}\,{\cal C}^{IJ}g_{J}\,\delta_{ij}\,,
\end{aligned}
\end{equation}
where ${\cal C}^{IJ}$ denotes the inverse of ${\cal C}_{IJ}$. In addition to this, the auxiliary field $D$ (which plays the role of a Lagrange multiplier) imposes a constraint on the scalars $X^I$,
\begin{equation}\label{eq:cubicconstraint}
C_{IJK}\, X^I X^J X^K=1\,,
\end{equation}
which implies that there are only $n$ independent scalars $\phi^x$, $x=1, \dots, n$. We can thus regard the $X^I$ as functions of the physical scalars, $X^I=X^I(\phi^x)$.  The expression for $D$, which will be needed when studying the higher-derivative theory, is found from the following combination of the equations of motion of the scalars,
\begin{equation}
X^I \frac{\delta \mathcal L_{2\partial}^{\rm off-shell}}{\delta X^I}=0\,.
\end{equation}
Making use of some of the expressions in \eqref{eq:solauxfields}, one finds
\begin{equation}\label{eq:D}
D = -\frac{1}{32}\left[R +\left(\frac{1}{3}\,\mathcal C_{IJ}- \frac{9}{4}X_I X_J\right) F_{\mu\nu}^I F^J{}^{\mu\nu} +2 \mathcal C_{IJ} \partial_\mu X^I \partial^\mu X^J-12\mathcal C^{IJ} g_I g_J +12 (g_I X^I)^2
\right]\,.
\end{equation}
 Finally, we substitute the expressions \eqref{eq:solauxfields} into \eqref{eq:offshellLagrangian2d} to recover the well-known bosonic supergravity Lagrangian for the propagating degrees of freedom \cite{Gunaydin:1984ak} (see also \cite{Gunaydin:1999zx, Ceresole:2000jd, Bergshoeff:2004kh}):\footnote{The complete theory can be found in  \cite{Bergshoeff:2004kh}. The dictionary between the fields and couplings here and in that reference is the following:
$$C_{IJK}^{\rm here} = \mathcal{C}_{IJK}^{\rm there}\,,\qquad X^I_{\rm here} = h^I_{\rm there}\,,\qquad a^{\rm here}_{IJ} = a^{\rm there}_{IJ}\,, \qquad
 A^I_{\rm here} = \sqrt{\tfrac{2}{3}} A^I_{\rm there}\,,\qquad   \tfrac{\sqrt3}{2} g_I^{\rm here} = (g\xi_I)^{\rm there} \,.
$$}
\begin{equation}\label{eq:2dSUGRA}
{\cal L}_{2\partial}=R-2{\cal V}-\frac{3}{2}\, a_{IJ}\, \partial_{\mu} X^I \partial^{\mu} X^J-\frac{3}{4}\, a_{IJ} F^I_{\mu\nu} F^{J}{}^{\mu\nu}
+\frac{1}{4}C_{IJK}\epsilon^{\mu\nu\rho\sigma\lambda}F^{I}_{\mu\nu} F^J_{\rho\sigma} A^{K}_{\lambda} \, ,
\end{equation}
where we have defined
\begin{equation}
a_{IJ}=3\,{X}_I \,{X}_{J}-\frac{1}{3}\,{\cal C}_{IJ}\,,
\end{equation}
and where the scalar potential ${\cal V}$ is given by
\begin{equation}
{\cal V}=\, -\frac{9}{4}\, (g_I X^I)^2-\frac{9}{2}\,{\cal C}^{IJ}g_I g_J\, .
\end{equation}
Defining the metric of the scalar manifold as
\begin{equation}\label{eq:scmetric}
{\mathfrak g}_{xy}=\frac{3}{2}\, a_{IJ}\partial_x X^I \partial_yX^J\, ,
\end{equation}
we can rewrite the two-derivative Lagrangian as
\begin{equation}\label{eq:action2dSUGRA} 
{\cal L}_{2\partial}=R-2{\cal V}-{\mathfrak g}_{xy}\, \partial_{\mu} \phi^x  \partial^{\mu} \phi^y-\frac{3}{4}\, a_{IJ}\,F^I_{\mu\nu} F^J{}^{\mu\nu}+\frac{1}{4}\,C_{IJK}\epsilon^{\mu\nu\rho\sigma\lambda}F^{I}_{\mu\nu} F^{J}_{\rho\sigma} A^{K}_{\lambda} \, .
\end{equation}
The tensors which are set to zero by the two-derivative equations of motion are:
\begin{eqnarray}\label{eq:2dEOMEinst}
{\cal E}_{\mu\nu}&=&R_{\mu\nu}-\frac{1}{2}\,g_{\mu\nu}\, \left(R-2{\cal V}\right) -T^{\rm{vectors}}_{\mu\nu}-T^{\rm{scalars}}_{\mu\nu}\,,\\[1mm]
\label{eq:2dEOMvec}
{\cal E}^{\mu}_{I}&=&\nabla_{\nu}\left(3\,a_{IJ}{F^J}^{\nu\mu}\right)+\frac{3}{4}\,C_{IJK}\epsilon^{\mu\nu\rho\sigma\lambda}F^{J}_{\nu\rho}F^{K}_{\sigma\lambda}\,, \\[1mm]
\label{eq:2dEOMsc}
{\cal E}_{x}&=&\nabla_{\mu}\left(2\,{\mathfrak g}_{xy}\partial^{\mu}\phi^y\right)-2\,\partial_{x}{\cal V} -\partial_{x}{\mathfrak g}_{yz} \partial_{\mu} \phi^y \partial^{\mu}\phi^z-\frac{3}{4}\,\partial_{x}a_{IJ} F^I_{\mu\nu}F^{J}{}^{\mu\nu}\, ,
\end{eqnarray}
where 
\begin{equation}
\begin{aligned}
T^{\rm{vectors}}_{\mu\nu}=\,&\frac{3}{2}\,a_{IJ}\left(F^{I}_{\mu\rho}\,{F^J}_{\nu}{}^{\rho}-\frac{1}{4}g_{\mu\nu}\,F^I_{\rho\sigma} F^J{}^{\rho\sigma}\right)\,,\\[1mm]
T^{\rm{scalars}}_{\mu\nu}=\,&\frac{3}{2}\, a_{IJ}\,\left(\partial_\mu X^I \partial_\nu X^J-\frac{1}{2}g_{\mu\nu}\,\partial_{\rho} X^I \partial^{\rho} X^J\right)\, .
\end{aligned}
\end{equation}

\subsection{Four-derivative corrections}
\label{sec:4der_action}

Our goal now is to obtain a four-derivative extension of ${\cal N}=2$, $D=5$ U(1)$_{R}$-gauged supergravity coupled to an arbitrary number of vector multiplets. To this aim, we modify the procedure followed in the two-derivative case adding the relevant four-derivative supersymmetric invariants. The off-shell Lagrangian will then contain two pieces,
\begin{equation}\label{eq:offshellLagrangian4der}
{\cal L}^{\rm{off-shell}}={\cal L}^{\rm{off-shell}}_{2\partial}+\alpha \, {\mathcal L}^{\rm{off-shell}}_{4\partial}\,,
\end{equation}
where $\alpha$, which will be our expansion parameter, is by definition the dimensionful part of the four-derivative coupling constants, hence $[\alpha]={\text{length}}^2$.  Before specifying the form of ${\mathcal L}^{\rm{off-shell}}_{4\partial}$, let us explain the general procedure we are going to follow to integrate out the auxiliary fields at linear order in $\alpha$. 

Let us denote by $\Phi_{\rm{aux}}$ all the auxiliary fields except for the combination of the scalars $X^I$ that is not dynamical, which is treated separately for the sake of clarity. The solution to the corrected equations of motion for the auxiliary fields derived from \eqref{eq:offshellLagrangian4der} is in general of the form,
\begin{equation}\label{eq:Psi}
\Phi_{\rm{aux}}\left(\Psi\right)=\Phi^{(0)}_{\rm{aux}}\left(\Psi\right)+\alpha \,\Phi^{(1)}_{\rm{aux}}\left(\Psi\right)\,,
\end{equation}
where $\Psi\equiv \{g_{\mu\nu}, A^I, \phi^x\}$ denotes the dynamical fields. If  ${\mathcal L}^{\rm{off-shell}}_{4\partial}$ depends on $D$, the cubic constraint of the very special geometry \eqref{eq:cubicconstraint} will receive corrections. Let us assume a generic modification
\begin{equation}
\left(C_{IJK}+\alpha\, C^{(1)}_{IJK}\right){X}^I {X}^J {X}^K=1\, ,
\end{equation}
where $C^{(1)}_{IJK}$ is an arbitrary symmetric tensor. Denoting by $X^{(0)}{}^{I}$ the scalars satisfying the original constraint $C_{IJK}X^{(0)}{}^{I}X^{(0)}{}^{J}X^{(0)}{}^{K}=1$, we have that
\begin{equation}
X^{I}=X^{(0)}{}^{I}+\alpha \,X^{(1)}{}^I\, , 
\end{equation}
with $X^{(1)}{}^I$ obeying the constraint
\begin{equation}
C_{IJK} X^{(0)}{}^{I}X^{(0)}{}^{J}X^{(1)}{}^{K}=-\frac{1}{3}\,C^{(1)}_{IJK}X^{(0)}{}^{I}X^{(0)}{}^{J}X^{(0)}{}^{K}\equiv -\frac{1}{3}\, {\cal C}^{(1)}\, ,
\end{equation}
whose solution is 
\begin{equation}\label{eq:sol_cc}
X^{(1)}{}^{I}=-\frac{1}{3}\, {\cal C}^{(1)} \,X^{(0)}{}^{I}\, , \hspace{1cm} X^I=X^{(0)}{}^{I}\left(1-\frac{\alpha}{3}\, {\cal C}^{(1)}\right)\, .
\end{equation}
Substituting the expressions for the auxiliary fields \eqref{eq:Psi} and the solution to the cubic constraint \eqref{eq:sol_cc} into the two-derivative off-shell Lagrangian, we get 
\begin{equation}
\begin{aligned}
{\cal L}^{\rm{off-shell}}_{2\partial}\Big|_{\{X^I(\Psi), \Phi_{\rm{aux}}(\Psi)\}}=\,&{\cal L}^{\rm{off-shell}}_{2\partial}\Big|_{(0)}+\alpha\, \Phi^{(1)}_{\rm{aux}}\frac{\delta{\cal L}^{\rm{off-shell}}_{2\partial}}{\delta \Phi_{\rm{aux}}}\Bigg|_{(0)}-\frac{\alpha}{3}\,{\cal C}^{(1)}\, X^{(0)}{}^{I}\, \frac{\delta{\cal L}^{\rm{off-shell}}_{2\partial}}{\delta X^I}\Bigg|_{(0)}\\[1mm]
=\,&{\cal L}^{\rm{off-shell}}_{2\partial}\Big|_{(0)}\,,
\end{aligned}
\end{equation}
up to boundary and ${\cal O}(\alpha^2)$ terms. The subscript $(0)$ in the above equation means evaluation using the zeroth-order expressions for the auxiliary fields $\Phi_{\rm{aux}}\to \Phi^{(0)}_{\rm{aux}}$ and for the scalars $X\to X^{(0)}{}$. Therefore, the first term yields the same result as before: the two-derivative Lagrangian \eqref{eq:2dSUGRA}. Instead, the second and  third term vanish, as they contain the two-derivative equations of motion for the auxiliary fields. Let us remark in particular that the combination of the scalar equations that appears in the third term is precisely the one that yields the equation of motion for the Lagrange multiplier $D$, \eqref{eq:D}. This is a consequence of the fact that the solution to the modified cubic constraint is in general of the form \eqref{eq:sol_cc}. 

We have just justified that we can make use of the zeroth-order expressions for the auxiliary fields and for the cubic constraint in the two-derivative off-shell Lagrangian. Thus, the final four-derivative on-shell Lagrangian ${\cal L}$ will be given by
\begin{equation}
{\cal L}\equiv {\cal L}^{\rm{off-shell}}\Big|_{\{X^I(\Psi), \Phi_{\rm{aux}}(\Psi)\}}=\left({\cal L}^{\rm{off-shell}}_{2\partial}+\alpha \,{\cal L}^{\rm{off-shell}}_{4\partial}\right)\Big|_{(0)}\,.
\end{equation}
We emphasize that when following the procedure just outlined we will be writing the resulting Lagrangian in terms of the constrained scalars $X^{(0)}{}^{I}$, which differ from those appearing in the parent off-shell theory, \eqref{eq:sol_cc}. 

 In what follows we apply this procedure including a specific combination of four-derivative off-shell invariants to be specified along the way. Then we will argue that this choice of  invariants suffices to obtain the most general four-derivative effective action, at least for our present purposes. To avoid the clutter, we will not write the superscript $(0)$ in the scalars $X^{(0)}{}^{I}$; we will denote them simply as $X^I$, keeping in mind that they satisfy the same constraint as in the original two-derivative theory. In addition, we will ignore terms involving the two-derivative equations of motion, as those can be removed with perturbative field redefinitions without affecting the rest of the terms. 
 
We find instructive to first consider the ungauged limit. As a matter of fact, the gauging does not enter into the four-derivative part of the Lagrangian. In turn, it gives rise to a set of two-derivative corrections.

\subsubsection{Ungauged limit}

In principle, given our purposes in this section, we must now add the most general linear combination of off-shell four-derivative invariants, see \cite{Ozkan:2024euj} for a recent review on this topic. On general grounds we expect three of them, corresponding (for instance) to the supersymmetrizations of the $R_{\mu\nu\rho\sigma}R^{\mu\nu\rho\sigma}$, $R_{\mu\nu}R^{\mu\nu}$ and $R^2$ terms. In the context of matter-coupled supergravity theories, a complete basis for these invariants has been constructed in the off-shell formulation based on the dilaton Weyl multiplet \cite{Ozkan:2013nwa}, but only in the ungauged limit. Indeed, our main motivation to work with the off-shell formulation based on the Standard Weyl multiplet is that the four-derivative invariants have been also constructed in the gauged case. On the downside, only the supersymmetrization of the Weyl-squared term \cite{Hanaki:2006pj} and of the $R^2$ \cite{Ozkan:2013nwa} are known. However, this will not be a problem for our purposes here as we can argue that in the ungauged case all the supersymmetric invariants based on Ricci curvature can be eliminated with perturbative field redefinitions. The reason  is that a term such as $R_{\mu\nu}R^{\mu\nu}$ or $R^2$ can be substituted by a series of terms involving the two-derivative Einstein equations plus other four-derivative terms made out of the matter fields exclusively. Perturbative field redefinitions allow us to eliminate the terms proportional to the two-derivative equations of motion. What remains must be supersymmetric on its own, but we notice that any of these terms involves Ricci curvature. Since on general grounds we expect no supersymmetric invariant can be constructed combining just the matter fields, the conclusion is that the remaining contribution must vanish.\footnote{This is something that we have explicitly checked for the complete basis of invariants based on the dilaton Weyl multiplet in the context of minimal supergravity.}

Let us then specify ${\cal L}^{\rm{off-shell}}_{4\partial}$ to be a linear combination of the Weyl-squared invariant ${\cal L}^{\rm{off-shell}}_{C^2}$ and the Ricci-scalar-squared invariant ${\cal L}^{\rm{off-shell}}_{R^2}$, whose explicit expressions are provided in eqs.~\eqref{eq:offshell_Weyl_squared} and \eqref{eq:offshell_R^2}, respectively.
\begin{equation}
{\cal L}^{\rm{off-shell}}_{4\partial}={\cal L}^{\rm{off-shell}}_{C^2}+{\cal L}^{\rm{off-shell}}_{R^2}\, .
\end{equation}
Next we show that the $R^2$ invariant yields a trivial contribution, as argued above. After using the expressions for the auxiliary fields derived in the previous section \eqref{eq:solauxfields} and the expression for $D$ given in \eqref{eq:D}, we find that it reduces to
\begin{equation}
{\mathcal L}^{\rm{off-shell}}_{R^2}\Big|_{(0)}\,=\,{\zeta}_I X^I \left(-\frac{3}{8}\,R+\frac{8}{3}\, T^2+4 D\right)^2\Big|_{(0)}\,=\,\frac{1}{9}{\zeta}_I X^I {\cal E}^2\, ,
\end{equation}
where ${\cal E}={\cal E}_{\mu}{}^{\mu}$ is the trace of Einstein equations. Therefore, this term can be directly ignored as it is effectively of order ${\cal O}(\alpha^2)$. This provides explicit evidence in favour of our previous claim, according to which the most general four-derivative effective Lagrangian can be obtained by just considering the Weyl-squared invariant, or equivalently any other supersymmetric invariant containing $R_{\mu\nu\rho\sigma}R^{\mu\nu\rho\sigma}$. Therefore, 
\begin{equation}
{\cal L}_{4\partial}\,=\,{\mathcal L}^{\rm{off-shell}}_{4\partial}\Big|_{(0)}\,=\,{\mathcal L}^{\rm{off-shell}}_{C^2}\Big|_{(0)}\, .
\end{equation}
After integration by parts, use of Ricci identities and ignoring terms which can be removed with field redefinitions (without affecting the rest), we get
\begin{equation}\label{eq:Weyl2}
\begin{aligned}
 {\cal L}_{4\partial} &=\lambda_{M}X^{M} C_{\mu\nu\rho\sigma}C^{\mu\nu\rho\sigma}+{\mathtt D}_{IJ}\,C_{\mu\nu\rho\sigma}{F^I}^{\mu\nu}{F^J}^{\rho\sigma}+{\mathtt E}_{IJKL} \, F^I_{\mu\nu} F^J{}^{\mu\nu} \,F^K_{\rho\sigma} F^L{}^{\rho\sigma} \\[1mm]
&\!\!\!+ \widetilde{\mathtt E}_{IJKL} \, F^{I}_{\mu\nu} F^{J}{}^{\nu\rho} \,F^{K}_{\rho\sigma} F^{L}{}^{\sigma\mu}+\frac{27}{8}\lambda_{M}X^{M} \left(\partial_{\mu}X_I\, \partial^{\mu}X^I\right)^2+{\mathtt G}_{IJKL}\,\partial_{\mu} X^I \partial^{\mu} X^J \, F^K_{\rho\sigma} F^L{}^{\rho\sigma}\\[1mm]
&\!\!\!+12\lambda_M X^M\partial_{\mu}X_I \partial_{\nu}X_J  F^{J}{}^{\mu\rho}F^{I}{}^{\nu}{}_{\rho}-\frac{2}{3} \lambda_I {\mathcal F}^{\mu\alpha}{\mathcal F}^{\nu}{}_{\alpha}\nabla_{\nu}\nabla_{\mu}X^{I}+\frac{1}{4}\lambda_I  X_J\, \epsilon^{\mu\nu\rho\sigma\lambda}\nabla_{\alpha}F^{[I}_{\mu\nu}F^{J]}_{\rho\sigma}{\mathcal F}_{\lambda}{}^{\alpha}\\[1mm]
&\!\!\!+{\mathtt H}_{IJKL}\, \epsilon^{\mu\nu\rho\sigma\lambda}{F^I}_{\mu\nu} {F^{J}}_{\rho}{}^{\alpha} {F^K}_{\sigma\alpha} \partial_{\lambda} X^L+\frac{1}{2}\lambda_I \epsilon^{\mu\nu\rho\sigma\lambda} R_{\mu\nu\alpha\beta}R_{\rho\sigma}{}^{\alpha\beta}{A}^I_{\lambda} \\[1mm]
&\!\!\!-\lambda_M X^M\left(\frac{2}{3}R_{\mu\nu}{\mathcal F}^{\mu\alpha}{\mathcal F}^{\nu}{}_{\alpha}-\frac{1}{6}R\, {\mathcal F}^2\right)+\frac{2}{3}\lambda_M X^M \nabla_{\mu}{\mathcal F}^{\mu\nu}\nabla_{\rho}{\mathcal F}^{\rho}{}_{\nu}\\[1mm]
&\!\!\!+{\mathtt W}_{IJ}\,\epsilon^{\mu\nu\rho\sigma\lambda}F^I_{\mu\nu}F^J_{\rho\sigma}\nabla^{\alpha}{\mathcal F}_{\alpha\lambda}\,,
\end{aligned}
\end{equation}
where $\lambda_I$ is a dimensionless coupling, ${\cal F}$ is defined as $\mathcal {F}=3X_I F^I $, and the different couplings appearing in the Lagrangian read
\begin{equation}\label{eq:couplings}
\begin{aligned}
{\mathtt D}_{IJ}=\,&3\lambda_{I} X_{J}-\frac{9}{2}\lambda_{M}X^{M}X_IX_J\,,\\[1mm]
{\mathtt E}_{IJKL}=\,& \frac{3}{16}\lambda_{M}X^{M}\left(\frac{1}{2}a_{IJ}a_{KL}+3a_{IJ}X_K X_L- 9X_I X_J X_K X_L\right)-\frac{3}{8}a_{IJ}X_K\lambda_L\,,\\[1mm]
\widetilde{\mathtt E}_{IJKL}=\,&\frac{81}{8}\lambda_{M}X^{M}X_I X_JX_K X_L-\frac{9}{2}X_I X_JX_K \lambda_L\,,\\[1mm]
{\mathtt G}_{IJKL}=\,&\frac{9}{8}\lambda_{M}X^{M}a_{IJ}\left(a_{KL}+3X_K X_L\right)-\frac{9}{4}a_{IJ}X_K\lambda_L-6\lambda_{M}X^{M}{a}_{IK}{a}_{JL}\,,\\[1mm]
{\mathtt H}_{IJKL}=\,& -\frac{21}{4}\lambda_J \,a_{IL} \,X_K-\frac{3}{4}\lambda_I\, a_{JL}\, X_K\,,\\[1mm]
{\mathtt W}_{IJ}=\,&-\frac{1}{8}\lambda_I X_J+\frac{3}{4}\lambda_M X^M X_I X_J\,.
\end{aligned}
\end{equation}

\subsubsection{Including the gauging}

The gauged case is more subtle, as it contains an additional length scale set by the effective cosmological constant or, equivalently, the gauging parameters $g_I$. This is precisely what allows for corrections to the two-derivative terms, since $\alpha g_I g_J$ is dimensionless. These will play indeed a crucial role in this story, as they will eventually account for the corrections to the cubic `t Hooft anomaly coefficients, $k^{(1)}_{IJK}$. 

A main consequence of these two-derivative corrections is that the reasoning used in the ungauged case to argue that supersymmetric invariants just containing Ricci curvature yield a trivial contribution (namely, removable with suitable field redefinitions) does not  work anymore. However, it should be true that their contributions reduce to corrections to the two-derivative terms. This is exactly the logic that was used in the minimal supergravity case to argue that the effective action presented  in \cite{Cassani:2022lrk} was the most general one, even if we did not use the complete basis of off-shell four-derivative invariants, as only two of them were known. Recently, the third one, which consists of the supersymmetrization of the $R_{\mu\nu}R^{\mu\nu}$ term, has been constructed in \cite{Gold:2023ymc,Gold:2023ykx}. After integrating out the auxiliary fields perturbatively, this gives rise to a four-derivative effective action which depends on three parameters: one controlling the four-derivative corrections and two controlling only two-derivative corrections. However, one can show that a combination of the parameters controlling the two-derivative corrections is unphysical, as it can be absorbed by a constant rescaling of the metric. The resulting action obtained after performing such field redefinition precisely matches the one of \cite{Cassani:2022lrk}.

Given this, we consider the following off-shell Lagrangian,
\begin{equation}\label{eq:finaloffshellLagrangian}
{\cal L}^{\rm{off-shell}}\,=\,{\cal L}_{2\partial}^{\rm{off-shell}}\Big|_{C_{IJK} \to C_{IJK}+\alpha {\widetilde \lambda}_{IJK}}+\alpha \,{\cal L}^{\rm{off-shell}}_{C^2}\,,
\end{equation}
where ${\widetilde \lambda}_{IJK}$ is an arbitrary symmetric constant tensor of mass dimension 2. Splitting the first term in \eqref{eq:finaloffshellLagrangian} into its zeroth- and first-order contributions, we have:
\begin{equation}
{\cal L}_{2\partial}^{\rm{off-shell}}\Big|_{C_{IJK} \to C_{IJK}+\alpha {\widetilde \lambda}_{IJK}}= {\cal L}_{2\partial}^{\rm{off-shell}}+\alpha \,\Delta{\cal L}^{\rm{off-shell}}_{2\partial}\,,
\end{equation}
where ${\cal L}_{2\partial}^{\rm{off-shell}}$ is the one in \eqref{eq:offshellLagrangian2d} and 
\begin{equation}\label{eq:2ndinvariantoffshell}
\begin{aligned}
\Delta{{\cal L}}^{\rm{off-shell}}_{2\partial}=\,&8\, {\widetilde \lambda}\left(\frac{1}{32} R+D+\frac{26}{3}T_{\mu\nu}T^{\mu\nu}\right)+\frac{1}{4} \,{\widetilde \lambda}_{IJ} F^I_{\mu\nu}F^J{}^{\mu\nu}+\frac{1}{2}\,{\widetilde \lambda}_{IJ}\,{\partial}_{\mu}X^I {\partial}^{\mu}X^J\\[1mm]
&-{\widetilde \lambda}_{IJ}\, Y^I{}^{ij}Y^J{}_{ij}-8 {\widetilde \lambda_I}  F^I_{\mu\nu}T^{\mu\nu}+\frac{1}{4}{\widetilde \lambda}_{IJK}\epsilon^{\mu\nu\rho\sigma\lambda}F^I_{\mu\nu} F^J_{\rho\sigma}A_\lambda^K\, ,
\end{aligned}
\end{equation}
where we have defined
\begin{equation}\label{eq:tildel_contracted}
{\widetilde\lambda}_{IJ}=6\,{\widetilde\lambda}_{IJK}X^K\,, \hspace{1cm} {\widetilde\lambda}_{I}=3\,{\widetilde\lambda}_{IJK}X^JX^K\,, \hspace{1cm}  {\widetilde\lambda}={\widetilde\lambda}_{IJK}X^I X^JX^K\,.
\end{equation}

Let us integrate out the auxiliary fields.  The Weyl-squared invariant now gives additional two-derivative corrections with respect to the ungauged case:
\begin{equation}
{\cal L}^{\rm{off-shell}}_{C^2}\Big|_{(0)}=\,{\cal L}_{4\partial}+\Delta {\cal L}^{C^2}_{2\partial}\,,
\end{equation}
where ${\cal L}_{4\partial}$ is still given by \eqref{eq:Weyl2} and
\begin{equation}\label{eq:DeltaL2dC2} 
\begin{aligned}
\Delta {\cal L}^{C^2}_{2\partial}=\,\,&6 \,\lambda_M X^M \left[{\cal V} + 3\left (g_M X^M\right)^2\right]^2- 9\lambda_M X^M \left[{\cal V}+3\left (g_M X^M\right)^2\right] a_{IJ} \,\partial_{\mu} X^I \partial^{\mu} X^J  \\[1mm]
& +f_{IJ} \,F^I_{\mu\nu}F^J{}^{\mu\nu} - \frac{3}{2}\lambda_{(I}g_Jg_{K)}\,\epsilon^{\mu\nu\rho\sigma\lambda}F^I_{\mu\nu} F^J_{\rho\sigma}A_\lambda^K\,,
\end{aligned}
\end{equation}
where
\begin{equation}
\begin{aligned}
f_{IJ}=\,&-\frac{3}{2}\left[{\cal V}+3\left ( g_M X^M\right)^2\right]\left[\lambda_M X^M \left(3X_I X_J + a_{IJ}\right) - 2\lambda_{(I}X_{J)} \right]+ 36 \lambda_K \,{\cal C}^{KL}\,g_L g_{(I} X_{J)}\\[1mm]
&  - 6 \lambda_M X^M\, g_I g_J\,.
 \end{aligned}
\end{equation}
In turn, the contribution from the second invariant $\Delta{{\cal L}}^{\rm{off-shell}}_{2\partial}$ is
\begin{equation}\label{eq:2ndinvariant}
\begin{aligned}
\Delta{\widetilde{\cal L}}_{2\partial}\equiv&\ \Delta{{\cal L}}^{\rm{off-shell}}_{2\partial}\Big|_{(0)}=\,\frac{{\widetilde \lambda}}{4} \,\left[R-6{\cal V}-18 \left(g_I X^I\right)^2\right]-9 \,{\widetilde \lambda}_{IJ} \,{\cal C}^{IK}{\cal C}^{JL}g_Kg_L\\[1mm]
&+\frac{1}{4}\left[{\widetilde \lambda}_{IJ} + 3{\widetilde \lambda} \left(3X_I X_J +\frac{1}{4}a_{IJ}\right)-6 {\widetilde \lambda}_{I} X_J\right]F^I_{\mu\nu} F^J{}^{\mu\nu}\\[1mm]
&+\frac{1}{2}\left({\widetilde\lambda}_{IJ}+\frac{9}{4} \,{\widetilde \lambda} \,a_{IJ}\right)\partial_{\mu} X^I \partial^{\mu} X^{J}+\frac{1}{4}{\widetilde \lambda}_{IJK}\epsilon^{\mu\nu\rho\sigma\lambda} F^I_{\mu\nu} F^J_{\rho\sigma}A^K_{\lambda}\,.
\end{aligned}
\end{equation}
Thus, the complete Lagrangian is
\begin{equation}\label{eq:4Der_Lagr}
\begin{aligned}
 {\cal L}=\,&\,R-2{\cal V}-\frac{3}{2}\, a_{IJ}\, \partial_{\mu} X^I \partial^{\mu} X^J-\frac{3}{4}\, a_{IJ}\,F^I_{\mu\nu} F^J{}^{\mu\nu}
+\frac{1}{4}C_{IJK}\epsilon^{\mu\nu\rho\sigma\lambda}F^{I}_{\mu\nu} F^{J}_{\rho\sigma} A^{K}_{\lambda} \\
&+\alpha \left({\cal L}_{4\partial} +\Delta {\cal L}^{C^2}_{2\partial}+\Delta {\widetilde{\cal L}}_{2\partial}\right) \,,
\end{aligned}
\end{equation}
where ${\cal L}_{4\partial}$, $\Delta {\cal L}^{C^2}_{2\partial}$ and $\Delta {\widetilde{\cal L}}_{2\partial}$ are given in eqs.~\eqref{eq:Weyl2}, \eqref{eq:DeltaL2dC2} and \eqref{eq:2ndinvariant}, respectively.

Some comments are in order. First, we expect that this effective Lagrangian captures, for particular choices of ${\widetilde\lambda}_{IJK}$, the corrections that one would obtain when considering any other basis of off-shell invariants. In particular we have checked this for the $R^2$ invariant of \cite{Ozkan:2013nwa}, as well as for the ``off-diagonal'' invariants constructed in \cite{Ozkan:2016csy}. However, let us emphasize that \eqref{eq:2ndinvariant} generalizes all of them, as the correction to the gauge Chern-Simons term is controlled by a (symmetric) tensor ${\widetilde \lambda}_{IJK}$, which is the most general possibility.  This is crucial as it will allow us to match any correction to the cubic anomalies of the dual field theories. 

Second, we note that when ${\widetilde \lambda}_{IJK}\propto C_{IJK}$, the correction from the second invariant \eqref{eq:2ndinvariant} reduces to a correction of Newton's constant, after a suitable constant rescaling of the metric is performed. This is exactly what happens for the supergravity theory dual to ${\cal N}=4$ SYM. In addition, for this theory the corrections coming from the Weyl-squared invariant also trivialize, as the anomaly matching imposes $\lambda_I=0$, as a consequence of the fact that $k_I=0$ at all orders in the large-$N$ expansion.

Finally, we note that the Lagrangian \eqref{eq:4Der_Lagr} can be further simplified using perturbative field redefinitions. In particular we can use them to remove the last four terms in \eqref{eq:Weyl2}, as well as all the two-derivative corrections to the Ricci scalar term. When doing this, however, we will modify some of the couplings to the remaining terms. We relegate the details of this procedure to appendix~\ref{app:field_redefinitions} and simply present the final form of the action (with the couplings updated) in the next subsection.

\subsection{Final form of the four-derivative effective Lagrangian}
\label{sec:final_Lagr} 

After implementing suitable perturbative field redefinitions --- see appendix~\ref{app:field_redefinitions} --- to reduce the number of terms in \eqref{eq:4Der_Lagr}, we arrive to the following final Lagrangian:
\begin{equation}\label{eq:finalL}
\begin{aligned}
{\cal L}=\,&\, R-2{\cal V}-\frac{3}{2}\, {a}_{IJ}\, \partial_{\mu} {X}^I \partial^{\mu} {X}^J-\frac{3}{4}\, {a}_{IJ}\,F^I_{\mu\nu} F^J{}^{\mu\nu}
+\frac{1}{4}{ C}_{IJK}\epsilon^{\mu\nu\rho\sigma\lambda}F^I_{\mu\nu} F^J_{\rho\sigma} A^K_{\lambda}\\
&+ \alpha \left({\cal L}_{4\partial}+{\Delta{\cal L}}_{2\partial}\right)\,,
\end{aligned}
\end{equation}
where
\begin{equation}\label{eq:4_der_terms_final}
\begin{aligned}
\mathcal L_{4\partial} &= \, \lambda_M X^M \,{\cal X}_{\rm{GB}} +D_{IJ} \,C_{\mu\nu\rho\sigma}\, F^{I\mu\nu} F^{J\rho\sigma} + E_{IJKL}\,F^I_{\mu\nu} F^J{}^{\mu\nu}\, F^K_{\rho\sigma} F^L{}^{\rho\sigma}
\\[1mm]
+&\,{\widetilde E}_{IJKL}\, F^I_{\mu\nu}F^{J\,\nu\rho} \,F^K_{\rho\sigma} F^{L\,\sigma\mu} + I_{IJKL}\,\partial_{\mu} X^I \partial^{\mu} X^J\, \partial_{\nu} X^K \partial^{\nu} X^L+ H_{IJKL}\, \partial_{\mu} X^I  \partial^{\mu} X^J\, F^K_{\rho\sigma}F^L{}^{\rho\sigma}
\\[1mm]
+&\,{\widetilde H}_{IJKL}\, \partial_\mu X^I \partial^\nu X^J \, F^{K\,\mu\rho}F^L_{\nu\rho}-\,6X_I X_J\lambda_K \,  F^{I}{}^{\mu\alpha} F^{J}{}^\nu{}_\alpha\, \nabla_\nu\partial_\mu X^K 
\\[1mm]
+&\,\frac{3}{4}\lambda_{[I}X_{J]}X_K\,\epsilon^{\mu\nu\rho\sigma\lambda}\nabla_\alpha F^I_{\mu\nu}F^J_{\rho\sigma} F^K_\lambda{}^\alpha + W_{IJKL}\, \epsilon^{\mu\nu\rho\sigma\lambda} F^I_{\mu\nu} F^J_\rho{}^\alpha F^K_{\sigma\alpha} \,\partial_\lambda X^L 
\\[1mm]
+&\,\frac{1}{2}\lambda_I\,\epsilon^{\mu\nu\rho\sigma\lambda} R_{\mu\nu\alpha\beta} \,R_{\rho\sigma}{}^{\alpha\beta}A_\lambda^I\,,
\end{aligned}
\end{equation}
being ${\cal X}_{\rm{GB}} = R_{\mu\nu\rho\sigma}^2 - 4 R_{\mu\nu}^2 + R^2$ is the Gauss-Bonnet combination. The four-derivative couplings are given by
\begin{equation}
\begin{aligned}
D_{IJ}=& \,3\lambda_{I} X_{J}-\frac{9}{2}\lambda_{M}X^{M}X_IX_J \,,
\\[1mm]
E_{IJKL} =&\,\lambda_M X^M \left(-\frac{27}{16}X_I X_J X_K X_L - \frac{9}{8}a_{IJ} a_{KL} + \frac{39}{16}a_{IJ} X_K X_L - \frac{3}{4}a_{IK} a_{JL} + \frac{9}{4}a_{IK} X_J X_L \right) \\[1mm]
&- \frac{3}{4}a_{J[I}\lambda_{L]}X_K -\frac{9}{8}\lambda_I X_J X_K X_L 
\,,\\[1mm]
{\widetilde E}_{IJKL}=&\, \lambda_M X^M \left( \frac{81}{8}X_I X_J X_K X_L + 6a_{IJ}a_{KL} + \frac{3}{2}a_{IK} a_{JL} -9X_I X_J a_{KL} - \frac{9}{2}X_I X_K a_{JL}\right)-\\[1mm]
&- \frac{9}{4}X_I X_J X_K \lambda_L -\frac{3}{4}a_{IK} X_J \lambda_L
\,,\\[1mm]
I_{IJKL}=&\, \lambda_M X^M \left( \frac{3}{2}a_{IJ}a_{KL} + 6a_{K(I}a_{J)L}\right)
\,,\\[1mm]
H_{IJKL}=& \,-\frac{3}{2}\lambda_M X^M a_{IJ} a_{KL} - 6\lambda_M X^M a_{IK} a_{JL} + \frac{45}{8}\lambda_M X^M a_{IJ} X_K X_L - \frac{9}{4}a_{IJ} X_K \lambda_L 
\,,\\[1mm]
{\widetilde H}_{IJKL}=&\,\lambda_M X^M\left(12a_{IL}a_{JK} + 6a_{IK}a_{JL}+ 12a_{IJ}a_{KL} - 9a_{IJ}X_K X_L\right)\,\,, \\[1mm]
W_{IJKL}=& -\frac{21}{4}\lambda_J a_{IL} X_K - \frac{3}{4}\lambda_J X_I a_{KL} + 3\lambda_M X^M \left(2a_{IJ} a_{KL} -3X_I X_J a_{KL} \right)
\,.
\end{aligned}
\end{equation}
Finally, $\Delta {\cal L}_{2\partial}$ contains all the two-derivative corrections: 
\begin{equation}
\begin{aligned}
{\Delta {\cal L}}_{2\partial}=\,&-2\Delta {\cal V}+\left\{\frac{1}{2}\left({\widetilde\lambda}_{IJ}+3{\widetilde \lambda} \,a_{IJ}\right) - 3\lambda_M X^M \left[4{\cal V}+9\left (g_M X^M\right)^2\right]a_{IJ} \right\}\,\partial_{\mu} X^I  \partial^{\mu} X^J \\[1mm]
& -\frac{3}{4}\Delta a_{IJ} F^I_{\mu\nu} F^J{}^{\mu\nu}+\frac{1}{4}\left({\widetilde \lambda}_{IJK}- 6\lambda_{(I}g_Jg_{K)}\right)\epsilon^{\mu\nu\rho\sigma\lambda} \,F^I_{\mu\nu} F^J_{\rho\sigma}A_\lambda^K\,,
\end{aligned}
\end{equation}
where ${\Delta}{\cal V}$ is the correction to the scalar potential, whose explicit expression reads
\begin{equation}
\begin{aligned}
\Delta {\cal V}=\,&-\lambda_M X^M \left[\frac{4}{3}\,{\cal V}^2 +18\left (g_M X^M\right)^2{\cal V}+27\left(g_M X^M\right)^4\right]+\frac{1}{3}\,{\widetilde \lambda}\,{\cal V}+\frac{9}{4}\,{\widetilde \lambda}\left(g_I X^I\right)^2\\[1mm]
&+\frac{9}{2} \,{\widetilde \lambda}_{IJ} \,{\cal C}^{IK}{\cal C}^{JL}g_Kg_L\,,
\end{aligned}
\end{equation}
and
\begin{equation}
\begin{aligned}
\Delta a_{IJ}=\,&2\left[{\cal V}+ 3\left (g_M X^M\right)^2\right]\left[\lambda_M X^M \left(3X_I X_J + a_{IJ}\right) - 2\lambda_{(I}X_{J)} \right]+8\lambda_M X^M\, g_I g_J\\[1mm]
& -48 \lambda_K \,{\cal C}^{KL}\,g_L g_{(I} X_{J)}+\frac{2}{3}\lambda_M X^M {\cal V} \left(a_{IJ}-2X_IX_J\right)\\[1mm]
&-\frac{1}{3}{\widetilde \lambda}_{IJ} - {\widetilde \lambda} \left(3X_I X_J +\frac{1}{3}\,a_{IJ}\right)+2{\widetilde \lambda}_{I} X_J\,.
\end{aligned}
\end{equation}

We conclude observing that the contribution from the invariant controlled by the coupling ${\widetilde \lambda}_{IJK}$ can be cast as the original two-derivative Lagrangian,
\begin{equation}\label{eq:tilded_Lagr}
{\cal L}|_{\lambda_I=0}= R-2{\widetilde{\cal V}}-\frac{3}{2}\,  {\widetilde a}_{IJ}\, \partial_{\mu} {\widetilde X}^I \partial^{\mu} {\widetilde X}^J-\frac{3}{4}\, {\widetilde a}_{IJ}\,F^I_{\mu\nu} F^J{}^{\mu\nu}
+\frac{1}{4}{\widetilde C}_{IJK}\epsilon^{\mu\nu\rho\sigma\lambda}F^{I}_{\mu\nu} F^{J}_{\rho\sigma} A^{K}_{\lambda}\, ,
\end{equation}
with a shifted Chern-Simons coupling
\begin{equation}\label{eq:subtildeaction1}
{\widetilde C}_{IJK}\,=\, C_{IJK}+\alpha \,{\widetilde \lambda}_{IJK}\,.
\end{equation}
In \eqref{eq:tilded_Lagr}, the tilded quantities are defined as in the two-derivative theory just using ${\widetilde C}_{IJK}$ instead of $C_{IJK}$. In particular the scalars ${\widetilde X}$ satisfy the cubic constraint with ${\widetilde C}_{IJK}$, which means that they are given in terms of $X^I$ by
\begin{equation}\label{eq:subtildeaction2}
{\widetilde X}^I=X^I\left(1-\frac{\alpha}{3}\,{\widetilde \lambda}\right)\, .
\end{equation}

\section{Supersymmetric black hole action and holographic match}\label{sec:onshellaction}

Given a holographic $\mathcal{N}=1$ SCFT$_4$, one may expect that the higher-derivative five-dimensional gauged supergravity reproducing its 't Hooft anomalies admits an asymptotically AdS$_5$ supersymmetric black hole solution whose on-shell action matches the large-$N$ expansion of the formula~\eqref{eq:index_asympt}. In this section we prove the validity of such expectation at linear order in the four-derivative corrections for the specific model dual to the $\mathbb{C}^3/\mathbb{Z}_\nu$ quiver theories, in the case of equal angular velocities, $\omega_1=\omega_2$.

We begin  in section \ref{sec:holographicdictionary} by providing the dictionary between the SCFT 't Hooft anomaly coefficients and the supergravity couplings. We will also specify the choices of couplings $\lambda_I$ and $\widetilde\lambda_{IJK}$ that we will consider. 
In section \ref{sec:blackhole} we review the black hole solution of the two-derivative theory and its thermodynamics; then we take the supersymmetric (and extremal) limit. We next turn to the four-derivative theory: in section \ref{sec:boundaryterms} we discuss the boundary terms to be included in order to holographically renormalize the theory, and in section \ref{sec:setupaction} we revisit the argument showing that we do not need the corrected solution to evaluate the four-derivative action at linear order in the corrections. In section~\ref{sec:resultsandmatch} we present our final result for the supersymmetric on-shell action, matching the large-$N$ expansion of~\eqref{eq:index_asympt}. 

\subsection{Holographic dictionary for anomaly coefficients}\label{sec:holographicdictionary}

The precise holographic dictionary between the field theory 't Hooft anomaly coefficients and the supergravity Chern-Simons couplings can be obtained by equating the anomalous variation of the respective partition functions under a transformation induced by the charges $Q_I$ \cite{Witten:1998qj}.\footnote{Equivalently, we could take the formal exterior derivative of the supergravity Chern-Simons terms and  compare the resulting six-form with the SCFT anomaly polynomial~\eqref{anomaly_poly}. }

We will use a hat to indicate the field theory background fields on $\mathcal{M}_4$ and $i,j,k$ for the spacetime indices on $\mathcal{M}_4$.\footnote{These should not be confused with the SU(2) indices $i=1,2$ used in section~\ref{sec:fourder_action}, which however will not appear in the present section.} We denote by $J^k_I$ the SCFT current associated with $Q_I$, by $\hat A^I_k$ the background gauge field that canonically couples to it, transforming as
\be
\label{gauge_var_SCFT}
\delta_{\Lambda} \hat{A}^I_k = \partial_k\Lambda^I\,,
\ee 
and by $\hat F^I_{jk}= 2\partial_{[j}\hat A^I_{k]}$ its field strength. Then the variation of the SCFT partition function reads (in Lorentzian signature),
\begin{equation}\label{variation_ZCFT}
\begin{aligned}
\delta_{\Lambda} \log Z_{\rm CFT} \,&=\,  -i \int_{{ \cal M}_4}\diff^4 x\, \hat e \, \Lambda^I \,\nabla_k  J^k_I \\[1mm]
\,&=\, - \frac{i}{{96\pi^2}} \int_{{ \cal M}_4}\diff^4 x\, \hat e \, \Lambda^I \left( k_{IJK} \,\hat\epsilon^{\,ijkl}\hat {F}^J_{ij} \hat {F}^K_{kl} \, - \,  \frac{1}{8}\,k_I\,\hat\epsilon^{\,ijkl}\hat {R}_{ijab} \hat {R}_{kl}{}^{ab}\right)\,.
\end{aligned}
\end{equation}

This is to be compared with the corresponding variation of the gravitational partition function.
 In the classical saddle-point approximation, the gravitational partition function is given by the renormalized supergravity action evaluated on-shell. For the action given in section~\ref{sec:final_Lagr}, the non-invariant sector made of the Chern-Simons terms yields under the variation \eqref{gauge_var_SCFT},\footnote{When we apply the Stokes theorem and pass from the bulk to the boundary, we introduce an unusual minus sign. This is because the orientation we use for $\mathcal{M}_4 = \partial \mathcal{M}$ is opposite to the orientation induced from the bulk by contracting the bulk volume form with the vector $\frac{\partial}{\partial r}$ normal to the boundary. Concretely, in this section the positive orientation in the bulk is given by $\epsilon_{tr123}>0$, while the positive orientation in the boundary is given by $\epsilon_{t 123}>0$, where $1,2,3$ label the coordinates for the spatial slices of the boundary.}
\begin{equation}\label{variation_Zgrav}
\begin{aligned}
\delta_{\Lambda} \log Z_{\rm grav} &\simeq i\, \delta_{\Lambda} S\\[1mm]
\,=&\, -\frac{i}{16 \pi G g^3}\int_{\partial {\cal M}} \diff^4x \,\hat{e}\, \Lambda^I \left( \frac{1}{4}\, C^{\rm (\alpha)}_{IJK} \,\hat{\epsilon}^{\,ijkl}\hat{F}^J_{ij}\hat{F}^K_{kl} \,+\, \frac{\alpha\lambda_I g^2}{2}\,\hat{\epsilon}^{\,ijkl}\hat{R}_{ijab}\hat{R}_{kl}{}^{ab}\right)\,,
\end{aligned}
\end{equation}
where we have identified the supergravity gauge potentials $A^I$ restricted to the conformal boundary $\partial \mathcal{M}$ with the field theory background gauge potentials $\hat A^I$ on $\mathcal{M}_4 =\partial\mathcal{M}$ via 
\be
A^I|_{\partial \mathcal{M}} \,=\, \frac{\hat A^I}{g} \,,
\ee
 $g$ being a parameter with the dimensions of a mass, which will later appear as the inverse radius of the two-derivative AdS solution. Here it is needed in order to make the mass dimensions consistent.\footnote{It follows that the gravitational electric charges and the respective electrostatic potentials will also carry an extra factor of $g$ compared to the corresponding field theory quantities. This will play a significant role when we will take the  limit $g\to 0$ in section~\ref{sec:ungaugedlimit}.}
We have also introduced the corrected gauge Chern-Simons coupling,
\be\label{correctedCIJK}
C^{\rm (\alpha)}_{IJK} \,=\,  C_{IJK} -6 \alpha \lambda_{(I} g_J g_{K)} + \alpha \widetilde\lambda_{IJK} \,.
\ee

Comparing \eqref{variation_ZCFT} and \eqref{variation_Zgrav}, we obtain the desired holographic dictionary:
\begin{align}
k_{IJK} \,&=\,  \frac{ 3\pi }{2Gg^3} \,C^{(\alpha)}_{IJK}  \,, \label{dict_cubic_anom} \\[1mm]
k_I \,&=\, -\frac{24\pi}{Gg}\, \alpha\lambda_I\,. \label{dict_linear_anom}
\end{align}
The first relation can be split into leading-order and correction terms as:
\be
k^{(0)}_{IJK} \,=\,  \frac{ 3\pi }{2Gg^3} \,C_{IJK} \,,\qquad   k^{(1)}_{IJK} \,=\,   \frac{ 3\pi\alpha }{2Gg^3} \,  \big(-6  \lambda_{(I} g_J g_{K)} +  \widetilde\lambda_{IJK}\big)\,.
\ee

Using the dictionary above, we can rephrase the SCFT formula \eqref{eq:index_asympt} in gravitational variables as
\be\label{eq:gravityactionprediction}
I \, =\,    \frac{ \pi }{4G} \, \frac{C^{(\alpha)}_{IJK} \varphi^I_{\rm gr}\varphi^J_{\rm gr}\varphi^K_{\rm gr}}{\omega_1\omega_2} + \frac{\pi}{G}\, \alpha\lambda_I \varphi_{\rm gr}^I \,\frac{\omega_1^2+\omega_2^2-4\pi^2}{\omega_1\omega_2} \,,
\ee
with
\be\label{eq:constraint_gravity}
\omega_1+ \omega_2  -  \frac{3}{\sqrt2} \, g_I  \varphi_{\rm gr}^I \,=\, \pm 2\pi i \,,
\ee
where $\varphi^I_{\rm gr}=g^{-1}\varphi^I$ and in the linear constraint we have used the identification 
\be\label{eq:rel_rI_gI}
 r_I \,= \, \frac{3}{2\sqrt2} \frac{g_I}{g}  \,,
 \ee
 which is derived in Appendix~\ref{app:sugradual}. The precise way the  supersymmetric chemical potentials $\varphi_{\rm gr}^I$ and $\omega$ should be evaluated on the gravity side will be specified below.
In the following we check the expectation that there exists a corrected supersymmetric AdS$_5$ black hole solution whose on-shell action matches \eqref{eq:gravityactionprediction} in the model dual to the $\mathbb{C}^3/\mathbb{Z}_\nu$ quiver theories of section~\ref{sec:orbifold_section}, for the case of equal angular velocities.

\subsection{Specialization to the orbifold theories}\label{sec:special_relations}
 
Since we eventually want to match the gravitational action with the $\mathbb{C}^3/\mathbb{Z}_\nu$ quiver theories  in section~\ref{sec:orbifold_section} for generic $\nu$, we will take the supergravity Lagrangian \eqref{eq:finaloffshellLagrangian} with $n =2$ vector multiplets (thus ignoring the baryonic symmetry available for even $\nu$). This contains Abelian vector fields $A_\mu^I$, $I=1,2,3$, coupling to the three global symmetries of the dual quantum field theories, and is known as U(1)$^3$ model. 

As a first thing, we observe that the general assumption we have made in~\eqref{relation_k3_k1}, which is in particular satisfied by the orbifold theories discussed in section~\ref{sec:orbifold_section}, translates via the dictionary above into the following relation,
\be
C^{(\alpha)}_{IJK} \,=\,   C_{IJK}  - 18 \alpha  \lambda_{(I} g_J g_{K)}  \,,
\ee
Comparing with \eqref{correctedCIJK}, this means that we need to choose
\be\label{choice_tildelambda}
\widetilde\lambda_{IJK} \,=\,   - 12   \lambda_{(I} g_J g_{K)} \,,
\ee
demonstrating that in order to discuss this  class of theories both couplings $\lambda_{I}$ and $\widetilde\lambda_{IJK}$ are needed. Notice that with this choice the prediction \eqref{eq:gravityactionprediction} for the on-shell action can be rewritten in the simpler form
\be\label{eq:gravityactionpredictionBis}
I \, =\,    \frac{ \pi }{4G} \, \frac{C_{IJK} \varphi^I_{\rm gr}\varphi^J_{\rm gr}\varphi^K_{\rm gr}}{\omega_1\omega_2} - \frac{2\pi}{G}\, \alpha\lambda_I \varphi_{\rm gr}^I \,\left[ 1 \mp 2\pi i \left(\frac{1}{\omega_1}+\frac{1}{\omega_2}\right) \right]\,.
\ee

We now focus more specifically on the gravity dual of the $\mathbb{C}^3/\mathbb{Z}_\nu$ quiver theories of section~\ref{sec:orbifold_section} for generic $\nu$. Using the dictionary above as well as information from appendix~\ref{sec:amaxim_section}, the SCFT anomaly coefficients \eqref{tHooft_coeff_orbifolds} translate into the following supergravity couplings,
\be
C_{IJK}  =  \frac{1}{6}|\epsilon_{IJK}|\,, \qquad\ \widetilde\lambda_{IJK} =   - 12   \lambda_{(I} g_J g_{K)} \,,\qquad\ \lambda_I = \frac{ g_I}{8\sqrt2 \,g}\,, \qquad   I=1,2,3\,,
\ee
with 
\be
g_I = \frac{\sqrt{2}}{3}\,g \,,
\ee 
and
\be
\frac{\pi}{2Gg^3}= \nu N^2\,,\qquad\quad \alpha g^2 = \frac{1}{4N^2}\,.
\ee
Also, the Weyl anomaly coefficients  \eqref{a_c_orbifolds}  can be expressed in gravitational units as
\be
\aa  \,=\,   \frac{\pi}{8Gg^3} \,  \left(1  -3\alpha g^2   \right)  \,,\qquad\quad
\cc \,=\, \frac{\pi}{8Gg^3} \,  \left(1  - 2 \alpha g^2 \right) \,.
\ee

\subsection{The two-derivative black hole solution}\label{sec:blackhole}

A general asymptotically AdS black hole solution of the model specified above has three independent electric charges $Q^{\rm gr}_I$ and two angular momenta $J_1$, $J_2$. Here we will focus on the case of equal angular momenta $J_1 = J_2\equiv J$. At the two-derivative level, the supersymmetric and extremal solution in this regime has been derived in~\cite{Gutowski:2004yv}, while the non-supersymmetric, thermal solution has been found in~\cite{Cvetic:2004ny}. The corrections to the solution introduced by the four-derivative terms are not known, however as we will see this still allows us to obtain the on-shell action at first order in the corrections.

The action of the two-derivative ${\rm U}(1)^3$ model reads
\begin{equation}\label{eq:u1cubemodelaction}
\begin{aligned}
S\, =\, \frac{1}{16\pi G}\int \diff^5x\, e\,& \bigg[  R + 4g^2\sum_{I=1}^3\left( X^I\right)^{-1}-\frac{1}{2}\partial\vec\phi^{\,2} -\frac{1}{2}\sum_{I=1}^3\left(X^I\right)^{-2}\,F^I_{\mu\nu} F^{I\,\mu\nu}\\
&\ +\frac{1}{24}\epsilon^{\mu\nu\rho\sigma\lambda}\, |\epsilon_{IJK}\, |  F^I_{\mu\nu} F^J_{\rho\sigma} A^K_\lambda
\bigg]\,,
\end{aligned}
\end{equation} 
where $A^I\,,\,I=1,2,3$, are Abelian gauge fields, $\vec \phi = \left(\phi_1,\phi_2\right)$ are two real scalar fields and
\begin{equation}
X^1 = {\rm e}^{-\frac{1}{\sqrt{6}}\phi_1 - \frac{1}{\sqrt{2}}\phi_2}\,,\quad\quad X^2 =  {\rm e}^{-\frac{1}{\sqrt{6}}\phi_1 +\frac{1}{\sqrt{2}}\phi_2}\,,\quad\quad X^3= {\rm e}^{\frac{2}{\sqrt{6}}\phi_1}\,,
\end{equation}
satisfy the constraint $X^1 X^2 X^3=1$.
This model is a consistent truncation of type IIB supergravity on $S^5/\mathbb{Z}_\nu$, where the Abelian gauge fields arise as KK vectors gauging the  U(1)$^3\subset$ SO(6) isometries of $S^5$. The action \eqref{eq:u1cubemodelaction} follows from the general expression \eqref{eq:2dSUGRA}  by taking
\begin{equation}\label{eq:conventionsveryspecialgeometry}
C_{IJK} = \frac{1}{6}|\epsilon_{IJK}|\,,\qquad\quad a_{IJ} =\frac{1}{3\left(X^I\right)^2}\delta_{IJ}\,,\qquad\quad g_I = \frac{\sqrt{2}}{3}\,g\,,\qquad\quad I,J,K=1,2,3\,.
\end{equation}
The scalar potential $\mathcal V = -2g^2\sum_{I=1}^3\left(X^I\right)^{-1}$ is extremized for
\be
\bar X^I=1  \quad \Rightarrow\quad   \bar X_I = C_{IJK}\bar X^J\bar X^K = \frac{1}{3}   \,,\qquad\quad I=1,2,3\,.
\ee 
Throughout this section, we denote by a bar the AdS$_5$ value of the scalar fields and functions thereof. It follows that the corresponding AdS$_5$ solution has radius $g^{-1}$. The gauging parameters can be expressed in terms of the vacuum value of the scalars as
\begin{equation}\label{eq:gaugingparameters}
g_I = \sqrt{2}\,g\,\bar X_I\,.
\end{equation}
This relation shows that the AdS$_5$ vacuum solution is supersymmetric, see the derivation of~\eqref{scalars_AdS_sol} in the appendix.

The metric, gauge, and scalar fields for the asymptotically AdS$_5$ black hole solution can be expressed in a non-rotating frame at infinity using the coordinates $\left(t,r,\theta,\phi,\psi\right)$ as
\begin{equation}\label{eq:metric5d}
\text ds^2 = \left(H_1 H_2 H_3\right)^{1/3}\left[ -\frac{r^2 Y}{f_1}\text dt^2+ \frac{r^4}{Y}\text dr^2+\frac{r^2}{4}\left(\sigma_1^2 + \sigma_2^2\right)+\frac{f_1}{4r^4H_1H_2H_3}\left(\sigma_3 -\frac{2f_2}{f_1}\text dt\right)^2
\right]\,,
\end{equation}
\begin{equation}\label{eq:scalars5d}
A^I = \left(\frac{2m}{r^2H_I}s_I\,c_I + z_I\right)\text dt + \left(\frac{m\,a}{r^2 H_I}\left(c_I\, s_J\,s_K-s_I\,c_J\,c_K\right)\right)\sigma_3\,,\quad\quad X^I = \frac{\left(H_1 H_2 H_3\right)^{1/3}}{H_I}\,,
\end{equation}
where the indices $I,J,K$ in $A^I$ are never equal, and the constant parameters  $z_I$ will be fixed later as  gauge choices. The solution is given in terms of the SU(2) left-invariant one-forms parametrized by the Euler angles on $S^3$, $\theta \in [0,\pi]$, $\phi\in[0,2\pi]$ and $\psi\in[0,4\pi]$,
\begin{equation}
\sigma_1 = \cos\psi \, \text d\theta + \sin\psi\sin\theta \, \text d\phi \,,\quad \sigma_2 = -\sin\psi \, \text d\theta + \cos\psi \sin\theta \, \text d\phi\,,\quad \sigma_3 = \text d\psi + \cos\theta \, \text d\phi\,.
\end{equation}
We have also introduced the following radial functions,
\begin{align}
H_I(r) &= 1+\frac{2m \,s_I^2}{r^2}\,, \nn \\[1mm] 
f_1(r)&= 4 a^2 m^2 \left[2\,s_1 s_2 s_3 \left(c_1c_2c_3-s_1s_2s_3\right)-s_1^2 s_2^2-s_1^2 s_3^2-s_2^2 s_3^2\right]+2 a^2 m\, r^2+H_1H_2 H_3\,  r^6, \nn \\[1mm]
f_2(r) &= 2m\,a\left(c_1c_2c_3-s_1s_2s_3\right)r^2+ 4m^2as_1s_2s_3\,,\nn\\[1mm]
f_3(r)&= 4 g^2a^2 m^2 \left[2\, s_1 s_2 s_3 \left(c_1c_2c_3-s_1s_2s_3\right)-s_1^2 s_2^2-s_1^2 s_3^2-s_2^2 s_3^2\right]+2 a^2 m \left(g^2 r^2+1\right)\,, \nn \\[1mm]
Y(r) &= f_3(r) + g^2 r^6H_1H_2H_3 + r^4 - 2m\, r^2\,.
\end{align}
The parameters $s_I$ and $c_I$ are shorthand notations for $s_I = \sinh \delta_I$ and $c_I =\cosh\delta_I$. Therefore, the black hole depends on five independent parameters $(m,\delta_{1},\delta_2,\delta_3,a)$, roughly corresponding to the five independent conserved charges (mass, three electric charges and angular momentum). The black hole has a Killing horizon, whose location, denoted by $r_+$, is given by the largest positive root of $Y(r)$.

The thermodynamical chemical potentials of the non-extremal black hole are the angular velocity $\Omega$, the electrostatic potentials $\Phi^I$ and the inverse Hawking temperature $\beta$. The angular velocity is read from the condition that the Killing vector $\xi = \partial_t + \Omega\,\partial_\psi$ becomes null at the Killing horizon. The electrostatic potentials are defined by the gauge invariant combination $\Phi^I = \iota_\xi A^I|_{r_+}-\iota_\xi A^I|_{+\infty}$. Their expressions are
\begin{equation}\label{eq:thermalchemicalpotentials2d}
\Omega = 2\frac{f_2(r_+)}{f_1(r_+)}\,,\quad\quad \Phi^I = \frac{2m}{r_+^2 H_I(r_+)}\left(s_I c_I + \frac{1}{2}a\,\Omega\left(c_I \,s_J \,s_K - s_I\, c_J \,c_K\right)\right)\,,\quad I\neq J\neq K\,.
\end{equation}
The inverse Hawking temperature is identified with the  period $\beta$ of the compactified Euclidean time $\tau = it$; this is fixed by regularity of the Euclideanized solution to
\begin{equation}\label{eq:inversetemperature2d}
\beta = 4\pi\,r_+ \sqrt{f_1(r_+)}\left(\frac{\text d Y}{\text dr}\,(r_+)\right)^{-1}\,.
\end{equation}
Regularity of the Euclidean solution at $r_+$ also fixes the gauge constants $z_I$ introduced in \eqref{eq:scalars5d} through the requirement that the gauge field has no component along the shrinking direction identified by the Killing vector at the horizon, that is
\begin{equation}
\iota_\xi A^I|_{r_+}=0 \quad\implies\quad z_I = -\Phi^I\,.
\end{equation}
As a consequence, the electrostatic potentials can be read off from the asymptotic boundary value of the gauge fields. The same holds for the angular velocity upon changing the angular coordinate $\psi$ so that the connection encoded in the $g_{t\psi}$ component of the metric vanishes at $r_+$.

The conserved charges read~\cite{Cvetic:2005zi}
\begin{equation}\label{eq:conservedcharges2d}
\begin{aligned}
E \,&=\, \frac{\pi}{4G}\,m\left(3 + g^2 a^2+ 2 s_1^2 + 2s_2^2 + 2s_3^2\right)\,,\\
J \,&=\, \frac{\pi}{2G}\,m\,a\left(c_1c_2c_3-s_1s_2s_3\right)\,,\\
Q^{\rm gr}_I \,&=\, \frac{\pi}{2G}\,m\,s_I\,c_I\,,
\end{aligned}
\end{equation}
while the Bekenstein-Hawking entropy is given by
\begin{equation}\label{eq:entropy2d}
\mathcal S \,=\, \frac{\pi^2}{2G}\sqrt{f_1(r_+)}\,.
\end{equation}

The expression for the Euclidean on-shell action of the two-derivative theory \eqref{eq:u1cubemodelaction} for the black hole solution \eqref{eq:metric5d}, \eqref{eq:scalars5d} can be found in~\cite{Cassani:2019mms}. We will distinguish between the action obtained there using holographic renormalization with a minimal subtraction scheme, that we denote by $I_{Z}$, and the action $I$ of interest here. These are related as $I = I_{Z}-I_{\rm AdS}$, where $I_{\rm AdS}$ denotes the contribution of the AdS vacuum, which in the scheme used in~\cite{Cassani:2019mms}  reads
\begin{equation}
I_{\rm AdS} = \beta E_{\rm AdS}\,\,,\quad\quad E_{\rm AdS} = \frac{3\pi}{32g^2\,G}\,,
\end{equation}
where $E_{\rm AdS}$ is the energy of the AdS$_5$ vacuum solution, dual to the field theory Casimir energy.  
Similarly, we will use $E = E_{\rm Z}-E_{\rm AdS}$, where $E_{\rm Z}$ is the energy computed using holographic renormalization as in~\cite{Cassani:2019mms}. The reason for this subtraction is that we want to describe the gravity dual of the superconformal index $\mathcal{I}$ as opposed to the SCFT partition function $Z$ computed via the path integral. The two are related by the contribution of the vacuum in the path integral, that is $Z = \rme^{-I_{\rm vac}}\, \mathcal{I} $, which in the saddle point approximation $Z\sim \rme^{-I_{Z}}$, $\mathcal{I}\sim \rme^{-I}$ becomes  $ I= I_{Z} - I_{\rm vac}$. Note that while $I_{Z}$ is sensitive to the scheme used, $I$ is not, since any finite counterterm that is added to our choice of boundary terms would contribute in the same way to the black hole action and the AdS action.

It can be shown that the following quantum statistical relation holds, 
\begin{equation}\label{eq:qsr}
I= \beta E -\mathcal S - \beta\Omega J -\beta\Phi^I Q^{\rm gr}_I\,,
\end{equation} 
showing that $I$ is the Legendre transform of the entropy and therefore, since the latter is a  function of the conserved quantities $(E,J,Q_I)$, should be seen as a function of the  chemical potentials, $I=I\left(\beta,\Omega,\Phi^I\right)$. This leads to the interpretation of the Euclidean solution with action $I$ as a saddle of a grand-canonical partition function, $Z_{\rm grand}\simeq \rme^{-I}$.
The interpretation is in agreement with the fact that the on-shell action with Dirichlet boundary conditions is a function of the conformal boundary values of the bulk fields~\cite{Papadimitriou:2005ii}, and that these just encode the chemical potentials once regularity of the Euclidean solution is imposed.

\paragraph{Turning on the baryonic charge.} Recall from section~\ref{baryonic_sec} that for even $\nu$ the $\mathbb{C}^3/\mathbb{Z}_\nu$ quivers also have a baryonic symmetry. It is thus natural to ask if the gravity dual to these models admits a black hole solution where the charge corresponding to the baryonic charge is also turned on.
At the two derivative level, this should solve the equations of motion of a  five-dimensional supergravity theory effectively capturing all cubic 't Hooft anomalies of the dual SCFT, including those involving the baryonic symmetry. This theory has to include the vector field that gauges the baryonic symmetry in the bulk and that should uplift to the massless twisted sector of type IIB string theory on $S^5/\mathbb{Z}_\nu$. We would thus consider a supergravity model featuring three vector multiplets and an Abelian gauging, with gauge coupling constants $g_1=g_2=g_3\neq 0$ and $g_B = 0$. The supersymmetric and extremal two-derivative solution may be closely related to the one of~\cite{Gutowski:2004yv,Kunduri:2006ek} with this choice of gauging parameters: the solution given in these references should just be adapted to take into account that $C^{IBB}$ (the supergravity dual of $k^{(0)IBB}$)  is not obtained by raising the indices of $C_{IBB}$ (the dual of $k^{(0)}_{IBB}$) via the Kronecker delta, as it is assumed there. On the other hand, the more general non-extremal solution is currently not known. We leave for future work the study of this solution and its possible uplift.

\subsection{Supersymmetric thermodynamics}\label{sec:BPSlimit}

The supersymmetric and extremal black hole~\cite{Gutowski:2004yv} develops and infinitely long AdS$_2$ throat in its near-horizon geometry, for this reason in order to obtain a BPS\footnote{By BPS we mean a regime in which the black hole is both supersymmetric and extremal. BPS quantities are denoted with a ``$*$''.} 
 black hole thermodynamics it is convenient to turn on some temperature as a regulator. We will follow the method of~\cite{Cabo-Bizet:2018ehj} and take the limit in two steps: first we impose the supersymmetry condition (but not the extremality one), evaluating the black hole on-shell action together with all related thermodynamic quantities in this regime, and only at the end we take the extremal limit. The advantage of this method is that the supersymmetric action turns out to be independent of the temperature, hence the extremal limit is smooth. Moreover, the supersymmetric non-extremal solutions appear to precisely capture the saddles of the dual superconformal index. The price to pay is that these configurations correspond to a complexified section of the original solution. 
At the two-derivative level, the supersymmetric limit of black hole thermodynamics for the solution under study has been discussed in~\cite{Cassani:2019mms}. We review here the main steps as a preparation to the four-derivative case, which will follow the same pattern.

The black hole is supersymmetric when the parameter $a$ satisfies the condition~\cite{Cvetic:2005zi}
\begin{equation}\label{eq:susycondition}
a\,=\,\frac{1}{g\,\text t_1 \text t_2\text t_3}\,,
\end{equation}
where we have defined $\text t_I \equiv \rme^{\delta_I}$.
As a consequence of this condition, the black hole charges \eqref{eq:conservedcharges2d} satisfy the linear relation
\begin{equation}
E \,=\, 2g\,J + \bar X^I Q^{\rm gr}_I\,.
\end{equation}
This relation does not automatically imply extremality, namely the vanishing of Hawking temperature $\beta^{-1}$. Following~\cite{Cassani:2019mms}, to study the extremal limit $m\rightarrow m^*$, it is more convenient to trade the parameter of the (Euclidean) black hole $m$ for the position of the outer horizon $r_+$, by solving $Y(r_+)=0$. Since this is a third order equation in $m$, its solutions are better manageable after a suitable change of coordinate, such that
\begin{equation}\label{eq:bpsradialcoord}
r^2\,=\, \frac{\zeta^2}{g^2}-2m\,s_1^2\,.
\end{equation}
In terms of the new coordinate $\zeta$, $Y(\zeta_+)=0$ becomes of second order in $m$ and can be easily solved as
\begin{equation}\label{eq:defmplus}
\begin{aligned}
&\frac{m_\pm}{m_*} =\\
=&\frac{\zeta_+^4\left( 2\left(\text t_1^4+1\right)\text t_2^2\,\text t_3^2-\text t_1^2\left(1+\text t_2^2\,\text t_3^2\right)\left(\text t_2^2+\text t_3^2\right)\right)+2\zeta_+^2\left(-2 + \text t_2^2\,\text t_3^2\left(1 + \text t_1^4\right)\right)-4\pm\left(\zeta_*^2-\zeta_+^2\right)\Upsilon}{2\zeta_*^2\left(-1-2\text t_2^2\,\text t_3^2+\text t_1^2\left(\text t_1^2\,\text t_2^2\,\text t_3^2+\text t_2^2+\text t_3^2\right) + \zeta_+^2\left(\text t_1^2-\text t_2^2\right)\left(\text t_1^2-\text t_3^2\right)\text t_2^2\,\text t_3^2\right)}
\end{aligned}
\end{equation}
where
\begin{equation}
\Upsilon= \frac{2}{\zeta_*^2}\sqrt{4+4\zeta_+^2\left( 1-\text t_1^4\,\text t_2^2\,\text t_3^2\right)+\zeta_+^4\,\text t_1^4\left(\text t_2^2-\text t_3^2\right)^2}\,,
\end{equation}
and
\begin{equation}\label{eq:extremalcondition}
\zeta_*^2 = \frac{2}{-1+\text t_2^2\,\text t_3^2}\,\,,\quad\quad m_* = \frac{4\text t_1^2\,\text t_2^2\,\text t_3^2}{g^2\left(-1+\text t_1^2\text t_2^2\right)\left(-1+\text t_1^2\text t_3^2\right)\left(-1+\text t_2^2\text t_3^2\right)}\,.
\end{equation}
Notice that as $\zeta_+$ becomes sufficiently close to $\zeta_*$, $m_\pm$ become imaginary due to the square-root in $\Upsilon$. Therefore, for general $\zeta_+$ we have identified a family of supersymmetric, complexified and non-extremal Euclidean solutions. In the extremal limit, that is obtained by sending $\zeta_+\rightarrow \zeta_*$, the solutions $m_\pm$ become real and equal to each other, and coincide with the BPS value $m_*$ in \eqref{eq:extremalcondition}. 
The supersymmetric non-extremal Lorentzian solution has in general closed time-like curves~\cite{Cvetic:2005zi}, but it turns out that the condition for avoiding them is to take the parameter $m$ to its extremal value \eqref{eq:extremalcondition}. Therefore, the BPS solution is regular also in Lorentzian signature.\footnote{The BPS solution was first derived in~\cite{Gutowski:2004yv}. The latter depends on three independent parameters $\mu_I$ , $I=1,2,3$, related to the $\delta_I$ as
$
\mu_I = \frac{2}{\rme^{2(\delta_J + \delta_K)}-1}\,,
$
where $I,J,K$ can only assume different values. The BPS metric is regular if 
$
\mu_I >0\,,$ and $ 4\mu_1\mu_2\mu_3\left( 1+ \mu_1 + \mu_2 + \mu_3\right) > \left( \mu_1 \mu_2 + \mu_1 \mu_3 + \mu_2\mu_3\right)^2\,.
$
} 
The position of the BPS horizon can be expressed in terms of the original $r$ coordinate as
\begin{equation}\label{eq:rstar}
r_*^2 =\frac{2}{g^2} \frac{1-\text t_1^2\,\text t_2^2-\text t_2^2\,\text t_3^2-\text t_1^2\,\text t_3^2+2\text t_1^2\,\text t_2^2\,\text t_3^2}{\left(-1+\text t_1^2\text t_2^2\right)\left(-1+\text t_1^2\text t_3^2\right)\left(-1+\text t_2^2\text t_3^2\right)}\,.
\end{equation}
In the supersymmetric and extremal limit the chemical potentials \eqref{eq:thermalchemicalpotentials2d}, \eqref{eq:inversetemperature2d} become 
\begin{equation}
\begin{aligned}
&\beta^* = +\infty\,,\quad\quad \Omega^* = 2g\,\,,\quad\quad \Phi^*{}^I  = 1\,.
\end{aligned}
\end{equation}

After imposing the supersymmetry condition \eqref{eq:susycondition}, the quantum statistical relation \eqref{eq:qsr} becomes
\begin{equation}\label{eq:susyqsr}
I  \,=\, -\mathcal S - \omega J - \varphi_{\rm gr}^I Q^{\rm gr}_I\,,
\end{equation} 
where we have introduced the supersymmetric chemical potentials 
\begin{equation}\label{eq:susychempot}
\omega= \beta\left( \Omega-\Omega^*\right)\,,\quad\quad \varphi^I_{\rm gr} = \beta\left(\Phi^I-\Phi^*{}^I\right)\,.
\end{equation}
Using  \eqref{eq:susycondition}, one can check that these satisfy the linear constraint
\begin{equation}\label{eq:linearconstraint_grav}
\omega- g\, (\varphi^1_{\rm gr} + \varphi^2_{\rm gr} + \varphi^3_{\rm gr})\,=\, \pm 2\pi i\,,
\end{equation}
which ensures the correct (anti)periodicity of the Killing spinor. The sign choice  corresponds to the two branches for the supersymmetric solution given by sign choice in \eqref{eq:defmplus}. The supersymmetric on-shell action takes the simple form
\begin{equation}
I\,=\, \frac{\pi}{G}\frac{\varphi^1_{\rm gr}\varphi^2_{\rm gr}\varphi^3_{\rm gr}}{\omega^2}\,,
\end{equation}
which is independent of the inverse temperature $\beta$. Recalling \eqref{eq:conventionsveryspecialgeometry}, these findings agree with~\eqref{eq:gravityactionprediction}, \eqref{eq:constraint_gravity}, setting $\alpha=0$ and $\omega_1=\omega_2=\frac{\omega}{2}$ there.

\subsection{Corrected boundary terms}\label{sec:boundaryterms}

We now turn to the evaluation of the four-derivative corrections to the Euclidean on-shell action.
This is a sum of three different contributions
\begin{equation}
I_{ Z} \,=\, I_{\rm bulk} + I_{\rm GH} + I_{\rm count}\,,
\end{equation}
where each term has an expansion in $\alpha$. The bulk action is the Euclidean version of the one given in section \ref{sec:final_Lagr}. We denote by $I_{\rm GH}$ (from Gibbons-Hawking) the terms required to guarantee that the variational problem with Dirichlet boundary conditions on all fields is well posed, while $I_{\rm count}$ is a set of covariant boundary counterterms removing the divergences of the action.

Regarding $I_{\rm GH}$, the only terms that we need are the standard Gibbons-Hawking term for the two-derivative action, and the one associated to the Gauss-Bonnet combination of four-derivative  curvature terms~\cite{Myers:1987yn}. Putting these together, $I_{\rm GH}$ reads
\begin{equation}\label{eq:Igh}
\begin{aligned}
I_{\rm GH}=& -\frac{1}{8\pi G}\int_{\partial\mathcal M} \text d^4 x\sqrt{h}\,K + \\
&+\frac{\alpha}{4\pi G}\int_{\partial\mathcal M} \text d^4 x\sqrt{h}\, \left(\lambda_I X^I\right) \left(\frac{1}{3}K^3- K K^{ij}K_{ij} + \frac{2}{3}K_{ij}K^{jk}K_k{}^i+2 \mathcal G_{ij}K^{ij} \right)\,,
\end{aligned}
\end{equation}
where $K_{ij}$ is the extrinsic curvature of the boundary, $\mathcal G_{ij} = \mathcal R_{ij}- \frac{1}{2}h_{ij} \mathcal R$ is the Einstein tensor of the induced boundary metric $h_{ij}$, with $\mathcal R_{ij}$ built out of $h_{ij}$. Additional boundary terms that are produced by studying the variational problem of the bulk action are either vanishing under Dirichlet boundary conditions or sufficiently suppressed in asymptotically locally AdS spacetimes (see~\cite{Cassani:2023vsa} and references therein for a more thorough analysis). We emphasize that an advantage of having performed the field redefinitions described in Appendix~\ref{app:field_redefinitions}, is that we can use a basis of curvature-square invariants comprising the Gauss-Bonnet combination, for which the appropriate boundary term is known.

We next discuss the counterterm action $I_{\rm count}$. Schematically, this is a sum of three contributions,
\begin{equation}
I_{\rm count} \,=\, I^{(0)}_{\rm count} + I^{(1)}_{\rm count} + \widetilde I^{\,(1)}_{\rm count}\,,
\end{equation}
where $I_{\rm count}^{(0)}$ comprises the counterterms for the $\alpha=0$ theory. If we restrict  to  solutions with no non-normalizable modes for the scalar fields, such as the solution \eqref{eq:metric5d}, \eqref{eq:scalars5d}, then for $I_{\rm count}^{(0)}$ it is sufficient to consider~\cite{Cassani:2019mms}
\begin{equation}\label{eq:Ict0}
I^{(0)}_{\rm count} =- \frac{1}{16\pi G}\int \text d^4x \sqrt{h}\,\left( \mathcal W + \Xi \, \mathcal R\right)\,,
\end{equation}
where 
\begin{equation}\label{eq:countertermsfunction}
\mathcal W = -6g\, \bar X_I X^I\,\,, \quad\quad\quad \Xi = -\frac{1}{2g}\,\bar X^I X_I\,.
\end{equation}
Turning to the corrections, $I^{(1)}_{\rm count}$ defines the counterterms needed to renormalize the higher-derivative terms coming from the terms that couple to $\lambda_I$, whereas $\widetilde I^{\,(1)}_{\rm count}$ is associated to the terms that couple to $\widetilde\lambda_{IJK}$. 
For what concerns $I^{(1)}_{\rm count}$, following the prescription in~\cite{Cremonini:2009ih,Cassani:2022lrk}, we expect that the counterterms are obtained through a shift of order $\alpha$ in the coefficients of the counterterms that are already present in the two-derivative theory. For the minimal supergravity theory, this has been demonstrated by applying the Hamilton-Jacobi method~\cite{Cassani:2023vsa}. We expect that the same prescription is valid here as well, since the divergences of the on-shell action come from $I_{\rm GH}$ rather than from the bulk integral, analogously to the case analysed in the mentioned references. Assuming this ansatz and demanding cancellation of the divergences we arrived at the following counterterm,
\begin{equation}\label{eq:Ict1}
\begin{aligned}
I_{\rm count}^{(0)} + I_{\rm count}^{(1)}=-\frac{1}{16\pi G}&\int_{\partial\mathcal M}\text d^4 x\sqrt{h}\left [\left(1-\frac{16}{3}g^2\alpha\,\lambda_I X^I\left(\bar X_K X^K\right)^2 \right)\mathcal W  \right.\\[1mm]
&\qquad\quad\ \qquad +\left( 1+ 8g^2\alpha\,\lambda_I X^I\left(\bar X_K X^K\right)^2\right)\Xi\,\mathcal R
\bigg ]\,.
\end{aligned}
\end{equation}
In order to obtain $\widetilde I_{\rm count}^{(1)}$, we recall that the theory including the corrections proportional to $\widetilde\lambda_{IJK}$ is equivalent to the two-derivative theory \eqref{eq:tilded_Lagr} together with the substitution rules \eqref{eq:subtildeaction1}, \eqref{eq:subtildeaction2}. Also, considering the supersymmetric AdS$_5$ condition \eqref{scalars_AdS_sol} in the theory \eqref{eq:tilded_Lagr}, we see that  the corresponding inverse AdS radius $\widetilde{g} = \frac{1}{\sqrt2} \,g_I\overline{\widetilde X}{}^I$  is mapped into
\begin{equation}\label{eq:tildeg}
\widetilde g \,=\, g\left( 1- \frac{\alpha}{3}\,\widetilde\lambda_{IJK}\bar X^I \bar X^J \bar X^K  \right)\,.
\end{equation}
The counterterms can, then, be derived starting from the two-derivative expression \eqref{eq:countertermsfunction}, now for the theory with tilded variables, and applying these substitution rules.
We find
\begin{equation}\label{eq:Ict2}
\begin{aligned}
I_{\rm count}^{(0)}+ \widetilde I_{\rm count}^{\,(1)} =&\, -\frac{1}{16\pi G}\int_{\partial\mathcal M} \text d^4 x\sqrt{h}\ \bigg[ \left(1-\frac{\alpha}{3}\,\widetilde\lambda -\alpha\,\widetilde\lambda_{IJK}\bar X^I \bar X^J\bar X^K \right)\mathcal W \\[1mm]
&\,-6g\,\alpha\,\widetilde\lambda_{IJK} X^I\bar X^J\bar X^K+ \left(1-\frac{2}{3}\alpha\,\widetilde\lambda\right)\Xi\,\mathcal R - \frac{\alpha}{6g}\,\bar X^I\,\widetilde\lambda_I\,\mathcal R
\bigg]\,,
\end{aligned}
\end{equation}
where $\widetilde\lambda_I$ and $\widetilde\lambda$ were defined in \eqref{eq:tildel_contracted}.
 
Putting everything together, we can give a set of counterterms for the bulk action \eqref{eq:finalL} that removes all $\sim r^4$ and $\sim r^2$ divergences from the on-shell action of asymptotically AdS solutions,
\begin{equation}\label{eq:generalcounterterms}
\begin{aligned}
I_{\rm count}
&= -\frac{1}{16\pi G}\!\int_{\partial\mathcal M} \!\!\text d^4 x\sqrt{h}\bigg\{\!\!\left[1-\alpha\left(\frac{16}{3}g^2\lambda_I X^I\left(\bar X_J X^J\right)^2+\frac{1}{3}\widetilde\lambda+\,\widetilde\lambda_{IJK}\bar X^I \bar X^J\bar X^K\right)\right] \mathcal W\quad\,\\[1mm]
& -6g\,\alpha\,\widetilde\lambda_{IJK} X^I\bar X^J\bar X^K+\left[\left(1+\alpha\,\Big(8g^2\lambda_I X^I\left(\bar X_J X^J\right)^2-\frac{2}{3}\,\widetilde\lambda\Big)\right)\Xi -\frac{\alpha}{6g}\bar X^I \widetilde\lambda_I\right]\mathcal R
\bigg\}\,.
\end{aligned}
\end{equation}

A rigorous derivation of these counterterms should be possible by adapting the arguments of~\cite{Batrachenko:2004fd,Cassani:2023vsa,Cremonini:2009ih} to the more involved case under study here. 
 However, the present effective prescription appears to give the correct set of counterterms for computing the supersymmetric action we are interested in, as we will discuss later.

\subsection{Setting up the computation}\label{sec:setupaction}

We want to evaluate the on-shell action at linear order in the four-derivative parameter $\alpha$. In order to do this, it is in fact not necessary to know the $\mathcal O(\alpha)$ corrected solution. The argument proving this claim can be found in~\cite{Reall:2019sah} for asymptotically flat solutions, and was adapted to asymptotically AdS spacetimes without scalars in~\cite{Cassani:2022lrk}. Here we revisit it specifying its conditions of validity in the presence of scalar fields. 

Schematically, the on-shell action has the form $I(\Psi)\equiv I^{(0)}(\Psi) + \alpha \, I^{(1)}(\Psi)$, where by $\Psi$ we denote the set of all fields, in this case $\Psi = (g_{\mu\nu} \,,\, A_\mu^I \,,\, X^I(\phi^x))$. When we include $\mathcal O(\alpha)$ corrections, a generic solution of the equations of motion is expressed as $\Psi = \Psi^{(0)} + \alpha\, \Psi^{(1)}$. We can, then, expand the action at first order in $\alpha$ around the leading-order solution,
\begin{equation}
I(\Psi) \,=\, I^{(0)}(\Psi^{(0)})+ \alpha\left( \partial_\alpha I^{(0)}(\Psi^{(0)}, \Psi^{(1)})+ I^{(1)}(\Psi^{(0)}) \right) + \mathcal O(\alpha^2)\,.
\end{equation}
As shown in~\cite{Cassani:2022lrk}, the only term that depends on the corrected solution, that is $\partial_\alpha I^{(0)}$, indeed vanishes when the two-derivative equations of motion are satisfied, up to a boundary term, that has the schematic form
\begin{equation}
\begin{aligned}\label{eq:bdry_term_corrections}
\partial_\alpha I^{(0)} \,\sim\, \int_{\partial\mathcal M_{r_{\rm bdry}}} \!\!\!\text d^4x&\,\bigg [\,\frac{\delta I^{(0)}}{\delta h^{ij}} \bigg |_{\Psi^{(0)}}\,h^{(1)}{}^{ij}+ \frac{\delta I^{(0)}}{\delta A^I_i}\bigg |_{\Psi^{(0)}}\, A_i^{(1)}{}^I+\\
&+\left(\mathtt n^\mu \frac{\delta I^{(0)}_{\rm bulk}}{\delta(\partial^\mu X^I)}+
\frac{\delta I^{(0)}_{\rm count}}{\delta X^I}\right)_{\Psi^{(0)}}\,\left(\delta^I{}_J- X^I X_J\right)X^{(1)}{}^J
\bigg]\,,
\end{aligned}
\end{equation}
where $r_{\rm bdry}$ is a radial cut-off that we introduce to regulate the divergences, such that the boundary $\partial\mathcal M_{r_{\rm bdry}}$ is located at $r=r_{\rm bdry}$ and $\mathtt n$ is the outward pointing normal vector. Eventually, the cutoff is removed by taking $r_{\rm bdry}\rightarrow +\infty$. The projector $\left(\delta^I{}_J- X^I X_J\right)$ ensures that only corrections satisfying the condition $X_I X^{(1)I}=0$ play a role (recall that  in our conventions the scalars satisfy the constraint $X_I X^I =1$ both at the two- and at the four-derivative level). 

The boundary term \eqref{eq:bdry_term_corrections} vanishes if we impose boundary conditions such that the $\mathcal O(\alpha)$ corrections  to the solution do not modify the leading-order asymptotic behaviour of the fields, so as not to modify the original Dirichlet boundary conditions. When scalar fields are present, we should distinguish whether their non-normalizable modes are turned on in the two-derivative solution or not, due to the different asymptotic behaviour in these two cases. We will focus on the  scalars of interest for this paper, which have squared mass $m^2 = -4g^2 $ and couple to operators with conformal dimension $\Delta=2$; their asymtptotic behavior can be found e.g.\ in~\cite{Bianchi:2001kw}, and is $\mathcal O(r_{\rm bdry}^{-2} \,\log r_{\rm bdry}^2)$ for the non-normalizable modes and $\mathcal O(r_{\rm bdry}^{-2})$ for the normalizable modes.

For solutions where the scalars have non-normalizable modes, the asymptotic behaviour of the fields is such that\footnote{In this case we need to use the general set of counterterms given in section 4 of~\cite{Cassani:2018mlh}.} 
\begin{equation}\label{eq:asymptoticboundarybehaviour}
\begin{aligned}
&\frac{\delta I^{(0)}}{\delta h^{ij}}\sim \mathcal O \left(r_{\rm bdry}^2\right)\,,\quad\quad \frac{\delta I^{(0)}}{\delta A_i}\sim \mathcal O \left(r_{\rm bdry}^0\right)\,, \\[1mm]
&\left(\mathtt n^\mu \frac{\delta I^{(0)}_{\rm bulk}}{\delta(\partial^\mu X^I)}+
\frac{\delta I^{(0)}_{\rm count}}{\delta X^I}\right)\left(\delta^I{}_J- X^I X_J\right)\ \sim\ \mathcal O\left(\frac{r_{\rm bdry}^2}{\log r_{\rm bdry}^2}\right)\,\,.
\end{aligned}
\end{equation}
As a consequence of \eqref{eq:asymptoticboundarybehaviour}, the boundary term vanishes if the corrected solution has the asymptotic behavior
\begin{equation}
h_{ij}^{(1)}\,=\, \mathcal O\big(r_{\rm bdry}^0\big) \,,\quad\quad A_i^{(1)}{}^I\,=\,\mathcal O\big(r_{\rm bdry}^{-2}\big)\,,\quad\quad X^{(1)}{}^I\,=\, \mathcal O\big(r_{\rm bdry}^{-2}\big)\,.
\end{equation}
As shown in~\cite{Cassani:2022lrk}, for the metric and the gauge fields this condition can always be met through a choice of gauge, and requires that the induced fields at the conformal boundary are not modified by $\mathcal O(\alpha)$ corrections. This is in harmony with the fact that the Dirichlet boundary conditions define a grand-canonical ensemble, where the chemical potentials, that can be read from the boundary values of the bulk metric and gauge fields after imposing regularity of the Euclidean solution, are kept fixed at their two-derivative expressions when including the corrections. 

However, in this work we are interested in two-derivative solutions where the scalars only have normalizable modes as in  \eqref{eq:scalars5d}, in which case one has the regular behaviour,
\begin{equation}
\left(\mathtt n^\mu \frac{\delta I^{(0)}_{\rm bulk}}{\delta(\partial^\mu X^I)}+
\frac{\delta I^{(0)}_{\rm count}}{\delta X^I}\right)\left(\delta^I{}_J- X^I X_J\right) \ \sim\ \mathcal O\left(r_{\rm bdry}^0\right)\,.
\end{equation}
Consequently, for the last term in \eqref{eq:bdry_term_corrections} to vanish it is sufficient that the vacuum value of the corrected scalars, which is of order $\sim r_{\rm bdry}^0$, is not modified, namely
\begin{equation}\label{eq:asymptbehaviour}
X^{(1)}{}^I= \mathcal O\left(r_{\rm bdry}^{-2} \log r_{\rm bdry}^2\right)\,.
\end{equation}

In order to understand if this condition is satisfied, we need to determine the $\mathcal{O}(\alpha)$ corrections to the value of the scalars in the supersymmetric AdS$_5$ solution. We thus set $A_\mu^I=0$, choose the AdS$_5$ metric, and take constant scalars $X^I =  \bar X^I + \alpha \, \bar X^{(1)}{}^I$. 
By studying the scalar field equations it is possible to show that the terms in the Lagrangian proportional to $\lambda_I$ do not correct the vacuum value of the scalars, whereas the terms depending on $\widetilde\lambda_{IJK}$ imply the following relation\footnote{We interpret this as a corrected supersymmetric AdS$_5$ vacuum condition. Indeed one obtains the same relation by starting from the two-derivative theory with tilded variables in \eqref{eq:tilded_Lagr}, writing the supersymmetric vacuum condition \eqref{scalars_AdS_sol} in this theory, that is $g_I = \sqrt 2\, \widetilde g\, \widetilde{C}_{IJK}\widetilde X^J\widetilde X^K$, and transforming the latter to the present variables using \eqref{eq:subtildeaction1}, \eqref{eq:subtildeaction2} and~\eqref{eq:tildeg}.}
\begin{equation}
g_I = \sqrt{2}\,g\left[\,\bar a_{IJ}\left(\bar X^J + \alpha \, \bar X^{(1)}{}^J\right)\left(1-\alpha \,\widetilde\lambda_{KLM}\bar X^K \bar X^L\bar X^M\right) +\alpha\,\widetilde\lambda_{ILM}\bar X^L \bar X^M \right]\,.
\end{equation}
Therefore, the corrections to the scalars in the supersymmetric vacuum have the form
\begin{equation}\label{corr_to_scalars}
\bar X^{(1)}{}^I= -\bar a^{IJ}\left(\delta_J{}^K-\bar X_J  \bar X^K \right)\widetilde\lambda_{KLM}\bar X^L \bar X^M\,.
\end{equation}
Hence, our condition \eqref{eq:asymptbehaviour} is satisfied only for a choice of coupling $\widetilde\lambda_{IJK}$ such that 
\begin{equation}\label{eq:secondinvariantcoefficient}
\left(\delta_I{}^J -  \bar X_I \bar X^J  \right)\widetilde\lambda_{JKL}\bar X^K \bar X^L=0\,.
\end{equation}

In fact the corrections we study, specified in \eqref{choice_tildelambda}, do satisfy \eqref{eq:secondinvariantcoefficient}. This proves that it is legitimate to use the two-derivative solution to evaluate the corrected action.

We conclude this section commenting on the field theory interpretation of \eqref{corr_to_scalars}, \eqref{eq:secondinvariantcoefficient}. Let us first consider just the effect of the invariant controlled by ${\widetilde \lambda}_{IJK}$. Using the holographic dictionary above, this means the anomaly coefficients $k_{I}$ vanish, while $k^{(1)}_{IJK}$ are left arbitrary, proportional to ${\widetilde \lambda}_{IJK}$. Then, one can show that $a$-maximization at next-to-leading order in the large-$N$ expansion (see appendix~\ref{amax_at_largeN}) imposes
\be
-2k^{(0)}_{IJK} {\bar s}^J \delta {\bar s}^{K}=(\delta_{I}{}^{J}-r_{I}{\bar s}^J)k^{(1)}_{JKL}{\bar s}^K{\bar s}^L\,.
\ee
This is precisely the field theory dual of \eqref{corr_to_scalars}, and from the above equation we see that \eqref{eq:secondinvariantcoefficient} corresponds to demanding that the coefficients $\bar s^I$ that determine the superconformal R-symmetry at leading-order do not receive corrections, $\bar s^I +\delta\bar s^I =\bar s^I$. In this case \eqref{eq:secondinvariantcoefficient} states that the coupling $\widetilde \lambda_{IJK}$ with one index projected on the flavour symmetries and the remaining two projected on the superconformal R-symmetry vanishes. 

Turning on the invariant controlled by $\lambda_I$ does not modify this discussion as its contribution to $k^{(1)}_{IJK}$ and $k_I$ automatically cancel each other when implementing $a$-maximization. This is reflected in supergravity by the fact that $\lambda_I$ does not correct the value of the scalars in the supersymmetric AdS$_5$ vacuum.

\subsection{Results for the supersymmetric on-shell action}\label{sec:resultsandmatch}

Despite the great simplification following from evaluating the corrected action on the uncorrected solution, the evaluation of the higher-derivative terms on the fields \eqref{eq:metric5d}, \eqref{eq:scalars5d} remains a technically challenging problem. We have not been able to perform a fully analytic computation of the four-derivative on-shell action in the most general case, so we have resorted to a mix of numerical and analytical checks of the result. 
This was done by assigning many different numerical values to the black hole parameters $\delta_I$, while keeping the remaining parameters $a$, $m$ (and $g$) analytic.
However,  in some limits we could carry out the full computation analytically: this was possible
 in the ungauged limit $g=0$ (to be discussed in section~\ref{sec:ungaugedlimit}) keeping all electric charges independent, and in the case where $\lambda_I =0$ and two of the three charges are set equal.

A non-trivial consistency check we have made on our result has been to verify that the Gibbs free energy $\mathcal G = I/\beta$, vanishes in our supersymmetric and extremal limit. This means that the action $I$ can (and will) be finite. An extremal but non-supersymmetric expression for $I$ would instead give a finite $\mathcal G$ as the $\mathcal{O}(\beta)$ terms on the right hand side of \eqref{eq:qsr} would fail to cancel between them. Here, it is important to recall from section~\ref{sec:blackhole} that our definition of $I$ is such that $I = I_{Z} - I_{\rm AdS}$, and that the AdS action to be subtracted receives $\mathcal{O}(\alpha)$ corrections.
 For the case at hands we find
\begin{equation}
I_{\rm AdS} \,=\, \frac{3\pi\beta}{32Gg^2}\left( 1+ \alpha\, \widetilde\lambda_{IJK}\bar X^I\bar X^J\bar X^K\right) \,=\, \frac{3}{4}g\beta\, \mathtt a\,.
\end{equation}
In the last equality we used~\eqref{a_holo_corrected} to rewrite the expression in terms of the corrected Weyl anomaly coefficient  $\mathtt a$ of the dual SCFT and the ratio $g\beta$ of the $S^1$ and $S^3$ radii on $\partial\mathcal{M}$. 

After this consistency check, in order to obtain the supersymmetric action, following the procedure highlighted in section \ref{sec:BPSlimit}, we substitute $a$ with its supersymmetric value \eqref{eq:susycondition}, and define the supersymmetric chemical potentials as \eqref{eq:susychempot}. Then, we change coordinate using \eqref{eq:bpsradialcoord}, and trade the dependence on $m$ with that on $\zeta_+$ via \eqref{eq:defmplus}.
Again, we find that the supersymmetric on-shell action in this regime has a complicated dependence on the parameters of the solution, however when written in terms of the supersymmetric chemical potentials \eqref{eq:susychempot} it takes a simple form. This is independent of the value of $\beta$, namely it is valid both before and after taking the limit $\zeta_+ \rightarrow \zeta_*$.  It reads 
\begin{equation}
\begin{aligned}
I 
\,&=\, \frac{\pi}{G}   \frac{\varphi_{\rm gr}^{1} \varphi_{\rm gr}^{2} \varphi_{\rm gr}^{3}}{\omega^{2}}     -\frac{2\pi}{G}\alpha\lambda_I \varphi_{\rm gr}^{I} \,\frac{\omega  \mp 8\pi i }{\omega}\,,
\end{aligned}
\end{equation} 
which perfectly agrees with the field theory prediction~\eqref{eq:gravityactionprediction} (in its simplified form~\eqref{eq:gravityactionpredictionBis}) after setting $\omega_1=\omega_2=\frac{\omega}{2}$ and using the identifications in section~\ref{sec:special_relations}.
It follows that the corrected  entropy for the supersymmetric extremal black hole under study is given by the Legendre transform of this expression, given in \eqref{eq:entropy_orbifolds_eqJ}.

\paragraph{More general validity of our results.}
We have also computed the supersymmetric on-shell action for a more general choice of higher-derivative couplings $\lambda_I$, $\widetilde \lambda_{IJK}$, consisting of those which do not require knowledge of the corrected solution to compute the action at $\mathcal O(\alpha)$. As explained in section \ref{sec:setupaction}, these are the corrections that satisfy \eqref{eq:secondinvariantcoefficient}. In these computations, we employ the same set of counterterms \eqref{eq:generalcounterterms} to remove the divergences.
Again, we have performed extensive numerical checks of the final form of the on-shell action.\footnote{Specifically, we define $\lambda_I = \alpha_1\,A_I$, $\widetilde\lambda_{IJK} = \alpha_2\, B_{IJK}$, for some constant parameters $\alpha_{1,2}$ and where the entries of $A_I$ and $B_{IJK}$ are randomly generated integers. The black hole parameters $\text t_I$ are chosen as randomly generated primes. The computation is still analytic with respect to the remaining black hole parameters and $g$.}  
For all choices that satisfy \eqref{eq:secondinvariantcoefficient}, we have obtained that the supersymmetric action can be written as
\begin{equation}
I \,=\, \frac{\pi}{G} \left(\frac16|\epsilon_{IJK}| - 6\alpha\,\lambda_{I}g_{J}g_{K} +\alpha\,\widetilde\lambda_{IJK} \right) \frac{\varphi_{\rm gr}^{I} \varphi_{\rm gr}^{J} \varphi_{\rm gr}^{K}}{\omega^{2}}  +\frac{2\pi}{G}\alpha\lambda_I \varphi_{\rm gr}^{I} \,\frac{\omega^{2}-8\pi^2}{\omega^{2}}\,.
\end{equation}  
This supports the expectation that the holographic match we have been discussing in this work should hold beyond the specific example considered.

\section{Supersymmetric on-shell action for asymptotically flat black holes}\label{sec:ungaugedlimit}

Supersymmetric non-extremal sections of black hole solutions, of the type first discussed in~\cite{Cabo-Bizet:2018ehj} for the asymptotically AdS case, have recently proven useful also in the context of microstate counting for asymptotically flat solutions in four \cite{Iliesiu:2021are,Hristov:2022pmo,H:2023qko,Boruch:2023gfn} and five~\cite{Anupam:2023yns} dimensions, where they are interpreted as Euclidean saddles of the gravitational path integral computing a string theory supersymmetric index. The role of higher-derivative corrections in this context has been emphasized very recently in~\cite{Chowdhury:2024ngg,Chen:2024gmc}. Motivated by these recent developments, in this section we consider asymptotically flat black holes in ungauged higher-derivative supergravity and compute the supersymmetric on-shell action. This follows essentially the same steps as the asymptotically AdS case discussed in the previous section. Therefore we keep the presentation short, emphasizing the relevant differences. 

The bulk Lagrangian contains two contributions, 
\begin{equation}
{\cal L}_{\rm ungauged}\,=\,{\cal L}_{2\partial}^{(g=0)}+\alpha \,{\cal L}_{4\partial}\,,
\end{equation}
where ${\cal L}_{2\partial}^{(g=0)}$ is the two-derivative Lagrangian \eqref{eq:2dSUGRA} with $g_I=0$ and ${\cal L}_{4\partial}$ is the one given in \eqref{eq:4_der_terms_final}. As in the previous section we choose the U(1)$^{3}$ model specified by $C_{IJK} = \frac{1}{6}|\epsilon_{IJK}|$, therefore $a_{IJ} =\frac{1}{3\left(X^I\right)^2}\delta_{IJ}$,  with $I=1,2,3$. The higher-derivative coupling constants $\lambda_I$ are left arbitrary; in a given string compactification scenario these should  match a specific set of $\alpha'$ corrections. By taking the $g\to 0$ limit of the solution given in eqs.~\eqref{eq:metric5d} and \eqref{eq:scalars5d}, one obtains a solution to the two-derivative equations of motion of the ungauged theory. The latter are simply obtained by setting $g_I=0$ in eqs.~\eqref{eq:2dEOMEinst}, \eqref{eq:2dEOMvec} and \eqref{eq:2dEOMsc}. This solution corresponds to the asymptotically flat rotating black hole carrying three electric charges found in \cite{Cvetic:1996xz}, in the limit where the two rotation parameters have been set equal.
In order to reach the supersymmetric locus, we first redefine the parameters of the solution as follows,
\begin{equation}\label{eq:ungaugedparams}
\text t_I = \frac{q_I}{\sqrt{m}}\,\,,\quad\quad a= \frac{b}{\sqrt{m}}\,,
\end{equation}
and then take the limit 
\begin{equation}\label{eq:ungaugedsusycondition}
m\rightarrow 0\,
\end{equation} 
while keeping the new parameters $q_I,\,b$ fixed.
In this limit the conserved quantities \eqref{eq:conservedcharges2d} remain finite and it is immediate to show that they satisfy the supersymmetric linear relation
\begin{equation}
E \,=\,  Q^{\rm gr}_1+Q^{\rm gr}_2+Q^{\rm gr}_3\,,
\end{equation}
which is the correct supersymmetric condition coming from the ungauged superalgebra. The angular momentum is non-vanishing and is proportional to the rotation parameter $b$. 
The position of the horizon is found by solving the equation $Y_{(g=0)}(r_\pm)=0$, which, after imposing supersymmetry, reads $r_\pm^4 + 2 b^2=0$. The solution is, therefore, non-extremal for $b\neq 0$, but the Lorentzian black hole is pathological and $r_+^2$ becomes imaginary. After Wick-rotating the time as $t=-i\tau$, one may also Wick-rotate the rotational parameter as $b\rightarrow  ib_E$, with $b_E$ real, which gives a family of Euclidean supersymmetric non-extremal black hole solutions with a real metric, capping off at $r^2 \to r_+^2= \sqrt 2 |b_E|$. In the extremal limit $r_+\rightarrow 0$ we recover the supersymmetric three-charge black hole of \cite{Cvetic:1996xz} in the static case (as we are taking the two rotation parameters equal).
The supersymmetric and extremal values of the chemical potentials \eqref{eq:thermalchemicalpotentials2d} read
\begin{equation}
\Omega^*=0\,\,,\quad\quad \Phi^{*\,I}= 1\,\,.
\end{equation}
The supersymmetric chemical potentials $\omega$, $\varphi^I_{\rm gr}$, which are defined as in the previous section, are given by
\begin{equation}
\omega=\beta(\Omega-\Omega^*)= \pm 2\pi i\,\,,\quad\quad \varphi^I_{\rm gr} = \beta \left(\Phi^I - \Phi^{*\,I}\right)= -\frac{\pi}{\sqrt{2}}\frac{q_J q_K}{q_I}\,\quad I\neq J\neq K\,,
\end{equation}
and as we can see they do not depend on $r_+$. Furthermore, the linear constraint satisfied by the supersymmetric chemical potentials in the AdS case simply reduces to
$\omega = \pm\, 2\pi i\,$.

Having specified the solution, we now turn to the computation of the Euclidean on-shell action. The ungauged four-derivative Euclidean action reads
\begin{equation}\label{eq:ungaugedaction}
I_{\rm ungauged} = -\frac{1}{16\pi G}\int\text d^5 x\,e\, {\cal L}_{\rm ungauged} + I_{\rm GH}\,,
\end{equation}
where $I_{\rm GH}$ is the generalized Gibbons-Hawking term defined in \eqref{eq:Igh} which renders the variational problem with Dirichlet boundary conditions well posed. The on-shell action \eqref{eq:ungaugedaction} diverges, just as the on-shell action of flat spacetime. To remove the divergences we implement the background subtraction approach of \cite{Gibbons:1976ue}. This amounts to the following limiting procedure. First, we evaluate the Euclidean on-shell action \eqref{eq:ungaugedaction} at a regulated boundary, $r=r_{\rm bdry}$. Then, we subtract the on-shell action of a flat background metric chosen so that the induced metric of the black hole and of the flat background match for large $r_{\rm bdry}$. Finally, we send the radial cutoff to infinity $r_{\rm bdry}\to \infty$ to obtain a finite Euclidean on-shell action. Following this procedure, we evaluate the Euclidean supersymmetric on-shell action of the solution we have just presented keeping $r_{+}$ finite, and working at linear order in the four-derivative corrections. Since this ungauged case is considerably simpler than the one studied in the previous section, we have been able to perform the whole computation analytically, finding that the supersymmetric on-shell action boils down the simple expression
\begin{equation}\label{eq:Iungauged}
I_{\rm ungauged}\,=\, -\frac{1}{4\pi G}\,\varphi^1_{\rm gr} \varphi^2_{\rm gr}\varphi^3_{\rm gr} +\frac{6\pi}{G}\alpha\,\lambda_I \varphi^I_{\rm gr}\,,
\end{equation}
which is nothing but the ungauged limit $g\to0$ of \eqref{eq:gravityactionprediction}, \eqref{eq:constraint_gravity} with $C_{IJK} = \frac{1}{6}|\epsilon_{IJK}|$ and $\omega_1=\omega_2=\pm \pi i$.

At this point it is straightforward to guess the expression for the supersymmetric on-shell action in the more general case where two independent rotation parameters are turned on in the asymptotically flat solution of \cite{Cvetic:1996xz}, also allowing for a generic $C_{IJK}$. In this case supersymmetry fixes $\omega_+ \equiv \omega_1+\omega_2 = \pm 2\pi i$, while leaving $\omega_- \equiv \omega_1-\omega_2$ free. The supersymmetric on-shell action should be obtained by taking the $g\rightarrow 0$ limit of \eqref{eq:gravityactionprediction}, \eqref{eq:constraint_gravity}, which gives
\begin{equation}\label{eq:SUSYungaugedaction}
\begin{aligned}
I_{\rm ungauged}\,&=\,\frac{\pi}{G}\, \frac{C_{IJK}\varphi^I_{\rm gr} \varphi^J_{\rm gr}\varphi^K_{\rm gr}}{\omega_+^2 - \omega_-^2}+\frac{2\pi}{G}\alpha\,\lambda_I \varphi^I_{\rm gr}\, \frac{3\,\omega_+^2+{\eta}\,\omega_-^2}{ \omega_+^2 - \omega_-^2}\,, \quad\quad \omega_+ = \pm 2\pi i\,,
\end{aligned}
\end{equation}
where $\eta=1$ in our case. We have checked by an explicit computation that this prediction is indeed correct for the case where the $\varphi^I_{\rm gr}$ are all equal, so that the two-derivative solution fits in minimal ungauged supergravity. The parameter $\eta$ has been introduced in order to compare with previous work in the literature, where the corrected black hole entropy has been obtained from the near-horizon extremal solution using Sen's entropy function formalism. Different expressions have been proposed due to different definitions adopted for the thermodynamic variables (in particular, the charges and the angular momenta), which are affected by the presence of non-gauge-invariant Chern-Simons terms in the Lagrangian, see e.g.\ \cite{Castro:2007ci,Banerjee:2011ts,Gupta:2021roy}. For instance, our correction to the supersymmetric on-shell action is in contrast with the entropy function provided in eq.~(1.3) of \cite{Gupta:2021roy}, which corresponds to our \eqref{eq:SUSYungaugedaction} with $\eta=0$.
We note however that in order to make a proper comparison we should identify the precise map between the thermodynamic variables we have been using, which are defined in the asymptotically flat solution, with those used in the near-horizon approach. We leave the clarification of this issue for future work.

We conclude this section by computing the corrected entropy via the Legendre transform of the supersymmetric on-shell action \eqref{eq:SUSYungaugedaction}. In order to do this we assume again  the validity of the gravitational analogue of \eqref{eq:cubic_relation}. Concretely, we demand the existence of a fully symmetric tensor $C^{IJK}$ such that
\begin{equation}
C^{IJK} C_{J(LM}C_{PQ)K} = \frac{1}{27}\delta^I{}_{(L}C_{MPQ)}\,.
\end{equation}
Following the same procedure as in section~\ref{sec:Legendre_transf_gen}, we find that the Legendre transform of \eqref{eq:SUSYungaugedaction} at leading order (i.e.\ for $\alpha=0)$ yields
\be
\begin{aligned}
 \mathcal{S}^{(0)}\,=&\, \,  {\rm{ext}}_{\left\{\Delta^I,\, \omega_{\pm},\, \Lambda\right\}}\left[  - I_{\rm ungauged}|_{\alpha=0}  -\omega_+ J_+-\omega_- J_- -\varphi^I_{\rm gr} Q_I-\Lambda\left(\omega_+ \mp 2\pi i\right)\right]\, \\[1mm]
=&\,\, 4\sqrt{\pi G}\, \sqrt{C^{IJK}Q_IQ_JQ_K - \frac{\pi}{4G}J_-^2} \,\mp\,  2\pi i J_+  \,,
 \end{aligned}
\ee
where $J_\pm = \frac{J_1\pm J_2}{2}$.
The reality condition $J_+ =0$ coincides with the extremality condition. This plays the same role as the more complicated non-linear constraint between the charges found in the asymptotically AdS case. Upon imposing this additional condition, the Legendre transform yields precisely the expression for the Bekenstein-Hawking entropy of the supersymmetric and extremal three-charge black hole found in~\cite{Cvetic:1996xz}, after using that $C^{IJK}Q_I Q_J Q_K=Q_1 Q_2 Q_3$, since $C^{IJK}= \frac{1}{6}|\epsilon^{IJK}|$ for the U(1)$^3$ model.

It turns out that the reality condition $J_+=0$ for the entropy is not modified once the four-derivative corrections are taken into account. Imposing it, we find that the corrected entropy is given by
\begin{equation}
\mathcal S = 4\sqrt{\pi \,G}\sqrt{C^{IJK}Q_I Q_J \left(Q_K + \frac{18\pi\alpha}{G}\,\lambda_K\right) - \frac{\pi}{4G}J_-^2\left( 1+ \frac{6\left(\eta + 3\right)\pi\alpha}{G}\frac{C^{IJK}Q_I Q_J \lambda_K}{C^{IJK}Q_I Q_J Q_K}\right)
}\,\,,
\end{equation}
where again we have $\eta =1$.

The result for the on-shell action given in this section provides further evidence that the five-dimensional black hole of \cite{Cvetic:1996xz} in the supersymmetric (extremal or non-extremal) limit allows for the interpretation as a saddle of a grand-canonical partition function computing a supersymmetric index. In particular, fixing the angular potential as $\omega_+ \equiv \beta \Omega_+= 2\pi i$ ensures that the gravitational partition function becomes a supersymmetric index independent of $\beta$, since it includes the term $\rme^{\beta\Omega_+ J_+}=\rme^{2\pi i J_+} =(-1)^F \rme^{-2\pi i J_-}$, where $F$ is the fermion number and $J_-$ commutes with the chosen supercharge, see e.g.\ the discussion in~\cite{Anupam:2023yns}. It would be interesting to further investigate on this relation with the supersymmetric index of a certain string theory, and the associated higher-derivative thermodynamics. We plan to come back to this in a future work.

\section{Conclusions}\label{sec:Conclusions}

In this paper we have considered asymptotically AdS$_5$ multi-charge black hole solutions to gauged supergravity including four-derivative corrections, and matched their supersymmetric on-shell action with the prediction from a flavoured Cardy-like formula in the dual SCFT$_4$. At the two-derivative level, the solutions considered belong to a ${\rm U}(1)^3$ gauged supergravity that is a consistent truncation of  type IIB supergravity  on $S^5$ or the orbifold $S^5/\Gamma$, dual to ${\rm SU}(N)$ $\mathcal{N}=4$ SYM or the $\mathbb{C}^3/\Gamma$ quiver theories. The corrections to the two-derivative action have been determined by imposing that the Chern-Simons terms  match the $1/N^2$ corrections to the SCFT 't Hooft anomaly coefficients, and then supersymmetrizing the Chern-Simons terms. In $\mathcal{N}=4$ SYM the corrections are very simple, as they just amount to the $N^2\to N^2-1$ shift in the cubic anomaly coefficient (here the correction is even exact). Therefore we mostly focused on the orbifold theories, which display more interesting corrections.  

  In the dimensional reduction from string theory, the corrections should arise from one-loop effects, which are known to generate Chern-Simons terms via the parity anomaly~\cite{Alvarez-Gaume:1984zst}. At least for some of the coefficients, including just the tower of supergravity KK modes in the loop may be enough; for the R-symmetry linear anomaly coefficient $k_R \sim (\aa-\cc)$ this was verified in~\cite{ArabiArdehali:2013jiu}. It would be nice to extend this study to the other supergravity couplings dual to 't Hooft anomalies, possibly in more general compactifications. Ideally, it should be possible, though very hard, to derive the whole corrected five-dimensional action presented in this paper by including quantum effects on top of the two-derivative consistent truncation. It may also be useful to recall that in field theory, the corrections to the 't Hooft anomaly coefficients can be understood as the consequence of decoupling a ${\rm U}(1)$ vector multiplet at each node of the quiver while going from the initial D3-brane theory with ${\rm U}(N)$ gauge factors to the infrared theory with ${\rm SU}(N)$ factors. One may ask if the corrections to the supergravity action can be understood via a similar decoupling phenomenon. In fact, in the case of type IIB supergravity on $S^5$ it was found in \cite{Bilal:1999ph} that the modes running in the loop responsible for the $N^2\to N^2-1$ corrections to the cubic Chern-Simons term are gauge modes; see also the discussion in~\cite{Beccaria:2014xda}. It would be interesting to study this effect more extensively, together with the general pattern of cancellation of perturbative quantum effects in the supersymmetric black hole background.

Explicit knowledge of the corrected solution was not needed in order to calculate the corrected black hole on-shell action, analogously to the minimal case studied in~\cite{Bobev:2022bjm,Cassani:2022lrk}.  Here we have specified the condition for this property to still hold in the presence of scalar fields, and interpreted it holographically as the superconformal R-symmetry not receiving corrections at next-to-leading order in the large-$N$ expansion.
The match of the action was limited to the simplifying case of equal angular momenta due to the intrinsic difficulty of dealing with higher-derivative terms in geometries with many parameters and little symmetry.  Extending it to the case of unequal angular momenta starting from the two-derivative solution given in~\cite{{Wu:2011gq}} should be conceptually straightforward, though computationally even more demanding. Another possible extension would be to turn on the baryonic charge in the black hole dual to $\mathbb{C}^3/\mathbb{Z}_\nu$ theories for even $\nu$ as discussed at the end of~section~\ref{sec:blackhole}.

We have also derived the corrected supersymmetric and extremal black hole entropy as a function of the conserved charges by taking the Legendre transform of the action, see section~\ref{subsec:entropy_orbifolds}. 
It would be interesting to perform a direct check of the expression for the corrected entropy, similarly to what was done  in~\cite{Cassani:2022lrk,Cassani:2023vsa,Cano:2024tcr} for the minimal gauged supergravity solution. This check is currently out of reach as it requires knowledge of the corrected solution, or at least of the corrected near-horizon extremal solution. The entropy could be computed from the Wald formula applied to the corrected solution, while the charges could be evaluated by generalizing the formulae of~\cite{Cassani:2023vsa} to the case where vector multiplet couplings are present. It should then be possible to express the entropy as a function of the charges and match the prediction we obtained in this paper.

Our results are expected to hold beyond the specific AdS/CFT dual pairs discussed here, and we have given some evidence in this sense.   It will be intriguing to explore further the expectation that, given a holographic $\mathcal{N}=1$ SCFT$_4$,  the higher-derivative five-dimensional gauged supergravity reproducing its 't Hooft anomalies admits an asymptotically AdS$_5$ supersymmetric black hole solution whose on-shell action matches the SCFT formula~\eqref{eq:index_asympt}. 
It will be also be interesting to investigate further the ungauged limit of our results and its relevance for the microstate counting of asymptotically flat black holes.

\section*{Acknowledgments} 
We would like to thank Marina David, Kiril Hristov, Noppadol Mekareeya, Mehmet Ozkan, Gabriele Tartaglino-Mazzucchelli, Sameer Murthy, Dan Waldram and Alberto Zaffaroni for discussions. 
DC is supported in part by the MUR-PRIN contract 2022YZ5BA2 - Effective quantum gravity. AR is supported by a postdoctoral fellowship associated to the MIUR-PRIN contract 2020KR4KN2 - String Theory as a bridge between Gauge Theories and Quantum Gravity, and partially by the INFN Sezione di Roma ``Tor Vergata''. AR further thanks the GSS group at the University of Padova for hospitality and the INFN Sezione di Padova for financial support.

\appendix

\section{$a$-maximization and its four-derivative 5d gravity dual}\label{sec:amaxim_section}

In this appendix, we discuss the gravity dual of $a$-maximization at first order in the higher-derivative corrections. We extend the analysis of~\cite{Tachikawa:2005tq,Hanaki:2006pj} by including the corrections controlled by $\widetilde\lambda_{IJK}$, which allows us to reproduce the most general corrections to the $\aa$ and $\cc$ Weyl anomaly coefficients.

\subsection{$a$-maximization and the large-$N$ expansion}\label{amax_at_largeN}

We start by relating the cubic and linear 't Hooft anomaly coefficients $k_{IJK}$ and $k_I$  to the Weyl anomaly coefficients $\aa$ and $\cc$ of the $\mathcal{N}=1$ SCFT, at first subleading order in the large-$N$ expansion. 

In any $\mathcal{N}=1$ SCFT, the Weyl anomaly coefficients can be expressed as~\cite{Anselmi:1997am}
\be\label{relacTrR}
\aa = \frac{3}{32}(3\,k_{RRR} - k_R)\,, \qquad \cc = \frac{1}{32}(9\,k_{RRR} - 5\, k_R)\,,
\ee
where $k_{RRR}={\rm Tr}\,Q_R^3$ and $k_R = {\rm Tr}\,Q_R$ are the cubic and linear 't Hooft anomaly coefficients for the superconformal R-symmetry $Q_R$.
The superconformal R-symmetry can be determined via the principle of $a$-maximization \cite{Intriligator:2003jj}. One starts from a trial R-charge $Q_R^{\rm trial}(s) = s^IQ_I$, where the coefficients $s^I$ in the linear combination of the charges need to respect the constraint 
\be\label{eq:constr_rs}
r_Is^I=1\,,
\ee 
so that the supercharge has the canonical R-charge $-1$ (recall our definition of $r_I$ in~\eqref{eq:comm_charges}). Then the values $\bar s^I$ such that $Q_R =\bar{s}^IQ_I $ is the exact superconformal R-charge are identified by maximizing the trial anomaly coefficient
\be\label{atrial}
\aa_{\rm trial}(s) \,=\,  \frac{3}{32}\left( 3k_{RRR}^{\rm trial}(s)  - k_R^{\rm trial} (s)\right) \,=\,    \frac{3}{32}\left(3k_{IJK}s^Is^Js^K   - k_Is^I\right)\,.
\ee

For a holographic $\mathcal{N}=1$ SCFT with a weakly-coupled gravity dual, we can study the problem of $a$-maximization in the large-$N$ expansion by recalling the expansion~\eqref{eq:expansion_k's} of the 't Hooft anomaly coefficients.
Suppose we have solved the maximization problem at leading order, i.e.\ we have found the values $\bar s^I$ such that the function $\aa^{(0)}_{\rm trial} = \frac{9}{32}k^{(0)}_{IJK}  s^Is^Js^K$ is maximized under the constraint $r_I s^I=1$ (here we are using the fact that $k_I$ is subleading in the large-$N$ expansion, which also gives $\aa^{(0)}=\cc^{(0)}$). When including the subleading terms  $k^{(1)}_{IJK}$ and $k^{(1)}_I$, the function $\aa_{\rm trial}$ in \eqref{atrial} will be maximized at some corrected values $\bar s^I + \delta s^I$.  However, the corrections $\delta s^I$ only contribute to the expression for the anomaly coefficient at quadratic order,  hence we immediately conclude that at first order,
\be\label{eq:maximized_a}
\aa  \,=\,   \frac{3}{32}\left(3k_{IJK} \bar s^I \bar s^J\bar s^K  - k_I \bar s^I\right)  + \ldots\,.
\ee
 This argument also allows us to express the anomaly coefficient $\cc$ at the same order,
\be\label{eq:c_for_maximized_a}
\cc  \,=\,   \frac{1}{32}\left(9k_{IJK} \bar s^I \bar s^J\bar s^K  - 5k_I \bar s^I\right) + \ldots\,.
\ee

Nevertheless, we find  it useful to implement $a$-maximization at first order in the corrections, as it provides some relations that will be needed in the main text. Before doing so, let us introduce the projectors on the vector subspaces parallel and orthogonal to any R-symmetry $Q_R^{\rm trial}(s)  = s^I Q_I$,
\be\label{projectors_Rsymm}
P_{\parallel\,I}{}^J = r_I s^J\,,\qquad\ P_{\perp\,I}{}^J = \delta_I{}^J - r_I s^J\,.
\ee  
When implementing these projections, we will denote by ``$R$'' the R-symmetry direction, and will append a tilde on all quantities whose indices are projected along the orthogonal flavour directions. For instance, we can decompose the charges  as $Q_I = r_I Q^{\rm trial}_R + \widetilde{Q}_I$, where the $\widetilde{Q}_I = P_{\perp\,I}{}^J  Q_J$ are flavour charges (as they commute with the supercharge $\mathcal{Q}$). Analogously, we can decompose the 't Hooft anomaly coefficients along the R-symmetry and flavour directions, for instance $P_{\perp\,I}{}^{I'} P_{\parallel\,J}{}^{J'} P_{\parallel\,K}{}^{K'} k_{I'J'K'}  =   \widetilde k_{ I RR}\, r_Jr_K$, and so on.

In order to study the extremization problem, it is convenient introduce a Lagrange multiplier $L$ imposing the constraint \eqref{eq:constr_rs}. Then, the ${\aa}_{\rm trial}$ function is given by
\be
{\aa}_{\rm trial}(s, L)\,=\, \frac{3}{32}\left(3k_{IJK}s^Is^Js^K   - k_Is^I\right)+L \left(r_I s^I-1\right)\,,
\ee
and the extremization equations read
\begin{eqnarray}
\frac{\partial {\aa_{\rm trial}}}{\partial s^I}\Big|_{s={\bar s}+\delta {\bar s}}&\,=\,&\frac{3}{32}\left[9k_{IJK}\left({\bar s}^J+\delta {\bar s}^J\right)\left({\bar s}^K+\delta {\bar s}^K\right)  - k_I\right]+L \,r_I=0\,,\\[1mm]
\label{eq:da/dlambda}
\frac{\partial {\aa_{\rm trial}}}{\partial L}\Big|_{s={\bar s}+\delta {\bar s}}&\,=\,&r_I \left({\bar s}^I+\delta {\bar s}^I\right)-1=0\, .
\end{eqnarray}
Contracting the first with ${\bar s}^I+\delta {\bar s}^I$ and using the second allows us to find the value of $L$. Substituting its value into the first equation, we arrive at
\begin{equation}\label{eq:extremization_a}
\left[\delta_{I}{}^{J}-r_I \left({\bar s}^J+\delta {\bar s}^J\right)\right]\left[9\,k_{JKL}\left({\bar s}^K+\delta {\bar s}^K\right)\left({\bar s}^L+\delta {\bar s}^L\right)-k_{J}\right]\,=\,0\,,
\end{equation}
which just says that
\begin{equation}
9\,{\widetilde k}_{IRR}= {\widetilde k}_{I}\, ,
\end{equation}
in agreement with \cite{Intriligator:2003jj}. Assuming the large-$N$ expansion of the anomaly coefficients reported in \eqref{eq:expansion_k's}, we find that \eqref{eq:extremization_a} at leading order in $N$ reduces to 
\begin{equation}\label{eq:extremization_a_leading_order}
{\widetilde k}^{(0)}_{IRR}\equiv\left[\delta_{I}{}^{J}-r_{I} {\bar s}^J\right]k^{(0)}_{JKL}{\bar s}^K{\bar s}^L\,=\,0\,,
\end{equation}
where the coefficients ${\bar s}^I$ satisfy the constraint $r_I{\bar s}^I=1$. 

Now we consider \eqref{eq:extremization_a} at next-to-leading order. First we note that, because of \eqref{eq:da/dlambda}, the $\delta {\bar s}^I$ satisfy
\begin{equation}
r_I\delta {\bar s}^I=0\,.
\end{equation}
Contracting \eqref{eq:extremization_a_leading_order} with $\delta{\bar s}^I$ and using the above constraint, we deduce
\begin{equation}\label{eq:kdsss}
{k}^{(0)}_{IJK}\delta{\bar s}^I{\bar s}^J{\bar s}^K\,=\,0\, ,
\end{equation}
which justifies why $\delta {\bar s}^I$ does not contribute to the first-order corrections of the anomaly coefficients $\aa$ and $\cc$, as we anticipated. Linearizing \eqref{eq:extremization_a} and using \eqref{eq:kdsss}, one finds that the coefficients $\delta {\bar s}^{I}$ satisfy the following equations
\be\label{eq:ds}
-18k^{(0)}_{IJK}\delta{\bar s}^J{\bar s}^{K}\,=\,\left(\delta_{I}{}^{J}-r_I {\bar s}^J\right)\left(9k^{(1)}_{JKL}{\bar s}^K{\bar s}^L-k^{(1)}_{J}\right)\,,
\ee
where only $n$ are independent, as the contraction with ${\bar s}^{I}$ yields a trivial identity.

In particular, this tells us that $\delta {\bar s}^I=0$ (provided the matrix $k^{(0)}_{IJK}{\bar s}^K$ is invertible) whenever the anomaly coefficients associated to the flavour directions are not corrected. This is precisely the case of the quiver theories considered in sections~\ref{sec:Legendre_transf_gen} and \ref{sec:orbifold_section}, whose anomaly coefficients satisfy \eqref{eq:exp_anomaly_coeff}.

\subsection{Supergravity dual}\label{app:sugradual}

It was demonstrated in~\cite{Tachikawa:2005tq} that the gravity dual of $a$-maximization at the two-derivative level is the extremization of the supergravity prepotential, with the trial $a$-function being proportional to $\frac{1}{({\rm prepotential})^{3}}$; this is equivalent to the conditions for a supersymmetric AdS$_5$ solution. In the following we briefly review the argument, applied to the  U(1) gauging with constant parameters $g_I$ we have been considering in this paper, and provide the $\aa$ and $\cc$ Weyl anomaly coefficients at linear order in the corrections. Our new ingredient is that we include the corrections controlled by $\widetilde\lambda_{IJK}$.

\paragraph{Supersymmetric AdS$_5$ vacuum.}
We start by briefly recalling the conditions for a supersymmetric AdS$_5$ solution in the two-derivative theory. We consider the supergravity theory~\eqref{eq:finalL}, with $\alpha=0$. The susy transformation of the gravitino reads 
\be
\begin{aligned}\label{gravitino_var}
\delta\psi_\mu^i \,&=\, \nabla_\mu \epsilon^i + \frac{3}{ 2\sqrt{2}} g_I A_\mu^I \delta^{ij}\epsilon_j + \frac{i }{2\sqrt{2}} g_IX^I \delta^{ij}\gamma_\mu\epsilon_j + \ldots  \,,
\end{aligned}
\ee
where the ellipsis denote additional terms that will not be important for the present discussion. 
In order to find an AdS solution we set  $A^I_\mu = 0$ and take  scalars with constant values $X^I = \bar X^I$. 
The gaugino variation requires  that the ``prepotential'' function $g_I X^I$ is extremized with respect to the unconstrained scalars $\phi^x$, $x=1,\ldots,n$, that is 
\be\label{eq_gaugino_var}
g_I  \left.\frac{\partial X^I}{\partial\phi^x} \right|_{X=\bar X} = 0\,.
\ee 
Recall that the constraint satisfied by the scalars can be written as $X_I X^I=1$, with $X_I = C_{IJK}X^JX^K$. 
Then~\eqref{eq_gaugino_var} is solved by requiring that the vectors $g_I$ and $\bar X_I$ are aligned, 
\be\label{scalars_AdS_sol}
g_I =  \sqrt 2\, g \bar X_I  \qquad \Rightarrow \qquad g = \frac{1}{\sqrt2} \,g_I\bar X^I\,.
\ee
The vanishing condition  of the gravitino variation then becomes
\be
\nabla_\mu \epsilon^i + \frac{i }{2} g\,  \delta^{ij}\gamma_\mu\epsilon_j \,=\,0\,,
\ee
which is the standard Killing spinor equation in an AdS spacetime with squared radius $\ell^2= 1/g^2$.

\paragraph{Dictionary with field theory quantities.} Studying the gravitino variation \eqref{gravitino_var}, one can see that the supersymmetry spinor parameter at the AdS boundary is only charged (with charge +1 in our conventions) under the symmetry gauged by the linear combination\footnote{See the analysis of~\cite{Cassani:2012ri}, where the bulk susy transformations were expanded near the boundary so as to identify the susy transformations of the non-dynamical conformal supergravity coupling to the field theory at the boundary. In order to compare with that reference, one should use $\delta^{ij}\epsilon_j=-\varepsilon^{ij}\epsilon^j$.
}
\be\label{rel_Acan_AI}
A^{\rm can} \,=\, \frac{3}{2\sqrt2}\, g_I A^I\,.
\ee 
It follows that the spinor parameter has charge  $\frac{3}{2\sqrt2} g_I $ under $A^I$, hence recalling \eqref{eq:comm_charges} we should identify
\be
 r_I =  \frac{3}{2\sqrt2} \frac{g_I}{g}  \,,
\ee
We are going to show below that in the two-derivative supergravity the coefficients $\bar s^I$ determining the superconformal R-symmetry are given by
\be\label{identif_s}
\bar s^I =   \frac{2\sqrt2}{3} \frac{g\bar X^I} {g_J \bar X^J}\,.
\ee
Notice that the identifications  above satisfy the condition $r_I \bar s^I=1$. Using the supersymmetry condition \eqref{scalars_AdS_sol}, the two expressions above can also be written simply as
\be\label{eq:rs_and_X}
 r_I =  \frac{3}{2}  \bar X_I \,,\qquad \quad
\bar s^I =   \frac{2}{3} \bar X^I\,,
\ee
hence the field theory quantities $r_I$ and $\bar s^I$ are related to each other by the field theory dual of the supergravity matrix $\bar a_{IJ}$.

One way to prove \eqref{identif_s} is to look at the supergravity definition of the conserved charges.
Indeed, recalling \eqref{rel_Acan_AI} and the supersymmetric AdS condition~\eqref{scalars_AdS_sol}, we can express
\be
A^I = \frac{2}{3g} \bar X^I A^{\rm can} + (\ldots)^I\,,
\ee
where $(\ldots)^I$ has vanishing contraction with $g_I$ and contains the combinations of gauge fields independent of $A^{\rm can}$. Then the superconformal R-charge is given by integrating the time  component of the R-current obtained by varying the renormalized on-shell action as:
\be
Q_R \,=  \  \chi \int_{\partial \mathcal{M}}\diff^4x\,\frac{\delta S}{\delta A^{\rm can}_t} \,=\, \frac{2}{3g} \bar X^I \chi \int_{\partial \mathcal{M}}\diff^4x\,\frac{\delta S}{\delta A^{I}_t}  \,= \, \frac{2}{3g} \bar X^I  Q_I^{\rm gr}\,,
\ee
where $\chi$ is a coefficient entering in the definition of the charges, that only appears in the intermediate steps. 
Identifying the relation  $Q_R = \frac{2}{3g} \bar X^I  Q_I^{\rm gr}$ thus obtained with the field theory relation $Q_R = \bar s^I Q_I$,  and noting that $Q^{\rm gr}_I = g \,Q_I$ (since $A^I = \hat A^I/g$) we conclude that we must take 
\be
\bar s^I \,=\, \frac{2}{3} \bar X^I  \,= \, \frac{2\sqrt2}{ 3} \frac{ g\bar X^I}{g_J\bar X^J}\,,
\ee
as we wanted to show.

Extending \eqref{identif_s} to hold outside the AdS fixed point, we have
that the leading-order trial anomaly coefficient can be expressed in gravitational variables as
\be
\aa_{\rm trial}^{(0)}(s) \,=\,\frac{9}{32} k^{(0)}_{IJK}s^Is^Js^K   \,=\, \frac{\pi}{2\sqrt2G(g_I X^I)^3}  \,,
\ee
where we used $
k_{IJK}^{(0)} \,=\,  \frac{ 3\pi }{2Gg^3} \, C_{IJK}$ from the dictionary~\eqref{dict_cubic_anom}. This is extremized at the  AdS solution, since the prepotential $g_I X^I$ is; one can also see that it is in fact maximized as a consequence of positive definiteness of the metric on the supergravity scalar manifold~\cite{Tachikawa:2005tq}. The value at the maximum is
\be
\aa^{(0)} = \frac{\pi}{8G g^3}\,,
\ee
which agrees with the result from the analysis of the holographic Weyl anomaly~\cite{Henningson:1998gx}.

\paragraph{Including the first-order corrections.} 
The same argument  used in section~\ref{amax_at_largeN} allows us to extend the result above to include the higher-derivative terms at linear order in $\alpha$.  Namely, as long as we work at first order, we do not need to extremize again, we just have to translate the corrected expressions \eqref{eq:maximized_a}, \eqref{eq:c_for_maximized_a}
in gravitational variables. For $\aa$ we obtain
\be
\aa  \,=\,   \frac{\pi}{8Gg^3} \,C^{(\alpha)}_{IJK} \bar X^I\bar X^J\bar X^K  + \frac{3\pi}{2Gg}\, \alpha\lambda_I \bar X^I     + \mathcal{O}(\alpha^2)\,,
\ee
where we used~\eqref{dict_cubic_anom},  \eqref{dict_linear_anom} for the anomaly coefficients  and \eqref{eq:rs_and_X} for the coefficients $\bar s^I$. Recalling the form of the corrected $C^{(\alpha)}_{IJK}$ in \eqref{correctedCIJK} and \eqref{scalars_AdS_sol}, this becomes
\be\label{a_holo_corrected}
\aa  \,=\,   \frac{\pi}{8Gg^3} \,  \left(1  + \alpha \widetilde\lambda_{IJK}\bar X^I\bar X^J\bar X^K \right)        + \mathcal{O}(\alpha^2)\,.
\ee
Notice that the four-derivative invariant controlled by $\lambda_I$ does not contribute to $\aa$. By setting $\widetilde\lambda_{IJK}=0$ we recover the findings of~\cite{Hanaki:2006pj}, where only the Weyl$^2$ invariant (i.e.\ the one controlled by our coefficients $\lambda_I$) was considered. Here we see that the corrections controlled by $\widetilde\lambda_{IJK}$, which were not included in the analysis of~\cite{Hanaki:2006pj}, also contribute to the result.
Finally, recalling that 
\be
\cc-\aa \,=\, -\frac{1}{16}{\rm Tr}R \,=\,  \frac{\pi}{Gg}\, \alpha\lambda_I  \bar X^I   + \mathcal{O}(\alpha^2)\,,
\ee 
we obtain the holographic expression for the corrected $\cc$,
\be\label{c_holo_corrected}
\cc \,=\, \frac{\pi}{8Gg^3} \,  \left(1  +  \alpha \widetilde\lambda_{IJK}\bar X^I\bar X^J\bar X^K   + 8\alpha g^2 \lambda_I  \bar X^I     \right)    + \mathcal{O}(\alpha^2)\,,
\ee
which depends on both types of corrections.

\paragraph{Examples.} When discussing specific examples of AdS/CFT dual pairs, we need to choose the coefficients controlling the higher-derivative corrections in our five-dimensional effective theory in such a way that they match their field theory counterpart, according to the dictionary we have derived. For instance, in order to describe the gravity dual of $\mathcal{N}=4$ SYM we must fix 
\be
\text{$\mathcal{N}=4$ SYM}\quad\longleftrightarrow\qquad \widetilde\lambda_{IJK} =  - g^2 C_{IJK}\,, \qquad \lambda_I=0\,,
\ee
where the overall numerical coefficient is immaterial as it can be set to any other non-zero value by redefining $\alpha$.
We see that the $\widetilde\lambda_{IJK}$ coefficient plays a crucial role here.
 In this way our formulae give the Weyl anomaly coefficients $\aa=\cc=  \frac{\pi}{8Gg^3} \,  (1  -   \alpha g^2 )$; these match the exact expressions $\aa=\cc=\frac{N^2-1}{4}$ upon identifying  $\frac{\pi}{2Gg^3}  = N^2$ and $  \alpha g^2 = 1/N^2$.
 
If instead we wish to describe the quiver theories with 't Hooft anomaly coefficients of the form \eqref{eq:exp_anomaly_coeff}, including the $\mathbb{C}^3/\Gamma$ orbifold theories of section~\ref{sec:orbifold_section}, then we should  take 
\be
   \text{quivers satisfying  \eqref{eq:exp_anomaly_coeff}}\quad\longleftrightarrow\qquad \widetilde\lambda_{IJK} \,=\,   - 12   \lambda_{(I} g_J g_{K)} \,,\qquad \lambda_I = \frac{ g_I}{8\sqrt2 \,g} = \frac18\bar X_I\,,
\ee 
where again we have arbitrarily chosen a convenient overall normalization.
In this case the holographic Weyl anomaly coefficients read
\be
\begin{aligned}
\aa  \,&=\,   \frac{\pi}{8Gg^3} \,  \left(1  -3\alpha g^2   \right)        + \mathcal{O}(\alpha^2)\,,\\[1mm]
\cc \,&=\, \frac{\pi}{8Gg^3} \,  \left(1  - 2 \alpha g^2 \right)        + \mathcal{O}(\alpha^2)\,
\end{aligned}
\ee
and the dictionary with the field theory quantities is $\aa^{(0)}=\frac{\pi}{8Gg^3}$ and $\nu = \frac{2\pi \alpha}{Gg}$. In particular, for the $\mathbb{C}^3/\mathbb{Z}_\nu$ orbifold theories, $\aa^{(0)}= \frac{\nu N^2}{4}$, hence  $\frac{\pi}{2Gg^3}= \nu N^2$ and $\alpha g^2 = \frac{1}{4N^2}$.

\paragraph{Consistency check from Weyl anomaly.} As a consistency check, we can compare the formulae for $\aa$ and $\cc$ given above with those obtained from the analysis of the holographic Weyl anomaly in the presence of four-derivative terms, see e.g.~\cite{Fukuma:2001uf}.

In order to determine $\aa$ and $\cc$, we can ignore all gauge fields and fix the scalars to the value they take in the two-derivative AdS$_5$ solution, $X = \bar X$. This value receives corrections at linear order in $\alpha$, however these only affect the Lagrangian at $\mathcal{O}(\alpha^2)$ since the scalar potential of the two-derivative theory, $\mathcal{V}(X(\phi))$, is extremized with respect to the physical scalars $\phi^x$.  Then the Lagrangian at first order in $\alpha$ takes the form
\be
e^{-1}\mathcal{L} = \frac{1}{16\pi G}\left( R - 2 ( \mathcal{V}(\bar X) + \alpha\mathcal{V}^{(1)}(\bar X)) +  \alpha_1 R^2+\alpha_2 R_{\mu\nu}R^{\mu\nu} + \alpha_3 R_{\mu\nu \rho\sigma}R^{\mu\nu \rho\sigma} \right)\,,
\ee
where $\mathcal{V}^{(1)}$ is the correction to the scalar  potential dictated by four-derivative supergravity, and $\alpha_1,\alpha_2,\alpha_3$ are some coefficients. In our Lagrangian given in section~\ref{sec:final_Lagr}, the only term quadratic in the Riemann curvature after setting to zero the gauge fields is the Gauss-Bonnet term, and the coefficients read 
\be
\alpha_1= \alpha \lambda_I \bar X^I\,, \quad \alpha_2= - 4 \alpha \lambda_I \bar X^I \,, \quad\alpha_3= \alpha \lambda_I \bar X^I\,. 
\ee
 Recalling that $\mathcal{V}(\bar  X)=-6g^2$ and defining $g_{\rm eff}^2$ such that   
 \be
 -6 g_{\rm eff}^2 = (\mathcal{V}(\bar X) + \alpha\mathcal{V}^{(1)}(\bar X)) = -6 g^2 \left (1 - 10 \alpha g^2  \lambda_I \bar X^I  -\frac23 \alpha  \widetilde \lambda_{IJK}\bar X^I\bar X^J\bar X^K\right) \,,
 \ee
 we end up with
\be
e^{-1}\mathcal{L} = \frac{1}{16\pi G}\left( R + 12 g_{\rm eff}^2 +  \alpha_1 R^2+\alpha_2 R_{\mu\nu}R^{\mu\nu} + \alpha_3 R_{\mu\nu\rho\sigma}R^{\mu\nu\rho\sigma} \right)\,.
\ee
One also finds the corrected AdS radius, 
\begin{equation}\label{eq:corrected_AdS_rad}
\ell = \frac{1}{g}\left(1+ 4\alpha\,g^2\,\lambda_I \bar X^I + \frac{1}{3}\alpha\, \widetilde\lambda_{IJK}\bar X^I \bar X^J \bar X^K \right)\,.
\end{equation}
The analysis now continues similarly to the minimal gauged supergravity case discussed in  Appendix~A of~\cite{Cassani:2022lrk}. One eventually arrives at the expressions for $\aa$, $\cc$ obtained in \eqref{a_holo_corrected}, \eqref{c_holo_corrected}. These agree with those given in eq.~(A.11) of \cite{Cassani:2022lrk} upon identifying $\lambda^{\rm there}_1 = \lambda_I\bar X^I$ and $\lambda_2^{\rm there} = \frac14 \widetilde{\lambda}_{IJK}\bar X^I\bar X^J\bar X^K$.

\section{Off-shell four-derivative invariants}\label{app:offshell_invariants}

We collect here the expressions for the off-shell four-derivative invariants that we use in section~\ref{sec:fourder_action}, which are taken from \cite{Ozkan:2013nwa}. The expression for the Weyl-squared invariant is
\be
\label{eq:offshell_Weyl_squared}
\begin{aligned}
{\cal L}_{C^2}^{\rm{off-shell}}=\,&8\,\lambda_I\left[\frac{1}{8}X^I C^{\mu\nu\rho\sigma}C_{\mu\nu\rho\sigma}+\frac{64}{3}X^I D^2+\frac{1024}{9}X^I D \,T_{\mu\nu}T^{\mu\nu}-\frac{32}{3}D\,T_{\mu\nu}F^{I}{}^{\mu\nu}\right.\\[1mm]
&-\frac{16}{3}\, X^I\, C_{\mu\nu\rho\sigma}T^{\mu\nu}T^{\rho\sigma}+2C_{\mu\nu\rho\sigma}T^{\mu\nu}F^{I}{}^{\rho\sigma}+\frac{1}{16}\,\epsilon^{\mu\nu\rho\sigma\lambda}A^I_{\mu}C_{\nu\rho\alpha\beta}C_{\sigma\lambda}{}^{\alpha\beta}\\[1mm]
&-\frac{1}{12}\epsilon^{\mu\nu\rho\sigma\lambda}A^I_{\mu}V^{ij}_{\nu\rho}V_{ij}{}_{\sigma\lambda}+\frac{16}{3}Y^I_{ij}V^{ij}_{\mu\nu}T^{\mu\nu}-\frac{1}{3}X^I V^{ij}_{\mu\nu}V_{ij}^{\mu\nu}+\frac{64}{3}X^I\nabla_{\nu}T_{\mu\rho}\nabla^{\mu}T^{\nu\rho}\\[1mm]
&-\frac{128}{3}X^IT_{\mu\nu}\nabla^{\nu}\nabla_{\rho}T^{\mu\rho}-\frac{256}{9}X^IR^{\nu\rho}T_{\mu\nu}T^{\mu}{}_{\rho}+\frac{32}{9}X^I R \,T_{\mu\nu}T^{\mu\nu}\\[1mm]
&-\frac{64}{3}X^I\nabla_{\mu}T_{\nu\rho}\nabla^{\mu}T^{\nu\rho}+1024 X^I T^{\mu\nu}T_{\nu\rho}T^{\rho\sigma}T_{\sigma\mu}-\frac{2816}{27}X^I \left(T_{\mu\nu}T^{\mu\nu}\right)^2\\[1mm]
&-\frac{64}{9}T^{\mu\nu}F^{I}_{\mu\nu}T_{\rho\sigma}T^{\rho\sigma}-\frac{256}{3}T_{\mu\rho}T^{\rho\lambda}T_{\nu\lambda}F^{I}{}^{\mu\nu}-\frac{32}{3}\epsilon^{\mu\nu\rho\sigma\lambda}T_{\rho\alpha}\nabla^{\alpha}T_{\sigma\lambda}F^I_{\mu\nu}\\[1mm]
&\left.-16\epsilon^{\mu\nu\rho\sigma\lambda}T_{\rho}{}^{\alpha}\nabla_{\sigma}T_{\lambda\alpha}F^I_{\mu\nu}-\frac{128}{3}X^I\epsilon^{\mu\nu\rho\sigma\lambda}T_{\mu\nu}T_{\rho\sigma}\nabla^{\alpha}T_{\lambda\alpha}\right]\,,
\end{aligned}
\ee
where $V_{\mu\nu}^{ij}=2\partial_{[\mu}V^{ij}_{\nu]}-2V^{k(i}_{[\mu}V_{\nu]}{}_{k}{}^{j)}$ is the field strength associated to the SU(2) connection $V_{\mu}^{ij}$.

The $R^2$ invariant is given by
\be
\label{eq:offshell_R^2}
\begin{aligned}
{\cal L}_{R^2}^{\rm{off-shell}}=\,&\zeta_I \left(X^I {\underline Y}_{ij}{\underline Y}^{ij}+2 {\underline X} \,{\underline Y}^{ij}Y_{ij}^I-\frac{1}{8}X^I {\underline X}^2 R-\frac{1}{4}X^I {\underline F}_{\mu\nu}{\underline F}^{\mu\nu}-\frac{1}{2}{\underline X}\,{\underline F}^{\mu\nu}F_{\mu\nu}^I\right.\\[1mm]
&+\frac{1}{2}X^I \partial_{\mu}{\underline X}\partial^{\mu}{\underline X}+X^I {\underline X}\nabla^2 {\underline X}-4 X^I {\underline X}^2\left(D+\frac{26}{3}T_{\mu\nu}T^{\mu\nu}\right)+4 {\underline X}^2 F^I_{\mu\nu}T^{\mu\nu}\\[1mm]
&\left.+8X^I {\underline X}\, {\underline F}_{\mu\nu}T^{\mu\nu}-\frac{1}{8}\epsilon^{\mu\nu\rho\sigma\lambda}A^I_{\mu}{\underline F}{}_{\nu\rho} {\underline F}{}_{\sigma\lambda}\right)\,,
\end{aligned}
\ee
where $\zeta_I$ is an arbitrary dimensionless constant and where the underlined fields are given by 
\be
\begin{aligned}
{\underline X}=\,&2N\,, \\[1mm]
{\underline Y}^{ij}=\,&\frac{1}{\sqrt{2}}\,\delta^{ij}\left(-\frac{3}{8}R-N^2-P_{\mu}P^{\mu}+\frac{8}{3}T_{\mu\nu}T^{\mu\nu}+4D-{V'}^{kl}_{\mu}{V'}_{kl}^{\mu}\right)+2P^{\mu}{V'}_{\mu}^{ij}\\[1mm]
&-\sqrt{2}\nabla^{\mu}{V'}_{\mu}^{k(i}\delta^{j)}{}_{k}\,,\\[1mm]
{\underline F}{}_{\mu\nu}=\,& 2\sqrt{2}\partial_{[\mu}\left(V_{\nu]}+\sqrt{2}P_{\nu]}\right)\, .
\end{aligned}
\ee
Let us emphasize that the scalars $X^I$ in this appendix do not coincide with the ones in the main text, as they satisfy a modified cubic constraint when going on-shell. We refer to the discussion at the beginning of section~\ref{sec:4der_action} for a detailed explanation. 

\section{Field redefinitions}\label{app:field_redefinitions}

Perturbative field redefinitions of the form
\begin{equation}
g_{\mu\nu}\to g_{\mu\nu}+\alpha \,\Delta_{\mu\nu}\,, \hspace{1cm} A^I_{\mu}\to A^I_{\mu}+\alpha \,\Delta^I_{\mu}\,,
\end{equation}
are very helpful to simplify the final form of the Lagrangian, permitting us to reduce the number of terms. This is due to the fact that they allow us to eliminate terms proportional to $\alpha$ and to the two-derivative equations of motion. Therfore, this is equivalent to using the two-derivative equations of motion in the part of the Lagrangian multiplied by $\alpha$. Doing so, one finds the following set of replacement rules:
\begin{equation}\label{eq:redriccisquare}
\begin{aligned}
R_{\mu\nu}R^{\mu\nu}\rightarrow&\, a_{IJ}a_{KL}\left(-\frac{7}{16}\, F^I_{\mu\nu}F^J{}^{\mu\nu}  F^K_{\rho\sigma}F^L{}^{\rho\sigma}+ \frac{9}{4}\,F^I{}_{\mu\nu}F^{J\,\nu\rho}F^K{}_{\rho\sigma}F^{L\,\sigma\mu}- \frac{3}{4}\, \partial_{\mu} X^I\partial^{\mu} X^J \, F^K_{\rho\sigma} F^L{}^{\rho\sigma}\right.\\[1mm]
&\left.+\frac{9}{2}\,\partial_\mu X^I \,\partial^\nu X^J \,F^{K\,\mu\rho}F^L_{\nu\rho}\right) + \frac{9}{4}a_{K(I}a_{J)L} \,\partial_{\mu} X^I \partial^{\mu} X^J \partial_{\nu} X^K \partial^{\nu} X^L \\[1mm]
&+ \frac{20}{9}{\cal V}^2 + \frac{1}{3}{\cal V}a_{IJ}\,F^I_{\mu\nu} F^J{}^{\mu\nu}+ 2{\cal V}a_{IJ}\, \partial_{\mu} X^I \partial^{\mu} X^J\,\,,
\end{aligned}
\end{equation}
\begin{equation}\label{eq:redscalarcurvsquare}
\begin{aligned}
R^2\rightarrow& \,a_{IJ}a_{KL}\left[\frac{1}{16}F^I_{\mu\nu} F^J{}^{\mu\nu}F^K_{\mu\nu} F^L{}^{\mu\nu} + \frac{9}{4}\,\partial_{\mu} X^I \partial^{\mu} X^J \left(\partial_{\nu} X^K \partial^{\nu} X^L+ \frac{1}{3}F^K_{\rho\sigma} F^L{}^{\rho\sigma}\right)\right]\\[1mm]
&+\frac{100}{9}\,{\cal V}^2 +10\,{\cal V} \,a_{IJ}\, \partial_{\mu} X^I  \partial^{\mu} X^J+ \frac{5}{3}\,{\cal V}\,a_{IJ}\,F^I_{\mu\nu} F^J{}^{\mu\nu}\,,
\end{aligned}
\end{equation}

\begin{equation}\label{eq:redricciFF}
\begin{aligned}
R^\mu{}_{\nu}\,\mathcal F_{\mu\rho}\mathcal F^{\nu\rho} \rightarrow &
\,-\frac{1}{4}\,a_{IJ}\, F^I_{\mu\nu} F^J{}^{\mu\nu}\mathcal F^2 + \frac{3}{2}\,a_{IJ}\,\mathcal F_{\mu\nu}\mathcal F^{\nu\rho}F^I_{\rho\sigma}F^{J\,\sigma\mu}\\[1mm]
&+\frac{3}{2}\,a_{KL}\, \partial_\mu X^K\, \partial^\nu X^L\, \mathcal F^{\mu\rho}\mathcal F_{\nu\rho}+\frac{2}{3}\,{\cal V}\,\mathcal F^2\,\,,
\end{aligned}
\end{equation}

\begin{equation}\label{eq:redscalarcurvFF}
\begin{aligned}
R\, \mathcal F^2 \rightarrow& 
\,\left(\frac{1}{4}\,a_{IJ} F^I_{\mu\nu} F^J{}^{\mu\nu} + \frac{3}{2}\,a_{IJ}\, \partial_{\mu} X^I \partial^{\mu}X^J+\frac{10}{3}{\cal V}\right)\mathcal F^2\,.
\end{aligned}
\end{equation}

\begin{equation}\label{eq:rednablaFsquare}
\begin{aligned}
\nabla_\mu\mathcal F^{\mu\nu}\, \nabla^\rho \mathcal F_{\rho\nu} \rightarrow& -\frac{1}{8}\,{\mathcal C}_{IK} {\mathcal C}_{JL} \left(F^I_{\mu\nu} F^J{}^{\mu\nu}F^K_{\rho\sigma} F^L{}^{\rho\sigma}-2 F^I_{\mu\nu}F^{J\,\nu\rho}F^K_{\rho\sigma} F^{L\,\sigma\mu}\right) \\[1mm]
&+ 9a_{IK}a_{JL}\,\partial_\mu X^I\, \partial^\nu X^J\, F^{K\,\mu\rho}F^L_{\nu\rho} - 3{\mathcal C}_{IJ} a_{KL}\, \epsilon^{\mu\nu\rho\sigma\lambda} F^I_{\mu\nu} F^J_\rho{}^{\alpha}F^K_{\sigma\alpha}\,\partial_\lambda X^L\,,
\end{aligned}
\end{equation}

\begin{equation}\label{eq:redepsilonFFnablaF}
\begin{aligned}
\epsilon^{\mu\nu\rho\sigma\lambda} F^I_{\mu\nu} F^J_{\rho\sigma}\,\nabla^\alpha\mathcal F_{\alpha\lambda}
\rightarrow& \, {\mathcal C}_{KL} \left(F^I_{\mu\nu} F^K{}^{\mu\nu}F^J_{\rho\sigma}F^L{}^{\rho\sigma}- 2\, F^I_{\mu\nu}F^{K\,\nu\rho}F^J_{\rho\sigma} F^{L\,\sigma \mu}\right) \\[1mm]
&+12a_{KL}\, \epsilon^{\mu\nu\rho\sigma\lambda} F^{(I}{}_{\mu\nu}F^{J)}{}_{\rho\alpha}F_\sigma^K{}^{\alpha}\,\partial_\lambda X^L \,.
\end{aligned}
\end{equation}
Using them we can remove the last four terms in \eqref{eq:Weyl2}, as well as the Ricci-squared terms. However, instead of removing the latter, we are going to fix the coefficient in front of them in a way such that the Weyl-squared invariant is completed to the Gauss-Bonnet term ${\cal X}_{\rm{GB}}=R_{\mu\nu\rho\sigma}R^{\mu\nu\rho\sigma}-4R_{\mu\nu}R^{\mu\nu}+R^2$, just for convenience. The only price to pay is that we have to update the couplings of the four-derivative terms in \eqref{eq:Weyl2} and of the two-derivative corrections in \eqref{eq:DeltaL2dC2}. Finally, a perturbative constant rescaling of the metric allows us to fix to 1 the coefficient in front of the Ricci scalar, so as to be in the Einstein frame. The resulting Lagrangian that is obtained upon implementing these field redefinitions is the one reported in section~\ref{sec:final_Lagr}.

\bibliographystyle{JHEP}
\bibliography{HighDer5d}
\end{document}